\documentstyle[12pt]{article}

\newcommand{\be}{\begin{equation}}
\newcommand{\ee}{\end{equation}}
\newcommand{\bea}{\begin{eqnarray}}
\newcommand{\eea}{\end{eqnarray}}

\renewcommand{\theequation}{\arabic{section}.\arabic{equation}}

\begin{document}
\begin{titlepage}

%\flushright{To Appear: }

\vspace{1in}

%%%%%
\hfill {\bf YITP-99-62}\\
%%%%

\begin{center}
\Large
{\bf RECENT DEVELOPMENTS IN STRING THEORY}

%\vspace{1.5in}
\vspace{1in}

\normalsize

\large{ Jnanadeva  Maharana }\\
E-mail maharana@iopb.res.in
\vspace{.5in}

Institute of Physics$^*$ \\
Bhubaneswar - 751005 \\
India \\      
\vspace{.1in}
and \\
\vspace{.1in}
Yukawa Institute for Theoretical Physics\\
Kyoto University\\
Kyoto 606\\
Japan
\end{center}

\vspace{.5in}

\baselineskip=24pt
\begin{abstract}
The purpose of this short review is to present progresses in string
theory in the recent past. There have been very important developments in
our understanding of string dynamics, especially the nonperturbative aspects.
In this context, dualities play a cardinal  role. The string theory
provides a deeper understanding of the physics of special class of black holes
from a microscopic point of view and has  resolved several
important questions. It is also recognized that M-theory provides a unified
description of the five perturbatively consistent string theories. 
The article covers some of these aspects
and high lights important progress made in string theory.
\end{abstract}

\vspace{.1in}

\noindent $*$ Permanent address \hfill {August, 1999}

\end{titlepage}

%\double 
\centerline{\Large  {Contents}}

\vspace{1.5in}
\noindent {\bf1.  Introduction} \hfill{2} \\

\noindent {\bf 2. Perturbative Aspects of String Theory} \hfill {7} \\

\noindent {\bf 3. Duality Symmetries in String Theory } \hfill {19} \\

\noindent {\bf 4. M-theory and Unified String Dynamics} \hfill {37} \\

\noindent {\bf 5. Black holes and String Theory} \hfill {44}  \\

\noindent { \bf 6. M-theory and the Matrix model} \hfill {51} \\

\noindent {\bf 7. Anti-de Sitter Space and Boundary Field Theory 
Correspondence } 
 \hfill {59}\\

\noindent {\bf 8. Cosmology and String Theory } \hfill {66} \\

\noindent { \bf 9. Summary and Conclusions} \hfill {73} \\

\noindent {\bf 10. References } \hfill {75} \\
\newpage

\section{Introduction}

\setcounter{equation}{0}

\def\theequation{\thesection.\arabic{equation}}

All along the progress in natural philosophy, curious minds have asked 
deep  questions pertaining to the fundamental  constituents of 
matter and creation and evolution of the cosmos. In the modern era,
physicists have endeavored to comprehend natural phenomena in 
terms of a simple set of principles. Therefore, the search has 
continued to discover the elementary  constituents of matter and
 identify the fundamental forces responsible for the
natural phenomena. It is accepted that there are four fundamental 
forces  : gravitation, the weak interaction, electromagnetism 
and the strong interaction. The unification of fundamental 
interactions has remained as one of the most outstanding challenge for
generations of physicists. In the latter half of this century, some
progress has taken place in this direction through the electroweak
unification scheme. The electroweak theory together with quantum
chromodynamics (QCD), referred to as the standard model, 
have been tested to
a great degree of accuracy. Thus, the standard model provides a very
good description of the `low energy physics', comprising of the 
spectrum of elementary particles and  their dynamics. The next 
step in fulfilling the dream of unification of forces were the 
schemes of grand unifications (GUT) which attempted to incorporate
the three fundamental interactions, leaving aside gravitational
interaction. The  QED has been tested to a great degree
of accuracy and two most important characteristics of that theory are
the invariance under local gauge transformations and renormalizability.
The electroweak theory and  QCD respect the principle of gauge 
invariance and are renormalizable. Moreover, it is well known that
the Einstein's theory of general relativity respects a local symmetry:  
invariance
under general coordinate transformations. However, the theory is not
renormalizable since the Newton's constant carries dimension 
of $(mass)^{-2}$, 
 unlike
the gauge coupling constants of the standard model  which
are dimensionless.\\
Although the standard model has successfully passed many stringent
experimental tests, it is recognized that one must seek for a more
fundamental theory. The standard model has many arbitrary parameters:
the gauge coupling constants, the coupling constants of the scalars,
Yukawa couplings of the Higgs bosons to fermions which are eventually
responsible for generating 
fermion masses, just to mention a few. Furthermore, when
one extrapolates the gauge coupling constants utilizing the 
renormalization group equations towards higher energy scale, there are
evidences that the three coupling constants tend to converge to a point
and it is natural to conclude that beyond that scale  there might be 
a unified description of the standard model. 
 Therefore, these observations lend support to the
proposal of GUTS put forward in early seventies. 
 As is well known, the existence of electroweak scale in
the TeV region and another unification scale in the neighbourhood of
$10^{16}$ to $10^{17}$ GeV leads to issues related to fine tuning 
of parameters,  known as gauge hierarchy problem.
The gauge hierarchy problem can be resolved in an elegant manner if
one envisages supersymmetric version of the standard model (moreover,
the convergence of gauge coupling constants in the unifying scale
is more favourable in supersymmetric theories; see Mohapatra's article
in this volume for details). The supersymmetric theories were constructed
so that bosons and fermions can belong to a supermultiplet. The supersymmetry
appeared in 2-dimensions in the construction of string theories. 
While attempts were being made to construct various types of   
grand unified theories, there
were developments in incorporating gravity into supersymmetric theories
which resulted in discovery of  supergravity theories. However, it 
was not possible to construct renormalizable field theories which
could unify the four fundamental forces.
It was being perceived by many physicists, in the beginning of 
eighties, that new radical ideas were required to unify the fundamental
interactions.\\
It is now accepted that string theory holds the promise of unifying all
the fundamental interactions. The progress   of the  string theory in 
diverse directions,  during the last fifteen years have been truly 
spectacular. The theory has not only has brought us nearer to the
dream of unification, but also has influenced  our understanding of 
various aspects of quantum gravity and has  revealed many beautiful 
features relevant to the 
nonperturbative aspects of field theories. \\
The string theory was invented to describe the dynamics of strongly
interacting particles. The vast amount of experimental data amassed 
from high energy accelerators, during fifties and sixties
 led to discovery of large number of hadronic 
resonances. One of the interesting  characteristics of those resonances
was that when one
plots squared of mass vs spins of these particles, 
families of the resonances
tend to lie on a straight line, known as the Chew-Frautschi plot. It was also
evident from the high energy  of scattering cross sections of hadrons 
that they follow a power law behaviour i.e. the crossed channel Regge
poles controlled cross sections at high energies. The 
duality relation in strong interactions, that is sum over 
direct channel resonances (from low energy  data) reproduces  the
Regge amplitude, was an important discovery  for construction of dual models. 
Veneziano \cite{gv} took crucial step  step when he proposed a four point  
amplitude which  satisfied requirements of duality and crossing symmetry. 
\be 
T(s,t) = B(s,t) + B(t,u) + B(u,s) \ee
where 
\be
B(s,t) = {{\Gamma (-\alpha (s)) \Gamma (-\alpha (t)) }\over {\Gamma (
-\alpha (s) - \alpha (t))}} \ee 
and  $\alpha (s) = \alpha _0 + {\alpha }'s $ is the parameterization of the 
linear  Regge trajectory. Here s,t and u refer to the Mandelstam variables; 
when we are in the center of mass frame, s is the squared of c.m. energy,  
t and u are  related to the c.m. scattering angle. The B-function has the integral
representation
\be 
B(s,t) = \int _0 ^1 dw w^{-1-\alpha (s)}(1-w)^{-1-\alpha (t)} \ee
Subsequently, generalized N-point amplitudes satisfying 
requirements of duality and crossing
symmetry were proposed by several authors \cite{multiv}
and one such amplitude is
\bea
F(p_1.....p_N) = | w_I - w_{II}| | w_{II}-w_{III}||w_{III
}-w_I | \int dw_1 ...dw_N \Pi _{i<j} | w_i -w_j | ^{2\alpha ' p_i 
.p_j}  \eea
$w_i$ are ordered cyclically, $w_I, w_{II}$ and $w_{III}$ are any three of
the variables of the set $\{ w_i \}$, but are held fixed. As in the  case of
4-point Veneziano amplitude, the full N-point amplitude is sum of all cyclically
inequivalent permutations. It was realized that it is possible to represent
the N-point function in a path integral form \cite{pathf}  
\bea
F_N \sim \int \Pi _{\mu ,\sigma} dX^{\mu}(\sigma) 
\int dw_1 ...dw_N exp (-{T\over 2}\int _{\sigma
_2 >0} d^2\sigma \partial _aX^{\mu}\partial ^a X^{\nu}\eta _{\mu \nu}) \Pi ^N 
e^{ip_I.X(w_I)} \eea
where $\partial _a X^{\mu} = {{ \partial X^{\mu}} \over{\partial \sigma ^a}}$, 
$\sigma ^1$ and $\sigma ^2$ are coordinates of a point in the upper plane,
$X^{\mu}(\sigma)$ are integrated over all functions of $\sigma$. The boundary
condition on $X^{\mu}$ is $\partial _2 X^{\mu} =0$ for $\sigma ^2 =0$. The
constant $T= {1\over {2\pi \alpha '}}$ was later on identified as the tension of
the string. Note the presence of $X^{\mu}(w_I)$; it is the value of $X^{\mu}(
\sigma ^1,\sigma ^2)$ on the line $\sigma ^1 =w, \sigma ^2 =0$. The connection
between dual amplitudes and dynamics of a relativistic string was recognized
by several authors independently \cite{strml}.  
Now, of course we know that this amplitude is obtained from an open string 
theory
and the action is that of a  string, there are vertex 
operator in the path integral formula 
and the open string boundary conditions are to be specified. Virasoro had
constructed another 4-point amplitude \cite{vir} 
fulfilling the requirement of duality and
crossing symmetry in sequel to Veneziano's paper and generalization of that
amplitude for N-particle scattering was derived with a path integral 
representation \cite{jshap}. 
It was realized that the Virasoro-Shapiro amplitude could be
obtained from a closed string theory. Finally, Nambu proposed the action
for the string so that one could start studying the dynamics of the string and
proceed to examine the consequences of its quantization.\\
The string theory as a theory of strong interaction dynamics was not free from
shortcomings.  While attempts were going on to rectify the pitfalls of
the theory and to construct new string theories  as models of strong 
interactions; QCD was proposed as the fundamental theory of strongly
interacting particle. The theory described interactions of the fundamental
constituents, quarks, of the hadrons with gluon as the carrier of the force.
Furthermore, the experimental data confirmed predictions of QCD steadily and
consequently; string theory as a theory of strong interaction was no longer
in the center stage.\\
In 1974, Joel Scherk and John Schwarz \cite{jsjs}
made a bold proposition that string theory
should be envisaged as a theory of gravity since the massless spin two particle
appears naturally in the closed string spectrum and this theory might be a
vehicle to achieve the goal of unification of the forces of Nature. 
If string theory were to incorporate the gravitational interaction, then the 
string tension should be order of the Planck scale in contrast to the
the tension of the original string which was of the order of one GeV, the scale 
of hadronic interaction determined from the slope of the Regge trajectories.
It was realized that one has to go up nineteen orders of magnitude in the energy
scale if the Scherk-Schwarz proposal was to be realized. At that time, this 
radical idea did not receive wide spread acceptance amongst theoretical 
high energy physicists. The crucial work of Green and Schwarz \cite{mgjs}
in the summer of
1984 led to conclusion that 10-dimensional super Yang-Mills theories coupled
to supergravity can be consistently constructed and  are free
from all anomalies \cite{wia} only for 
the gauge groups $SO(32)$ and $ E_8 \times E_8$.
The results of Green and Schwarz had profound impact on the field of high
energy physics. It was recognized that string theory could fulfill the cherished
dream of unifying fundamental forces. The construction of the heterotic string
theory \cite{gfour} 
was a very important break through towards realization of this goal since
it had the desired gauge groups i.e. $SO(32)$ or $E_8 \times E_8$, depending
on the construction one adopted. The ten dimensional theory had chiral fermions,
$N=1$ supergravity coupled to  supersymmetric Yang-Mills with appropriate 
gauge groups. 
Moreover, when the $E_8 \times E_8$ heterotic string theory was compactified
to four dimensions on a Calabi-Yau manifold, the resulting theory was shown to
possess several desirable features that one expected from some of the 
grand unified
theories. Furthermore, it was possible to demonstrate
 that the standard model gauge
groups $SU(3) \times SU(2) \times U(1)$ were  contained in such
four dimensional theories. Indeed, optimistically, one could feel that a 
unified theory was in sight and string theory was popularly named as the
`Theory of Everything'.\\
Let us recapitulate some of the essential features of string theory. 
The string is a one dimensional object which executes motion in spacetime. There
are, grossly speaking, two types of strings: open and closed strings. 
As the name
suggests, the ends of open strings are free (there are special types whose ends
might get stuck to some surfaces and they play very important roles too) and
it is required to satisfy suitable boundary conditions for the end points. The
closed string, by definition, has its both ends glued together, forming a loop.
It is well known that when a point particle evolves in spacetime, it traces out
a trajectory describing its history. In case of an open string, it sweeps a two
dimensional surface and similarly the closed string sweeps a  surface which is
that of a cylinder. The natural question is why we do not observe these strings
in high energy collisions. The answer to this question lies in the fact that 
the strings are much smaller in size than the present accelerators can probe. If
we could have accelerators which have energies of the order of $10^{19}$ GeV,
then it will be  possible to observe the dynamics of the strings directly and
test the predictions of string theory at the Planckian energies. In
contrast, the present day accelerators have energies of the order of TeV - 
almost 16 orders of magnitudes below the string scale.\\
The string has tension and it vibrates in an infinite number of modes. We 
identify each mode of the string with a particle. 
Of course, the string will have
the lowest mode and we identify that with a particular particle. The next mode
will correspond to an excited state and it is separated in energy from the 
lowest mode in suitable unit of string tension - separation between two 
neighboring levels is order of $10^{19}$ GeV (recall that  for the
hadronic models they excitations were on Regge 
trajectories and there the tension
was order of GeV). The string theories of interests to us contain massless 
particles in their lowest mode. For example, in 10-dimensional heterotic string
theory, we have graviton, antisymmetric tensor and dilaton together with
the super Yang-Mills multiplets corresponding to the gauge groups $SO(32)$ or
$E_8 \times E_8$ in its massless sector. Therefore, in the  low energy limit,
the string theory effectively reduces to a point particle  field theory (this is
when we want to describe physics  at the present day accelerator energy scales).
In other words, the zero slope limits of string theories correspond to known
field theories - superstring theories go over to supergravity theories in this
limit. \\
Now we give an outline of the rest of the article. Since  it is to appear in a
volume on `Field Theory', we shall avoid involved technical details. The field
has progressed in diverse directions and our strategy will be to adopt a course
to high light important developments. We shall attempt to present different
aspects of string theory in a pedagogical manner. In order to get across 
some issues, known examples from field theory will be presented. There has
been intense activities in this field since 1984, when it was recognized that
string theory is the most promising candidate for unification of forces of
Nature. It is not possible to cover all the important literatures in a vibrant
field like this within the frame work of this article. I apologize in advance
to all the authors whose works have not been cited. There are two books which
cover all the important aspects of string theory in detail besides 
several monographs and reprint collection volumes. The first one
\cite{book1}, in two volumes provides foundation for string theory and
includes the developments up to 1986. The second one \cite{book2} has laid
the emphasis on the progress made after the second superstring revolution.
I have listed some of the review articles written in the first phase of the
developments of string theory \cite{o1,o2,o3,o4,o5,o6}. There are a large
number of review articles written in recent time 
\cite{senr1,r1,r2,r3,r4,r5,r6,r7,r8,r9,r10,r11}  
The next section
deals with a brief review of the perturbative aspects of string theory to 
familiarize the reader with well known results. First, the string worldsheet
action is introduced and the  symmetries of the action are listed. A very quick
exposition is given to the solutions of the equations of motion and mode
expansions for the string coordinates and essentials of the Virasoro algebra
are recalled. The evolution of the string in its massless background in the
first quantized approach is discussed and the consequences of conformal 
invariance are noted. Section III deals with the symmetries of string theory.
The theory is  endowed with a rich symmetry structure in the 
target space besides
the worldsheet symmetries. We introduce the duality symmetries since they play
a very important role in our understanding of the string dynamics in various
spacetime dimensions and they unravel the intimate connections between different
string theories. The subsequent Section, IV, is devoted to to discuss the 
recent efforts to unify string theories. Besides duality symmetry, spatially
extended objects, generically called p-branes, which appear as solutions to the
effective action, are crucial to our understanding of string dynamics
and to test some of the duality conjectures. 
We introduce some of the salient features of these objects
and provide simple examples of the solutions. The raison de etre for M-theory
is presented. We give an example how compactification of M-theory provides
connections with the string theories and their various brane content. The fifth
Section deals with issues related  to black holes that appear in string theory.
Since string theory describes gravity, it is  expected that the theory will be
able to provide insights into deep questions in quantum gravity. Indeed, some
of the issues in the physics of the black holes have been resolved by string
theory. It is known for more than two decades that a black hole is characterized
by entropy and  the Hawking temperature from the thermodynamic analogies. 
Moreover, the seminal work of Hawking demonstrated that the black holes radiate
when quantum effects are taken into account. Recently, the black hole entropy
has been computed  as a microscopic derivation in the frame work of string 
theory. Furthermore, the absorption cross sections for incident particles and 
the distribution of Hawking radiation emitted from special class of stringy 
black holes have been evaluated from a microscopic theory. 
Section VI contains a brief account of the M(atrix) model. The M(atrix) model
proposal to describe M-theory has drawn considerable attention. Some of the 
calculations in this model give surprising agreements with results of 
supergravity theories. 
Moreover, when one considers compactification of the model
on torii the resulting theory can be related to supersymmetric Yang-Mills 
theories through duality. We discuss some of the features of the Maldacena
conjectures in Section VII. According to the conjecture, in a concrete form,
if one considers N coincident 3-branes of type IIB theory on $AdS_5 \times S^5$
then the correlation functions of supergravity on the $AdS_5$ get related to
correlation functions of the $N=4$ super Yang-Mills theory living on the
boundary of the $AdS_5$. This is a rapidly developing area and we shall be
contented with some of the simple examples. In the last section we present an 
overview of the field. We make a few remarks to convey the reader how the work
in string theory has influenced research in other branches of physics.

%%%%%%THE NEXT SECTION %%%%%%%

\section{Perturbative Aspects of String Theory }

\setcounter{equation}{0}

\def\theequation{\thesection.\arabic{equation}}

We have outlined the historical backgrounds and the developments
of string theory in its early phase in the previous section. In
this section, we shall present some of the essential features
of string theory such as its quantization, the perturbative spectrum
of theory and the supersymmetric version of string theory.\\
Nambu had proposed the action for a string in analogy with the
action for a relativistic point particle: the action for point particle
in an integral over an line element; the string action is expected to be
an integral over a surface.\\
The Nambu-Goto action \cite{nam} was introduced almost three decades ago and has the form
\bea
S_{NG} = - T\int d^2\sigma \sqrt { ({\dot X}.X')^2 - ({\dot X})^2(X')^2} \eea
where $\sigma$ and $\tau$ are the coordinates on the surface swept out by the
the string, called `worldsheet'; ${\dot X}^{\mu} ={{\partial X^{\mu}}
\over {\partial \tau}}$ and $X'^{\mu} = 
{{\partial X^{\mu}}\over {\partial \sigma}}$
and we shall follow this definition all along unless specified otherwise. The
equations of motion can be derived after specifying boundary conditions for the
types of string one is dealing with. One important point to be noted is that
the theory described by the above action satisfies two constraints
\be
\Pi .X' =0, ~~~~  {\Pi } ^2 + T{X'} ^2 = 0 \ee
where $\Pi _{\mu} = {{\delta L}\over {{\dot X}^{\mu}}}$ is the canonical
momentum of $X^{\mu}$ obtained from this action. We reserve the 
notation $P_{\mu}$
for the canonical momentum of the coordinate derived from the Polyakov action.
We shall elaborate significance of these constraints later.\\
However, this form of action was not very convenient to deal with the
quantization of string and an alternative form of action was proposed
by Polyakov \cite{polya}
\bea
\label{poly}
S= -{T\over 2}\int d^2\sigma {\sqrt {-\gamma}}\gamma ^{ab}
\partial _aX^{\mu}\partial _bX^{\nu}\eta_{\mu \nu} \eea
$\gamma _{ab}$ is the worldsheet metric, $\gamma ^{ab}$ is 
its inverse, $\gamma$ is
determinant of worldsheet metric and $\eta _{\mu \nu}$ 
is the flat space metric of
the target space. The variation of the action 
with respect to $\gamma ^{ab}$ results
in the worldsheet stress energy momentum tensor
\bea
T_{ab} = \partial _aX.\partial _bX -{1\over 2} \gamma _{ab} \gamma ^{cd}\partial _cX.
\partial _dX \eea
$T_{ab} =0$, since there is no kinetic term i.e. as the 
analogue of Einstein-Hilbert
piece, 
 $\int d^2\sigma R^{(2)}$ is a topological term. We can solve for $\gamma _{ab}$
from the above equation
\be
\gamma _{ab} = \partial _aX^{\mu}\partial _bX^{\nu}\eta _{\mu \nu} \ee
If we insert the above  expression for worldsheet 
metric into the Polyakov action,  then
we recover Nambu's action.\\
The action (\ref{poly}) has following symmetry properties.

\noindent (a) Two dimensional reparameterization invariance
\bea
\delta \gamma _{ab}=\xi ^c\partial _c\gamma_{ab}+\partial _a\xi ^c\gamma _{bc}+
\partial _b\xi ^c\gamma _{ac} \eea
and hence $\delta {\sqrt {-\gamma}}=\partial _a(\xi ^a {\sqrt {-\gamma}})$. The
string coordinate transforms as
\be \delta X ^{\mu} = \xi ^a \partial _aX^{\mu} \ee

\noindent Weyl invariance
\be \delta \gamma _{ab} = 2\Omega \gamma _{ab} , ~~~~~ \delta X^{\mu} = 0 \ee

\noindent Poincare invariance (in target space)
\be \delta X^{\mu} = \omega ^{\mu}_{\nu}X^{\nu} + a^{\mu}, ~~~~~ 
\delta \gamma _{ab} =0 \ee
where  $\omega _{\nu}^{\mu}$ are  antisymmetric parameters associated with 
the Lorentz transformation  and $a^{\mu}$ are  the parameters  of
translation.\\
Note that the Weyl invariance implies tracelessness of the two dimensional
energy momentum tensor for the classical theory. The quantum invariance of
this symmetry has far reaching consequences in string theory. 

If we make the orthonormal gauge choice for the worldsheet metric 
$\gamma _{ab}=
e^{2\Omega (\sigma ,\tau)} \eta _{ab}$ with $\eta _{ab}={\rm diag}(-1,+1)$ the,
 form
of Polyakov action simplifies since ${\sqrt {-\gamma}}\gamma ^{ab} = \eta ^{ab}$
in this gauge. The condition of the vanishing of $T_{ab}$ reduces to two
constraints
\be ({\dot X} \pm  X')^2 =0 \ee
These are the Virasoro constraints. They take the following form 
in the Hamiltonian
formalism
\be 
\label{ham}
P_{\mu}X'^{\mu} =0, ~~~~~~ H={1\over 2}(P^2 + TX'^2) =0 \ee
where $P_{\mu}$ is momentum conjugate to $X^{\mu}$ derived from Polyakov
action.
It is easy to see that the first constraint generates $\sigma$ translation 
on the
worldsheet, whereas latter being the canonical Hamiltonian generates $\tau$
translation.\\
The equation of motion for the string coordinates, in the light-cone variables
$\xi _+ = \tau +\sigma$ and $\xi _-=\tau-\sigma$, are given by
\be \partial _+\partial _-X^{\mu} =0 \ee
We note that the equation of motion is derived with following 
boundary conditions:
(i) $X^{\mu} (\tau , \sigma +2\pi) = X^{\mu}(\tau ,\sigma)$ for the closed
strings, and (ii) $X'^{\mu} =0$ for $\sigma =0$ and $\sigma =2\pi$ 
in the case of
open strings, when we apply  the variational method.\\
Let us illustrate the mode expansion for the closed string starting 
from the equation of
motion with periodic boundary condition in $\sigma$. We first note that the 
string coordinate
can be decomposed as  a sum of left-moving and right-moving coordinates.
\be X^{\mu}(\tau ,\sigma) = X^{\mu}_L(\tau +\sigma) + 
X^{\mu}_R(\tau -\sigma) \ee
Then the two can be expanded as follows:

\be
X^{\mu}_L(\tau +\sigma) = {{x^{\mu}\over 2}}+
{{{p^{\mu}}\over {4\pi T}}}(\tau +\sigma)
+{i\over {{\sqrt {4\pi T}}}}\sum {{{\bar \alpha }^{\mu}_m}\over m}
e^{-im(\tau +\sigma)}
\ee

\bea
X^{\mu}_R(\tau -\sigma) ={{x^{\mu}}\over 2} +{{p^{\mu}}\over {4\pi T}}
(\tau -\sigma)
+ {i\over {{\sqrt {4\pi T}}}}\sum {{\alpha ^{\mu}_m}\over m}
e^{-im(\tau -\sigma)} \eea

The sum is over all integer values of m ($m=0$ is excluded) in the above equations.
$\alpha ^{\mu}_m$ and ${\bar \alpha }^{\mu}_m$ are the Fourier modes
\footnote{We have  adopted a shortcut route and taken the $\alpha 's$ to be
time independent. The proper procedure would be, since we are yet to
quantize, to allow them to be $\tau$-dependant and determine their
equations of motion from the Poisson bracket with the Hamiltonian
which in turn would  determine the $\tau$-dependence of the $\alpha 's$.
Thus, the systematic steps would have been to take combination of
$X$ coordinates and $P$'s as one does in the case of harmonic oscillator
go through the appropriate steps on this occasion also.}. Since $
X^{\mu}_{L,R}$ are real, so are $x^{\mu}$ and $p^{\mu}$; the Fourier modes satisfy
\be (\alpha ^{\mu}_m)^* = \alpha ^{\mu}_{-m}, ~~({\bar \alpha} ^{\mu}_m)^* =
{\bar \alpha} ^{\mu}_{-m} \ee
from the reality condition. For the closed string case, the classical  Hamiltonian
is given by
\be H={1\over 2} (\sum  \alpha _m .\alpha _{-m}  
+ \sum {\bar \alpha }_m{\bar \alpha }_{-m} ) \ee
in terms of the Fourier modes. The constraint, $T_{ab}=0$, obtained from the
Polyakov action, takes the form $T_{--} = {1\over 2} (\partial _-X)^2 =0$ and
$T_{++}= {1\over 2} (\partial _+X)^2=0$, in terms of the light-cone coordinates, after
one has gone over to the ON gauge. It is more convenient to express these constraints
in terms of the Fourier modes introduced above and define the Virasoro generators
\be L_m = {1\over 2}\sum \alpha _{m-n}\alpha_n, ~~{\rm and}~~
{\bar L}_m={1\over 2}\sum {\bar \alpha}_{m-n}{\bar \alpha}_n \ee
And
\be H= L_0 + {\bar L}_0 \ee
We can obtain classical Poisson bracket relations amongst $L_m$, similarly for
the set ${\bar L}_m$, starting from the canonical Poisson bracket between
$X^{\mu}$  and $ P_{\mu}$.
\be [L_m ,L_n ]_{PB} = -i(m-n)L_{m+n} \ee
\be [L_m ,{\bar L}_n]_{PB} =0 \ee
the PB between $\{{\bar L}_m\}$ is same as for $\{L_m\}$. These are classical
Virasoro algebra.\\ 
%%%% REMARKS ABOUT LOCAL SYMMETRY BRS ETC %%%%%%%%%%%
We have noted earlier that the string theory is endowed with local symmetries
in the worldsheet and the action is that of D-scalar fields in $1+1$ dimensions,
since $\mu =0,1... D-1$ takes D values. When we proceed to quantize  this theory,
we encounter problems similar to the one faced in quantization of gauge theory.
In other words we have to fix the gauge here too. One can choose to work in a
noncovariant gauge which has the advantage of dealing with physical  degrees
of freedoms directly, but at the price of losing 
manifest Lorentz covariance. On the
other hand one can adopt covariant quantization prescription 
with all its elegance
and power. The light-cone quantization, although noncovariant, is very useful
and gives us a physical picture. As the first step, the 
classical constraints are
solved and one is left with less number of variables Recall that there were some
remnant symmetries after choosing conformal (ON) gauge:
$ \xi _{+}' = \lambda _1 (\xi _+)$ and $ \xi _{-} '=\lambda _2 (\xi _{-})$. One
can utilize this property to write
\be X^+ = x^+ +\alpha 'p^+ \tau \ee
Defining the light-cone string coordinated 
$X^{\pm} = X^0 \pm X^1 $ one can impose
the classical Virasoro constraints 
$({\dot X} \pm X')^2 =0$. Thus $X^-$ is determined
in terms of the rest of the (transverse) coordinates, $X^i$; and in this process
both $X^+$ and $X^-$ are totally eliminated and we are left with $\{ X^i \}$. 
Then
the oscillators of these coordinates will create the states which could be
identified with particles with physical degrees of freedom only. So it gives us
a physical 
picture of the states. However, as mentioned earlier and as is the case with
noncovariant gauge fixing in QED or Yang-Mills theories, 
the Lorentz invariance must
be checked explicitly. For the case of string theory, one is required 
 to construct
the generators of Lorentz transformations and ensure  
that the generators satisfy the algebra.
It is well known that this requirement is not 
fulfilled unless the string propagates
in 26-dimensional spacetime. On the other hand, if one adopts the covariant BRST
procedure, it is necessary to add the ghost term to the action and construct the
corresponding Virasoro generators for the ghosts. 
Thus the full Virasoro generator
is a sum of the oscillators coming from string coordinates and 
those from the ghosts.
When we compute the quantum Virasoro algebra, there is an anomaly of 26 
from the ghost sector 
which gets precisely canceled if the spacetime dimension is 26 since each  
bosonic degrees of freedom contributes a factor of one to the anomaly with
a sign just opposite to that coming from ghosts.\\
%%%%% END OF REMARKS %%%%%%%%%%%%%%
There are infinite tower of states in string theory. It is useful to arrange
them according to their oscillator levels. Notice that the worldsheet degrees
of freedom of the string are envisaged as a collection of infinite number 
of harmonic oscillators. If we consider creation operator of one of these
oscillators,  we could define a level such that the number of of units of
worldsheet momenta created by this operator while acting on the vacuum. If
we have a state, then the total oscillator level of that state is the
sum of the levels of all the oscillators acting on the Fock vacuum to create
this state.  For the free string, the coordinates can be decomposed
into left moving and right moving sectors. Therefore, one can define
left and right moving oscillator oscillator levels (same decomposition
is valid when we add fermionic degrees of freedom). Thus one can write
$L_0 = {1\over 2}(E+P)$ and ${\bar L}_0 ={1\over 2}(E-P)$, where E and P are
worldsheet energy and momentum respectively.  Therefore, $L_0$ and ${\bar L}_0
$ get contributions from the oscillators and from the Fock vacuum. We may
remark in passing that the momenta of the spacetime D-dimensional theory
($25+1$ for bosonic string and $9+1$ for superstring) are the ones conjugate to
the zero modes of the bosonic/and/or fermionic worldsheet theory. Therefore, 
the ground state of the closed bosonic string is a tachyon
satisfying the relation $\alpha '{\rm m}^2 = -4$, with $\alpha '={1\over {
2\pi T}}$. The first excited (massless) states of closed string are: \\
$\bullet$ Spin 2 state, $G_{\mu \nu}$, identified as graviton.\\
$\bullet$ An antisymmetric tensor field, $B_{\mu \nu}$. \\
$\bullet$ A scalar, $\phi$,  called dilaton.\\
They belong to the irreducible representation of the $SO(24)$ group. These
states are created by action of a single creation operator from the
left moving sector and another creation operator from the  right moving
sector. Therefore, they will have two target space Lorentz indices and one
can decompose them according to irreducible representations of the
corresponding rotation group.\\
%%%%%%% FERMIONIC STRINGS%%%%%%

Introduction worldsheet fermions has important consequences. In fact,  if one
demands worldsheet superconformal symmetry generalising from the bosonic
string coordinates to include fermionic degrees of freedom, then resulting
theory is the superstring. First we need to construct two dimensional
supergravity action. One needs to add to the action (\ref{poly}) the action
\be
\label{sup}
 -{T\over 2}\int d^2\sigma e\{ i\psi ^{\mu}\gamma ^0\gamma ^a\partial _a\psi _{\mu}
-i{\bar \lambda }_a\gamma ^b\gamma ^a\psi ^{\mu}\partial _b\psi _{\mu}
-{1\over 4}\psi ^{\mu}\gamma ^0\psi _{\mu}{\bar \lambda}_a\gamma ^b\gamma ^a\lambda _b
\} \ee

The notations are as follows \cite{fubini}: $\psi ^{\mu}$ are
worldsheet 
two component Majorana fermions, $e^i _a$ are the zweibeins
associated with the worldsheet
metric, e is its determinant. $\lambda ^a$ is the gravitino on
worldsheet satisfying
$\lambda _a^* = \lambda _a$. The gamma matrices in the
worldsheet have following
representations: $\gamma ^0 =\sigma _2, \gamma ^1=i\sigma _1$
and $\gamma _5=
\gamma ^0\gamma ^1 =\sigma _3$, $\sigma _i$ being the three Pauli matrices.
We shall go over to the superorthonormal gauge, where
the worldsheet metric is flat metric times a conformal factor 
(mentioned already) and
gravitino is chosen to be
$\lambda _a=\gamma _a \zeta$ where $\zeta$ is a constant
Majorana spinor. Then the action (\ref{sup})
takes a simple form and is expressed
in terms of the Weyl Majorana fermions (it is a free fermion theory now)
\bea
-{iT\over 2}\int d^2\sigma [\psi _+^{\mu}(\partial _{\tau} -\partial _{\sigma})
\psi _{+\mu}+\psi _-^{\mu}(\partial _{\tau} +\partial _{\sigma})\psi _{-\mu} ] \eea
with the definition of the chiral fermions:  $\psi _+={1\over 2}(1-\gamma _5)\psi$
and $\psi _-={1\over 2}(1+\gamma _5)\psi$, the spacetime index is suppressed. 
Now it is
evident that the fermion equations of motion will separated according to the
chiralities, as is expected for massless fermions.  The
worldsheet supersymmetry  transformations are
\be \delta X^{\mu}={\bar \epsilon} \psi ^{\mu} \ee
\be \delta \psi ^{\mu}= -i\gamma ^a\partial _aX^{\mu}\epsilon \ee
For the two component Majorana fermions; $\epsilon$ is the 
fermionic parameter associated with the supersymmetry transformation.  
The supercharge is the 
time component
of supercurrent integrated over $\sigma$ variable. The current is
\be J^a =\gamma ^b\partial _bX^{\mu}\gamma ^a\psi _{\mu} \ee
Next, one defines  the super Virasoro generators and compute the quantum algebra
and derive the condition for absence of anomaly. In case of the superstring the
critical dimension is ten in contrast to bosonic string where it was 26.\\
Now we shall consider a few points before discussing how spacetime supersymmetry
multiplets appear in the spectrum of the superstring. We had mentioned that the
bosonic string has a tachyon in its lowest level which will render the theory
unstable. Although, worldsheet supersymmetric theory moves in ten dimensional
spacetime, the super Virasoro algebra does not impose sufficient constraints
to remove the undesirable tachyon from the spectrum in general.\\
Notice from the fermionic equations of motion (we suppress the bosonic part
momentarily to focus attentions on fermions only) that there is some
freedom in the choice of the boundary condition as $\sigma$ goes over a
period of $2\pi$. The is due to the fact that the action remains invariant
under $\psi \rightarrow -\psi$ for fermions of  either chirality. The 
boundary conditions are:
\be \psi (\sigma +2\pi) = -\psi (\sigma) \ee
known as Neveu-Schwarz boundary condition is antiperiodic\cite{ns}.  
The periodic boundary
condition
\be \psi (\sigma +2\pi) =\psi (\sigma) \ee
is the Ramond condition \cite{r};  the indices are   suppressed for
notational convenience. The mode expansion for, say the holomorphic field, is

\be \psi _+^{\mu}(\tau +\sigma) = \Sigma _n \psi _n^{\mu} e^{-n(\tau +\sigma)}
                                                                          \ee
It is easy to see that for Ramond boundary condition, n must be integers.
When we impose NS (Neveu-Schwarz) boundary condition and expand the fermions
in Fourier modes, then n will take  half integer values.
We note that the NS fermions have no zero modes, whereas the Ramond fermions
have zero modes in the Fourier expansions. \\
Let us extend the arguments, we used for the bosonic string, for the
superstring and examine  their spectrum. The  aim is to get rid of the
tachyon and to construct states using  bosonic and fermionic operators
such that these states  transform like fermions and the resulting theory be
endowed with spacetime supersymmetry. We shall consider the light-cone
gauge so that physical degrees of freedom become transparent. In addition
to the condition $X^+ = x^+ +p^+\tau$, one imposes constraint
\be \psi ^+={\bar \psi}^+=0 \ee
for the NS fermions, when we have Ramond fermions, they can be set to zero
except for the zero modes. Now we look at the superconformal constraints
and solve for $X^-, \psi ^-, {\bar \psi}^-$ in terms of the rest of the
coordinates. Thus we can use the (physical) transverse oscillators of
both $X$ and $\psi$  to construct the physical states and keep in mind
the presence of appropriate zero modes. It follows from straight forward  
calculation that the ground state in the NS sector is tachyon.  
The next level obtained by operating $\psi ^i$ 
contains massless states. 
Thus we need to remove the tachyon as well as some of the unwanted states,
at the same time, keeping the massless spectrum in tact. Notice that worldsheet
fermions are anticommuting objects, although they create bosonic states
while operating on a state of the theory. This feature is not very desirable
as will be evident from the following example. Let us consider a specific
bosonic state of a superstring and then operate on it a worldsheet spinor,
$\psi ^{i}_+$,
obeying NS boundary condition. The resulting state will still be an integer
spin object even if we have operated by anticommuting operator; this is
rather unusual. We can think of a situation when odd number of NS operators
act on a bosonic state and obviously same situation will continue to prevail,
whereas for even number of such operators we face no problem since even number
of anticommuting operators can be grouped to behave like bosonic operators.
If we demand that all the states be even under $(-1)^F$, then half of states
which had above mentioned undesirable feature, are removed including the
tachyon.  This is the GSO \cite{gso} 
projection. Moreover, after the unwanted states have
been discarded from the spectrum, the remaining states of the theory belong
to the representations of spacetime supersymmetry when we consider full
spectrum of the superstring theory. Note
that the operator $(-1)^F$ is defined up to a sign ambiguity. If we choose
the sign convention that the first excited state has $(-1)^F =+1$ which
arises due to  action
of  $\psi ^i$, on the ground state, then we can fix $(-1)^F$
quantum number of the rest of
the states. In this sign convention, tachyon will carry quantum number -1.
There is another convention where tachyon has quantum number +1 and then the
massless, first excited state, carries quantum number -1.
The fermion numbers $F_L$ and $F_R$ can be introduced separately for
the left and right moving sectors respectively.
When one computes supercharge algebra with Ramond condition, the
zero modes of the fermions in supercharge give an anomaly term besides the
$L_0$ term (that is Hamiltonian) and  anomaly vanishes for $D=10$. Moreover,
the anticommutation relation of the R-zero modes  are like Dirac gamma
matrices carrying target space indices. One finds that massless states appear
in the R-sector and they satisfy Dirac equation. They transform as
ten dimensional spinors (S) or conjugate spinors (C).Since we are considering
the left moving sector here at the moment, S has $+1$ eigenvalue and
C has $-1$ eigenvalue under the $(-1)^{F_L}$. When we construct other
excited states on these states they turn out to be massive. In view of this
one need not apply GSO projection, no tachyon is to be removed.\\
When we combine the left and right moving sectors four combinations will
appear in the description of the closed string spectrum. NS-NS, NS-R, R-NS
and R-R; where the first sector is from left movers and second is from right
movers in the  above four combinations. Let us look at them one by one.\\
\noindent (i) NS-NS: The states are created due to the action of
the creation operators from the left and the right moving sector. They will
transform as tensors under 10-dimensional Lorentz transformation. 
After GSO projection is implemented,  the lowest lying
states are massless and they can be decomposed into three groups,
symmetric traceless, antisymmetric tensor and a scalar under the rotations.
\\
\noindent (ii) NS-R: The GSO projection, as discussed is $(-1)^{F_L} =1$
and one keeps the S representation of the R sector here. The the massless
states consist of spacetime spinors.\\
\noindent (iii) R-NS: Here the GSO projection on NS sector from right side
gives fermion number 1. We have the choice of keeping S spinor or the C
spinor and obviously the states are spinorial.\\
\noindent (iv) R-R: The fermionic operators act from both sides and
therefore, the resulting state will be bosonic in character. It will depend
what combination we decide to keep. For example, if one keeps S from left
side and $\bar C$ from right side the product decomposes into a vector and
a three index antisymmetric tensor (has to be antisymmetric - it arises from
anticommuting objects). These belong to the bosonic sectors of type IIA
theory. There is other combination which S from left
and ${\bar S}$ from right combine and their decomposition is a scalar,
2-form potential and 4-form (antisymmetric) potential whose field strengths
are self-dual in ten dimensions and these states  are bosonic sector of
type IIB theory.\\
We are in a position to classify string theories according to their
important characteristics.
There are two 10-dimensional theories which have $N=2$ supersymmetry in
target space. Their massless bosonic sectors are as follows: type IIA
has graviton, $G_{\mu \nu}$, antisymmetric tensor $B_{\mu \nu}$ and dilaton,
$\phi$, coming from the  NS-NS sector and a gauge potential $A_{\mu}$ and
three index antisymmetric tensor potential $C_{\mu \nu \lambda}$, 
coming from the R-R sector. These two theories have 32 generators of
supersymmetry; type IIA is called non-chiral theory whereas type IIB is
known as chiral theory. Although the bosonic fields coming from the
RR sectors in these two string theories are tensors of different ranks,
the total number of degrees of  freedom of these tensors in each of
the theories (A and B) are the same \cite{bachas} and this can be checked
by counting the physical degrees of freedom RR gauge fields of type
IIA and IIB. \\
Next, we introduce the heterotic string which is very attractive
when one tries to establish connection of string theory with the gauge
groups of the standard model.
The heterotic string, in ten dimensions, contains $N=1$
supergravity multiplet, super Yang-Mills gauge theory along with
 chiral fermions.
There are two possible choices for the gauge
groups: $SO(32)$ or $E_8 \times E_8$, in the construction of the heterotic
string. Therefore, heterotic string theory fulfills Green-Schwarz anomaly
cancellation condition.
  Moreover, when the theory is compactified
to four dimensions
on Calabi-Yau manifold, the resulting theory has many features of the standard
model and the gauge group $SU(3)\times SU(2) \times U(1) $ can be
 embedded in the
4-dimensional theory. Let us briefly discuss how the heterotic string is
constructed.
We discussed the closed bosonic string and noted that the string coordinates can
be decomposed to left movers and right movers and each can be  expanded in
Fourier modes. Moreover, the Virasoro generators are also separated into
two groups, one group is expressed in terms of oscillators of one
kind only (say left mover) and the other group of generators are expressed
in terms of the oscillators of the other types (left movers). 
When one computes the
quantum algebra, the anomaly free condition is imposed on each groups of
Virasoro generators. In case of a closed string with worldsheet supersymmetry,
same situation appears, because the fermion equations of motion is also
written in terms of equations of motion of the Weyl Majorana fermions.
If we were  interested in constructing  a string theory which satisfies requirements of
conformal invariance, we could have a left moving closed bosonic string
and a right moving superstring. The former will satisfy Virasoro algebra
and latter the super Virasoro algebra.\\
The triumph of the Heterotic string is that, when we look at the
massless spectrum of the theory, it has $N=1$ spacetime supersymmetry,
contains the appropriate  gauge groups ($SO(32)$ or $E_8 \times E_8$) as is
required for the consistency due to 
Green-Schwarz anomaly cancellation condition. Therefore, the closed
bosonic string has 16 of its spatial coordinates compactified so that
those coordinates themselves  are periodic. Furthermore, using the standard
techniques of $1+1$ dimensional field theory, the compact bosonic
coordinates  could be fermionised to
give 32 Weyl Majorana fermions which are left movers. Thus, we have 10
bosonic coordinates and their 10 super partners (in light-cone gauge 8 bosons
and 8 fermions) in the right moving sector and 10 bosonic coordinates and
32 fermions (from compact coordinates) on the left moving sector. Whenever,
we adopt NS boundary conditions for these fermions arising out of 
compactification, tachyon will appear in the spectrum. Of course,
by introducing GSO projection on the  right moving sector we shall have
spacetime supersymmetry. So far as right moving part is concerned, bosonic
states come from states with NS boundary condition and fermions arise
due to the Ramond boundary conditions. The choice of boundary conditions on the
left moving fermions (coming from compact directions),
 give rise to two different types of gauge groups.
(i) All the left moving fermions can satisfy R-type (periodic) 
boundary condition
or they can satisfy NS-type boundary conditions. Then there is GSO condition
which ensures that there are only states which have even number of these
fermions (only one type boundary condition). Thus  the massless bosonic spectrum
is given by symmetric second rank tensor field, antisymmetric tensor field
and a scalar together with 496 gauge bosons belonging to the adjoint
representation of $SO(32)$. (ii) The second possible choice of 
boundary condition
for the left moving fermions is to divide them to two groups containing
16 fermions. Now there are four choices of boundary conditions (a) All
satisfy R boundary conditions, (b) periodic (R) boundary condition is
imposed on both the groups, (c) all the fermions in first group (call it I)
have R boundary condition and the group II has NS antiperiodicity and finally
(d) group I belong to NS boundary condition and II are in R. The
GSO projection is such that it keeps even number of fermions from each group
in the spectrum in every sector. When one works out the bosonic spectrum,
it contains again second rank symmetric tensor, antisymmetric tensor of rank
two, the scalar, dilaton and 496 gauge bosons in the adjoint representation
of $E_8 \times E_8$.\\
There is another superstring theory, known as type I. A  simple way
to describe type I string is from the perspective of IIB theory. Consider the
parity operation $\cal P$ on the worldsheet such that the `spatial' coordinate
$\sigma \rightarrow -\sigma$ under $\cal P$. In type IIB theory, $\cal P$
exchanges left and right moving sectors. Now, if we demand that we retain only
those states which are invariant under $\cal P$, we get the type I string.
In the NS-NS sector, graviton and the dilaton survive; 
the antisymmetric tensor is removed. 
From the RR sector,  the only surviving field is the second rank
antisymmetric tensor. Moreover, there are Weyl Majorana fermions and a gravitino
surviving the operation giving rise to $N=1$ supergravity multiplet. The open
string states are also included in type I spectrum. In this case, the worldsheet
degrees of  freedom are same as in the closed string case. One imposes Neumann
boundary conditions on the bosonic coordinates and suitable boundary conditions
on worldsheet fermions. The gauge group that can get attached to the open string
is $SO(32)$ and thus there is corresponding super Yang-Mills theory besides the
states we mentioned above.\\
Thus there are five perturbatively consistent string theories. The scattering
of particles belonging to spectrum of a string theory can be described
by introducing vertex operators \cite{vxfv}. 
They are required to satisfy constraints
due to conformal or superconformal transformations. They must transform
as representations of Lorentz group, like a  wave function. In the first quantized
frame work, one can calculate scattering of these particles in a well defined
perturbation theory. It is one of the great virtues of the superstring theories that
all these calculations are ultraviolet finite. Therefore, we have five
different string theories in ten dimensions.\\
One of the most efficient ways to study properties of string theory is 
to investigate the evolution of a string in the background of its
massless  excitations and then  explore the consequences of
conformal invariance for such a situation. Let us consider
closed bosonic string in the background of its  massless excitations
such as graviton, 
 antisymmetric
tensor and dilaton. 
 The action (\ref{poly}) generalizes to
\bea
\label{bgr}
-{T\over 2}(\int d^2\sigma \{{\sqrt{-\gamma}}\gamma ^{ab}G_{\mu \nu}(X) 
+\epsilon ^{ab} B_{\mu\nu}(X)
\partial _aX^{\mu}\partial _bX^{\nu} \}+
{1\over 2}\int d^2 \sigma {\sqrt {-\gamma}}R^{(2)}\phi (X) ) \eea

Here $R^{(2)}$ is the scalar curvature of the worldsheet computed  with
$\gamma _{ab}$.
 The first two terms show the couplings of $G_{\mu \nu}$ and 
$B_{\mu \nu}$ to the
string coordinates. 
In close string theory there is a massless state which transforms as symmetric
second
rank tensor and it is identified as graviton and there is an antisymmetric
massless second rank tensor state.  The above action describes motion of the
string in the background of these massless states, $G_{\mu\nu}$ and
$B_{\mu\nu}$; the  last term is the coupling  of the string to the
massless scalar, the dilaton.
This is an action for a two dimensional $\sigma$-model
and we can interpret that $G_{\mu \nu}$ and
$B_{\mu \nu}$  play the role of coupling constants. At the
classical level the dilaton coupling breaks the conformal invariance explicitly.
However, it is important to explore the consequences of the quantum invariance
as we have seen that the quantum invariance principle imposes strong constraints
on the theory. There is a well defined procedure to compute the conformal
anomaly for such theories\cite{agfm}. 
One of the ways to ensure conformal invariance of the quantum theory is to
demand that the two dimensional energy momentum stress tensor has vanishing
trace. As is well known, the conformal anomaly is related to the corresponding
$\beta$-function of the theory.                                  
 Thus, vanishing of the 
$\beta$-functions
will ensure conformal invariance. Moreover, the beta functions can be computed
order by order in the $\sigma$-model perturbation theory; $\alpha '$ being  
 the expansion parameter. The relevant $\beta$-functions are:
\be
{{\beta _{\mu \nu}^G}\over {\alpha '}}= R_{\mu \nu} -{1\over 4}H_{\mu \rho \lambda}
H_{\nu}^{\rho \lambda}+\nabla _{\mu}\nabla _{\nu} \phi  \ee
\be
{{\beta _{\mu \nu}^B}\over {\alpha '}}= \nabla ^{\rho}[e^{-\phi} H_{\mu \nu \rho}] \ee
\be
\beta ^{\phi} = \Lambda +3\alpha '[(\nabla \phi)^2 -2\nabla ^{\mu}\nabla _{\mu}\phi
-R +{1\over {12}} H^2] \ee
The notations are as follows: $R_{\mu \nu}$ is the Ricci tensor for the target space
computed from the string frame metric $G_{\mu \nu}$. $\Lambda =  D-26 ~{
\rm or}~ D-10$
depending on whether we are dealing with a pure bosonic string or superstring
(if we deal with superstring the coupling of worldsheet fermions to the background
has to be taken into account), D being the spacetime dimension.
$ H_{\mu \nu \rho} = \partial _{\mu}B_{\nu \rho} +
{\rm {cycl. perm}} $, is the field strength of two form potential $B_{\mu \nu}$.
It might be worthwhile to point out that for the constant value of dilaton
the last term in (\ref{bgr}) is just the Euler character of the surface. When
we write the path integral form with the action, we see that the factor
$e^{-\chi {\phi _0}}$ comes out; where $\chi$ is the Euler character and $\phi _0$
is the constant value of the dilaton. In this light the string coupling constant
is defined as
\be g_{str} = e^{{\phi }_0 /2} \ee
Let us look for an action in the target space such that the variation of that
action with respect to the backgrounds $G_{\mu \nu}, B_{\mu \nu}$ and $\phi$
would reproduce the $\beta$-function equations we have obtained earlier. 
 We also know that
these $\beta$-functions must vanish (to the order in $\alpha '$ they are computed)
in order to respect conformal invariance  of the theory.
\\
The resulting action is
\be
S= \int d^Dx {\sqrt {-G}}e^{-\phi}[ R+(\partial \phi )^2 -{1\over 12} H^2] \ee
This action is called the tree level string effective action. Solutions of
the equation of motion of this action (same as solution to $\beta$-function
equation) correspond to admissible background configurations with respect
conformal invariance. In other words, every solution is an acceptable vacuum
of the string theory to lowest order in $\alpha '$ since the effective
action is obtained from the $\beta$-function equations keeping only lowest
order terms in $\sigma$-model perturbation theory. Therefore, if we find
solutions which correspond to cosmological situation with given G, B and
$\phi$, or a black hole solution, or a wormhole solution all these types
of geometries with the appropriate matter content, consistent with the
equations of motion, can be interpreted as string vacuum backgrounds.\\
So far we have been discussing the quantization of string theories and
examining the consequences of conformal invariance. Note that all the
consistent string theories are defined in spacetime dimensions higher
than four i.e. $D=10$. Therefore,
one must answer the question what these theories
have to do with the spacetime where we live. This issue has been taken up by 
Kaluza and Klein more than seven decades ago. The basic idea is rather simple.
In order to construct a unified theory of gravity and electrodynamics,
they considered an Einstein-Hilbert type action in 5-spacetime dimensions
which is invariant under general coordinate transformations in five
dimensions. Let us imagine that one of the dimensions, the 5th one, is
a circle of very small radius which could not be probed today using any particle
whole de Broglie wave length is comparable to the size of that circle.
Then we shall not be aware of this scale.
Let us assume, to a first approximation,  that the metric does
 not depend on the 5th coordinate.
Kaluza and Klein showed that the resulting theory looks like
Einstein theory and Maxwell theory in four dimensions. What was
general coordinate invariance in 5-dimensional theory, turned out to be
general coordinate transformation and Abelian gauge transformation
(of Maxwell theory) in four dimensions. Although, the original Kaluza-Klein
proposal had many short comings, the idea is very relevant for  
construction of  four dimensional theories starting from the 10-dimensional
string theories in the present context. We shall explore this
aspect and we shall see how duality symmetries arise for compactified
string theories. We shall set $T=1$ from now on, whenever, we shall
need to introduce the slope parameter/tension , we shall explicitly mention in
that context.
%% THIS IS DUALITY SECTION  
%%%%%%%
\section{Duality Symmetries in  String Theory}

\setcounter{equation}{0}

\def\theequation{\thesection.\arabic{equation}}

 One of the marvels of the string theory is its rich symmetry structure.
We have noticed how the conformal invariance imposes strong constraints
on the theory: when we consider flat spacetime the dimensionality
is fixed by this symmetry. On the other hand if we consider strings in
backgrounds, we get the equations of motion for them by demanding that
the corresponding $\beta$-functions must vanish. 
Moreover, there are local symmetries like
invariance associated with general coordinate transformation due to  the
presence of the graviton and an Abelian gauge symmetry since 
the antisymmetric tensor is also a part of the massless multiplet.\\
The duality symmetries play a crucial role in understanding various
features of string theory. Since string is an extended  object, there are
symmetries special to  string theory. Consider a particle whose motion is
on a circular path, the momentum is quantized in suitable units of
the inverse radius in order that the wave function maintains single
valuedness. However, in case of a string,  one of whose coordinate has 
geometry of a circle,  offers more interesting possibilities. In fact
a string theory with one spatial direction compactified as $S^1$ 
 of radius $R$ cannot be
distinguished from another theory whose coordinate is compactified on a
circle of radius $1\over R$.  
Let the compactified coordinate be
denoted by $Y(\sigma, \tau)$ with the periodicity condition
\be Y(\sigma,\tau) +2\pi R = Y(\sigma,\tau) \ee
Furthermore, the string coordinate is also periodic when $\sigma$
goes over $2\pi $ for the closed string. Since, the coordinate is
compact, zero momentum mode must be quantized to maintain single
valuedness of the wave function just as the case  in field theory. In
case of the string, the string can wind around the compact direction.
It will cost more energy if the string winds m-number time, because
it will have to stretch more. Therefore, the  effect due to windings has to be
taken into account too  while estimating energy levels \cite{wind}.
Thus the mode expansions for left and right moving sectors are:
\be Y_R = y_R +{\sqrt {1\over 2}}p_R(\tau -\sigma) +{\rm oscillators} \ee
\be Y_L = y_L +{\sqrt {1\over 2}}p_L(\tau +\sigma)+{\rm oscillators} \ee

The momentum zero modes $p_{R,L}$  will have the following form to be
consistent with what we said earlier
\be p_R ={{1\over {\sqrt 2}}}({n\over R} -R m), ~~{\rm and} ~~
p_L={{1\over {\sqrt 2}}}({n\over R}+Rm) \ee
The above equation states that in general the contribution of the Kaluza-Klein
mode is $1\over R$ times an integer and the winding mode is an integer
times the radius. The total momentum is just $P= {{1\over {\sqrt 2}}}(p_R+p_L),$
which is integral of momentum density over $\sigma$. The total Hamiltonian is
\be H = L_0 +{\bar L}_0 ={1\over 2}(p_L^2 +p_R^2)+{\rm oscillators} \ee
Now we consider the general case of toroidal compactification and present 
the derivation as was done in reference \cite{ms}. Let $G_{\alpha
 \beta}
{\rm and } B_{\alpha \beta}$ be constant backgrounds, $\alpha ,\beta =1,... d$,
and $Y^{\alpha}(\sigma,\tau)$ are the string coordinates. 
 The two-dimensional $\sigma$-model
 action containing these coordinates is
\be
I_{compact}={1\over 2} \int d^2 \sigma ~ \big[ G_{\alpha\beta} \eta^{ab}\partial_{a}
Y^{\alpha}
\partial_{b} Y^{\beta} + \epsilon^{ab} B_{\alpha\beta} \partial_{a}
Y^{\alpha}
\partial_{b} Y^{\beta} \big]\, \ee 
where $G_{\alpha\beta}$ and $B_{\alpha\beta}$ are constant backgrounds.
The coordinates are taken to satisfy the periodicity conditions
$Y^{\alpha} \simeq Y^{\alpha} + 2 \pi$. Here we take the compactification
radius to be unity for simplicity in calculations.   
For closed strings it is necessary that
\be 
Y^{\alpha} (2 \pi , \tau) = Y^{\alpha} ( 0, \tau) + 2 \pi m^{\alpha}\,
\ee
where the integers $m^{\alpha}$ are called winding numbers. It follows from
the single-valuedness of the wave function on the torus that the zero modes
of the canonical momentum, $P_{\alpha} = G_{\alpha\beta}
\partial_{\tau} Y^{\beta} + B_{\alpha\beta}
\partial_{\sigma} Y^{\beta}$, are also integers $n_{\alpha}$. Therefore
the zero modes of $Y^{\alpha}$ are given by
\be
Y^{\alpha}_0 = y^{\alpha} + m^{\alpha} \sigma + G^{\alpha\beta}
(n_{\beta} - B_{\beta\gamma} n^{\gamma}) \tau \,\ee 
where $G^{\alpha\beta}$ is the inverse of $G_{\alpha\beta}$.
The Hamiltonian is given by
\be
{\cal H} = {1\over 2} G_{\alpha\beta} ( \dot Y^{\alpha} \dot
Y^{\beta} + Y'^{\alpha} Y'^{\beta} )\, \ee 
where $\dot Y^{\alpha}$ and $Y'^{\beta}$ are
derivatives with respect to $\tau$ and $\sigma$, respectively.
Let us elaborate a little bit on the significance of what we have done with
respect to the compact coordinates. Since the coordinates $Y^{\alpha}$,
are compact, they satisfy.
eq.(3.7). Moreover, these coordinates can be expanded as usual in terms of
their zero modes and the oscillators.
However, for the discussion of T-duality, we focus our attentions on the zero
mode parts  and the contribution of these parts to the Hamiltonian, given
above.

Since $Y^{\alpha} (\sigma , \tau)$ satisfies the free wave equation, we can
decompose it as the sum of left- and right-moving pieces.
The zero mode of $P^{\alpha}=G^{\alpha\beta}P_{\beta}$
is given by $p_L^{\alpha}+p_R^{\alpha}$ where
\be
p^{\alpha}_L = {1\over 2}
[ m^{\alpha} + G^{\alpha\beta} (n_{\beta} - B_{\beta\gamma} m^{\gamma}) ] \ee

\be
 p^{\alpha}_R = {1\over 2}
[ - m^{\alpha} + G^{\alpha\beta} (n_{\beta} - B_{\beta\gamma}
m^{\gamma} ) ]      \ee 

The mass-squared operator, which corresponds to the zero mode of ${\cal H}$,
is given (aside from a constant) by

\be
(mass)^2 = G_{\alpha\beta} \big( p^{\alpha}_L p^{\beta}_L + p^{\alpha}_R
p^{\beta}_R \big) +
\sum^{\infty}_{m=1}\sum_{i=1}^d (\alpha^i_{- m} \alpha^{i}_m + \bar
\alpha^{i}_{- m}
\bar\alpha^{i}_m)   \ee 
As usual, $\{\alpha_m\}$
and $\{ \bar \alpha_m\}$ denote oscillators associated with
right- and left-moving coordinates,
respectively. Substituting the expressions for $p_L$ and $p_R$,
the mass squared can be rewritten as

\be
\label{zero}
(mass)^2 = {1\over 2} G_{\alpha\beta}
m^{\alpha} m^{\beta} + {1\over 2} G^{\alpha\beta} (n_{\alpha} -
B_{\alpha\gamma} m^{\gamma})(n_{\beta} - B_{\beta\delta} m^{\delta})
+\sum (\alpha^i_{- m} \alpha^i_m + \bar \alpha^i_{- m}
\bar \alpha^i_m) \, \ee 
It is significant that the zero mode portion of (\ref{zero}) can be
expressed in the form
\be  (M_0)^2 = {1\over 2}
(m \ \ n)  M^{-1} \pmatrix {m\cr n\cr}, \ee 
where $M$ is the $2d\times2d$ symmetric matrix expressed in terms of constant
backgrounds G and B 
\be
\label{mmatrix}
M = \pmatrix {G^{-1} & -G^{-1} B\cr
BG^{-1} & G - BG^{-1} B\cr} \ee
In order to satisfy
$\sigma$-translation symmetry, the contributions of left- and
right-moving sectors to the mass squared
must agree; $L_{0}=\bar L_{0}$. 
 The zero mode contribution to
their difference is
\be 
\label{diff}
G_{\alpha\beta} (p^{\alpha}_L p^{\beta}_L - p^{\alpha}_R p^{\beta}_R )
= m^{\alpha} n_{\alpha} ~    \ee 
Since this is an integer, it always can be compensated by oscillator
contributions, which are also integers.

Equation (\ref{diff})  is invariant under interchange of
the winding numbers $m^{\alpha}$ and the discrete momenta $n_{\alpha}$.
Indeed, the
entire spectrum remains invariant if we interchange
$m^{\alpha} \leftrightarrow n_{\alpha}$
 simultaneously let \cite{int} 

\be
(G - B G^{-1} B) \leftrightarrow G^{- 1} ~~~ {\rm
and} ~~~ B G^{- 1} \leftrightarrow - G^{- 1} B \,  \ee 
These interchanges precisely correspond to inverting the $2d\times2d$
matrix $M$. This is the spacetime duality
transformation generalizing the well-known duality $R\leftrightarrow
{1\over R}$ in the $d=1$ case discussed earlier.  The general duality
symmetry
implies that the $2d$-dimensional Lorentzian lattice
spanned by the vectors ${\sqrt 2}
(p^{\alpha}_L , \, p^{\alpha}_R)$ with inner product
\be
{\sqrt 2}~ (p_L , \, p_R) \cdot
{\sqrt 2}~ (p'_L , \, p'_R) \equiv 2G_{\alpha\beta} (p^{\alpha}_L p'^{\beta}_L
-
p^{\alpha}_R p'^{\beta}_R) = (m^{\alpha} n'_{\alpha} + m'^{\alpha}
n_{\alpha})\,  \ee 
is even and self-dual (\cite{narain}).i
For toroidally compactified string theory, the coordinates satisfy periodicity
condition and the conjugate momenta belong to the dual space and are quantized
in suitable units. Furthermore, one can define corresponding metric to introduce
the norm for the coordinates and their dual momentum vectors 
 and define an inner product also. For a class of  lattices the 
space of the coordinates (since the coordinates satisfy periodicity condition
it is
like crystals) is the same as the dual space, then the lattice
 is called self-dual. Of
special significance,  are the spaces where the length of the vector is even
(with the definition of norm). In that case we have even self-dual
lattice. These types of lattices are very important in construction of string
theories  with nonabelian gauge groups and to satisfy consistency
requirements of the theory.

The moduli space parametrized by $G_{\alpha\beta}$ and $B_{\alpha\beta}$
is locally the coset
$O(d, d)/O(d) \times O(d)$.
The global geometry requires also modding out the group of discrete symmetries
generated by $B_{\alpha\beta} \rightarrow B_{\alpha\beta} +
N_{\alpha\beta}$ and $G + B \rightarrow
(G + B)^{-1}$.  These symmetries generate the $O(d,d,Z)$ subgroup of
$O(d,d)$. An $O(d,d,Z)$ transformation is given by a $2d\times2d$ matrix $A$
having integral entries and satisfying $A^T \eta A = \eta$, where $\eta$
consists of off-diagonal unit matrices defined below. Under an $O(d,d,Z)$
transformation
\be \pmatrix {m \cr n} \rightarrow \pmatrix {m' \cr n'} =
A \pmatrix {m \cr n \cr}
\quad {\rm and} \quad M \rightarrow AMA^T\ \ee  
It is evident that
\be m\cdot n = {1\over 2}(m \ \ n)
  \eta \pmatrix {m
\cr n \cr}\  \ee
\be \eta=\pmatrix {0 & {\bf 1}\cr {\bf 1} & 0\cr} ,\ee 
which appears in eq.(\ref{diff}),
and $M_0^2$ in eq.(\ref{zero})  are preserved under these transformations.
Note that $\eta$ is symmetric $2d\times 2d$ matrix with off diagonal
elements which are d-dimensional unit matrices.
The crucial fact, already evident from the spectrum, is that toroidally
compactified string theory certainly does not share the full $O(d,d)$
symmetry of the low energy effective theory. It is at most invariant
under the discrete $O(d,d,Z)$ subgroup.\\
So far, in discussing issue compactifications, we have considered situations
when all the coordinates are compact. However, one can envisage the scenario,
when some of the string string coordinates are compactified and the rest are
noncompact. Furthermore, we treated the backgrounds to be constant; however,
in more realistic situations the backgrounds should be allowed to depend on
noncompact coordinates. This is the more interesting situation where we have
a ten dimensional string theory and six of its spatial coordinates are
compactified on a torus $T^6$ so that the resulting theory is reduced to a four
dimensional effective theory. We shall adopt the general prescription of
dimensional reduction \cite{ss,ms,hs}
 so that we can compactify an arbitrary number of 
dimensions so that the effective theory is defined in a lower spacetime
dimension, not necessarily four. This will be useful, since the duality
conjectures are in various spacetime dimensions and string
theories are related by the web of dualities in diverse dimensions.\\
The starting point is to consider the string effective action in $\hat D$
spacetime dimensions. The coordinates, metric and all other tensors in the
$\hat D$ dimensional space are specified with a `hat'. The coordinates in
D-dimensional spacetime are denoted by $x^{\mu}, \mu,\nu , etc$ are spacetime
indices. Therefore, $\hat D=D+d$.
The theory is compactified on a d-dimensional torus,
$T^d$, to D-dimension spacetime. The coordinates on the torus, sometimes
referred to coordinates of internal dimensions, are denoted as $y^{\alpha},
\alpha =1,... d$. The bosonic part of the action is given by 
\bea \hat S = \int d^{\hat D}x~ \sqrt{- \hat G} e^{-\hat\phi}
\big [\hat R
(\hat G) + \hat G^{\hat \mu \hat \nu} \partial_{\hat \mu} \hat\phi
\partial_{\hat \nu} \hat\phi  - {1 \over 12} ~ \hat H_{\hat \mu \hat \nu
\hat \rho} ~
\hat H^{\hat \mu \hat \nu \hat \rho}\big ].\eea
Note that $\hat S$ is the bosonic part of the string
effective action with backgrounds coming from NS-NS sector.
\noindent $\hat H$ is the field strength of antisymmetric tensor and $\hat
\phi$ is the dilaton. 
The backgrounds  are taken to be independent of the internal coordinates,
 $y^{\alpha}$ of the torus.
Consequently, any transformations of the coordinated ${y^{\alpha}, \alpha =1,
2,..d}$ does not affect the background fields and we recognize that there
are $d$ isometries. Furthermore, associated with these isometries, there
will be $d$ Abelian gauge fields since the $\hat D$-dimensional metric will
have components carrying a D-dimensional spacetime index and an internal
index $\alpha$. There will be components of the $\hat D$-dimensional metric
which will carry indices of the toroidal coordinates, say $\alpha, \beta$
and these will transform as scalars, often refer to as moduli. Similarly,
if we consider the components of the $\hat D$-dimensional antisymmetric
tensor field it will have $D\times D$ component antisymmetric tensor, $d$
Abelian gauge fields coming from spacetime and internal component and $d\times
d$ dimensional moduli  (antisymmetric) when considered from D-dimensional
point of view.

The metric $\hat G_{\hat \mu \hat \nu} $
 can be decomposed as
\bea \hat G_{\hat \mu \hat \nu} = \left (\matrix {{\cal G}_{\mu \nu} +
A^{(1)\gamma}_{\mu} A^{(1)}_{\nu \gamma} &  A^{(1)}_{\mu \beta}\cr
A^{(1)}_{\nu \alpha} & G_{\alpha \beta}\cr}\right ),\eea

\noindent where $G_{\alpha \beta}$ is the internal metric and 
${\cal G}_{\mu\nu}$,
the $D$-dimensional space-time metric, depend on the coordinates $x^{\mu}$.
Note the appearance of Abelian gauge fields $A^{(1)\alpha}$ due to the
presence of the isometries. We also expect same number of gauge fields
from the antisymmetric tensor ${\hat B}_{{\hat \mu}{\hat \nu}}$. Thus 
The dimensionally reduced action is,

\bea  S_{D}&= &\int d^Dx \sqrt {-g} e^{-\phi}
\bigg\{ R + g^{\mu \nu}
\partial_{\mu} \phi \partial_{\nu} \phi -{1\over 12}H_{\mu \nu \rho}
H^{\mu \nu \rho}\nonumber\\
& +& {1 \over 8} {\rm tr} (\partial_\mu M^{-1} \partial^\mu
M)- {1 \over 4}
{\cal F}^i_{\mu \nu} (M^{-1})_{ij} {\cal F}^{\mu \nu j} \bigg\}.
 \eea

\noindent Here $\phi=\hat\phi-{1\over 2}\log\det G$ is the
shifted dilaton.

\bea H_{\mu \nu \rho} = \partial_\mu B_{\nu \rho} - {1 \over 2} {\cal A}^i_\mu
\eta_{ij} {\cal F}^j_{\nu \rho} + ({\rm cyc.~ perms.}),\eea

\noindent ${\cal F}^i_{\mu \nu}$ is the $2d$-component vector of field
strengths
\bea {\cal F}^i_{\mu \nu} = \pmatrix {F^{(1) \alpha}_{\mu \nu}
\cr F^{(2)}_{\mu \nu \alpha}\cr} = \partial_\mu {\cal A}^i_\nu - \partial_\nu
{\cal A}^i_\mu \,\, ,\eea
\noindent $A^{(2)}_{\mu \alpha} = \hat B_{\mu \alpha} + B_{\alpha \beta}
A^{(1) \beta}_{\mu}$ (recall $B_{\alpha \beta}=\hat B_{\alpha \beta}$), and
the $2d\times 2d$ matrices $M$ and $\eta$ are defined as
\begin{eqnarray} M = \pmatrix {G^{-1} & -G^{-1} B\cr
BG^{-1} & G - BG^{-1} B\cr},\qquad \eta =  \pmatrix {0 & 1\cr 1 & 0\cr}
\, .\end{eqnarray}

Note that the elements of the matrix M, $G_{\alpha \beta} ~{\rm and} ~
B_{\alpha \beta}$ depend on spacetime coordinates $x^{\mu}$ in contrast to
the earlier case (\ref{mmatrix}) where those back grounds were taken to
be constant.
\noindent The action (3) is invariant under a global $O(d,d)$ transformation,

\bea M \rightarrow \Omega^T M \Omega,~~~  \Omega \eta \Omega^T = \eta,~~~ 
{\cal A}_{\mu}^i \rightarrow \Omega^i{}_j {\cal A}^j_\mu,~~ {\rm
where} ~~ 
\Omega \in O(d,d). \eea
\noindent and the shifted dilaton, $\phi$, remains invariant
under the $O(d,d)$ transformations.
Moreover,   $M\in O(d,d)$  and $M^T\eta M=\eta$. Thus if we
solve for a set of backgrounds, $M$,$\cal F$ and $\phi$,
satisfying the equations of motion they correspond to a vacuum
configuration of the string theory. The $O(d,d)$ symmetry is known as the
target space duality (or T-duality) symmetry, it is a stringy symmetry and there is no analogue of winding modes in ordinary field theory.  The 
symmetry  holds good order by order
in string perturbation theory. Therefore, predictions of T-duality can be 
tested within the frame work of perturbation theory. We remark in passing
that, if we had considered an effective action in $\hat D$ dimensions
with n Abelian gauge fields, the reduced action in $D$ dimensions will be
invariant under $O(d,d+n)$ symmetry. This is of importance, since in case of
the heterotic string, the ten dimensional action with 16 Abelian gauge fields
corresponding to the Cartan subalgebra of the nonabelian gauge groups of the
theory, when reduced to lower dimensions with exhibit the symmetry $O(d,d+n)$
we mentioned.\\
Thus if we have a set of background configurations it is possible to
generate another set of gauge inequivalent backgrounds by implementing suitable
$O(d,d)$ transformations. The new backgrounds will also satisfy the equations
of motion and they will be acceptable vacuum configurations. In fact the
$O(d,d)$ symmetry was discovered for non-constant backgrounds in the context
of cosmological solutions in string theory \cite{vc,mvc}, when the backgrounds
carried only time dependence. One could generate new cosmological solutions
through $O(d,d)$ transformations \cite{mvc1,gmv}. The applications of $O(d,d)$
transformations in the context of black holes was to generate new black hole
solutions was initiated by Sen \cite{senbh} and there is a 
vast literature in this subject \cite{senr1,youm}. \\
Next, we discuss S-duality in string theory. This symmetry relates a theory
in the weak coupling regime to a theory in the strong coupling domain. In
some  it is the same theory which gets related to  itself, like the
type IIB theory. In some other situations 
one theory gets related to another one: a 
familiar example is that heterotic string compactified on $T^4$ is 
related to type IIA theory compactified on $K_3$. A simple  example is the
Maxwell electrodynamics. The equations are invariant under ${\bf E}\rightarrow
{\bf B}$ and $\bf B \rightarrow -\bf E$.  
However, in the presence of sources, one
has to be careful. The usual Maxwell equations have only sources carrying
electric charges and then the equations are not symmetric under the above
duality transformations. Thus it is necessary to add sources carrying magnetic
charges to maintain electric-magnetic duality.
This led  Dirac to formulate the theory of magnetic monopoles. As is
well known, the existence of magnetic monopole in the theory leads to
the famous charge quantization condition: $e\cdot g =2\pi n$, where $e$
is the electric charge and $g$ is the magnetic charge. This relation has
profound implications; if the theory of electrically charged particles is
described by a small coupling constant (indeed fine structure constant
$\alpha = {1\over {137}}$), then the theory describing magnetic monopoles 
will have large value for such charges corresponding to 
strong coupling constant. In the case of gauge theories
with spontaneous symmetry breaking, magnetic monopoles  appear as 
classical solutions of nonlinear field equations \cite{t,p}.  
Note that the electric charge in
such theories are obtained from the Noether currents whereas, the magnetic
charge of 't Hooft-Polyakov monopoles are of topological nature.
The charges respect the Dirac
quantization condition. Furthermore, the massive gauge bosons (acquiring
mass through Higg's mechanism) have masses proportional to the gauge
coupling constant, whereas the monopole masses  are inversely proportional
to the gauge coupling constant (electric charge). Consequently, if the
gauge bosons are light in a SSB theory, the monopoles are heavy; indeed the
monopoles have the interpretation of being
 the solitons of the theory. One of the most fundamental
contributions to developments in S-duality came from the work of Montonen and
Olive \cite{mn}. According to them, 
we might envisage a dual formulation of fundamental
physics where the role of Noether charges  and topological charges are
interchanged. One can visualize that monopoles will appear as elementary
particles and the W-bosons will be solitonic  counter parts. In fact one
could check their mass formula $m^2=C(e^2+g^2)$; where C is related to VEV of 
Higgs in SSB theories. In fact W boson and photon satisfy this formula. 
If
a particle had been discovered with magnetic charge this relation could be
verified. Since it is symmetric under the interchange of e and $g$ and Dirac's
rule tells us that e and $g$ are related, one could formulate the theory in the
dual picture. However, the monopole mass obtained in SSB theory is a classical
one and it is subject to quantum corrections. Thus, Montonen-Olive idea
could not be consistently checked in usual field theories. There are special
types of supersymmetric field theories where there is no quantum correction to
the mass and furthermore, the W-bosons and monopoles belong to the same
multiplet. In such cases there is the possibility of checking this conjecture. 
\\
We recall that the Yang-Mills
theory also admits the introduction of the $\theta$ term in its action.
 Thus, gauge theories have two parameters, the Yang-Mills coupling
constant $e$ and the $\theta$ parameter. The latter couples to the
field strengths as follows:
\be 
-{{\theta e^2}\over {32 \pi ^2}}F_{\mu\nu}^a{\tilde F}^{\mu \nu}_a, \ee
 where ${\tilde F}^{a{\mu \nu}} = \epsilon ^{\rho \lambda}_{\mu \nu }F^a_{{\rho
\lambda}}$.  
 Note that this term is a surface term and does not contribute to classical
equations of motion and presence of this term does not affect renormalizability
 in the perturbation theory. It was noted by Witten \cite{wm} that in the 
presence of monopoles, this term shifts the allowed values of the electric
charge in the monopole sector. Thus we can have electrically charged,
magnetically charged particles and a third kind of particles carrying both
the charges. The Yang-Mills Lagrangian can be written  in the following
form after taking into account the effect of the $\theta$ term and 
introducing a complex coupling constant
$ \tau ={ \theta \over {2\pi}} +{{4i\pi }\over{e^2}}$ 

\be  {\cal L} = -{1\over{32\pi}}Im (\tau 
[F^{\mu \nu}_a+i{\tilde F}^{\mu \nu}_a ] 
[F _{\mu \nu} ^a+i{\tilde F}_{\mu \nu} ^a])  \ee 
Following qualitative argument tells us about the strong-weak duality group.
(i) When $\theta$ goes over its period $2\pi$ physics is the same. Thus, we
expect that the  theory be invariant when $\tau \rightarrow \tau +1$.
(ii) We also know that, under electric 
magnetic duality, $\tau \rightarrow -{1\over {\tau}}$
One can argue that, when $\theta$ is arbitrary, the duality group is generated
by these transformations. Thus, the duality group is identified to be
$SL(2,Z)$.
Therefore, in a  theory with $SL(2,Z)$ symmetry one could check the spectrum
with charged particles, monopoles and dyons.
The complex coupling constant $\tau$ is often referred to as modular
parameter or moduli.                                                       
 Moreover, when we discuss strong-weak duality  in the context of string
theory, dilaton and axion will be combined to define the moduli field. As
mentioned earlier, string theory does not admit any arbitrary parameters as
coupling constants. All the coupling constants appear as VEV of some scalar
fields, i.e. moduli. Therefore, very often, the term coupling constants and
moduli are used interchangeably in string theory.
\\
As mentioned earlier, the mass formulas are protected from quantum corrections
in supersymmetric theory. Moreover, some of the solitonic solutions in the
supersymmetric theories satisfy special properties: (i) They saturate the
BPS bound and (ii) these solutions preserve a part of the supersymmetry of
the original theory. These attributes play a very important part in testing
duality conjectures in field theory and in string theory. 
In order to illustrate the basic point, let us consider a two dimensional
example due to Witten and Olive \cite{ow}, where the field content is a
scalar field and Majorana fermion.  The Lagrangian density is
\be
{\cal L}={1\over 2}[(\partial _{\mu} \Phi)^2 +i{\bar \Psi}\gamma ^{\mu}
\partial _{\mu}\Psi -V^2( \Phi)-V'(\Phi){\bar \Psi}\Psi] \ee
The potential is arbitrary function of $\Phi$ and `prime' denotes 
derivative with respect to $\Phi$. As was the 
case in worldsheet supersymmetry,
we can work in terms of chiral components of fermions and the two super charges
are
\be Q_+=\int dx [(\partial _0 +\partial _1)\Phi\Psi _+-V(\Phi)
\Psi _-] \ee
\be Q_-=\int dx [(\partial _0-\partial _1)
\Phi\Psi _-+V(\Phi)\Psi _-]
\ee
In light-cone variables $Q_{\pm} ^2 =P_{\pm}$, with $P_{\pm}=P_0\pm P_1$ and 
it turns out that  $\{Q_+, Q_- \} =0$, in most of the case. However, 
careful analysis shows that the anticommutator, is proportional
to a surface integral
\be
\label{t}
\{Q_+,Q_-\}=2\int dx{{\partial }\over  {\partial x}}H(\Phi) \ee
and $H'(\Phi)=V(\Phi)$. This surface integral 
does not necessarily vanish when one
considers solitonic states. If we denote the R.H.S. of (\ref{t}) by the operator
T, then it can be evaluated for the case at hand. Now the algebra
of charges are different from usual case and one can write
\be
P_++P_-=T+(Q_+-Q_-)^2   \ee

\be  P_++P_-=-T+(Q_++Q_-)^2  \ee  
The R.H.S. of each equation above has a piece which is  a complete square and
we have $P_++P_- \ge |T| $. If we consider single particle of mass M 
and go to its
rest frame $P_{\pm}=M$;  we  arrive at 
\be M \ge |T| \ee
The bound will be equality when we have states, $|s\rangle $
such that $(Q_++Q_-) |s \rangle =0 ~~{\rm or }~~
(Q_+-Q_-)| s \rangle =0$. The bound on
M is the Bogomolny bound. The state  which saturates it is called a BPS state.
This bound also can be derived in a Lorentz covariant manner.  We note that, 
for the states saturating the BPS bound,  
only half of the supersymmetries are preserved. 
In string theory or field theories with 
large number of supersymmetries, the algebra of the charges for a set of 
charges $\{ Q_{\alpha} \}, \alpha =1,... N$, can be brought
to the form
\be \{ Q_{\alpha}, Q_{\beta} \}=\delta _{\alpha \beta} \ee
This will be possible if there are no states which are annihilated by
some of these charges and in that case, we shall get supermultiplets as usual.
However, just like the soliton case considered earlier, if there are 
states which will be annihilated by some charges then we shall have a situation
where
\be \{Q_a, Q_b \} = \delta _{ab},~~{\rm for}~~a,b =1,..M \ee
\be \{Q_{\alpha},Q_{\beta} \}=0, ~~{\alpha},{\beta}=M+1,...N  \ee
So we see that these states will be lower dimensional representations since
$M<N$.  Again, citing the example of two dimensional case, we can state the
general result that when there are soliton like states getting annihilated by
some of the supercharges, then the symmetric matrix $\{Q_{\alpha},Q_{\beta} \}$
will have some zero eigen values. The charges (analog of T) and masses get
related in the process. This is true for monopoles in 4-dimensional theories.
The string effective action is defined  in 10 dimensions and one can
seek solutions for extended objects in space and there are BPS states in this
regime too.\\
Let us compactify the heterotic string effective action on $T^6$ to come
to a four dimensional theory. As mentioned earlier, the T-duality group is
$O(6,22)$ with scalars parameterizing the moduli ${{O(6,22)}\over {O(6)\times
O(22)}}$, 28 gauge bosons, graviton ${\cal G}_{\mu \nu}$ and antisymmetric
tensor $B_{\mu \nu}$.
The four dimensional effective action for the heterotic string, following
the prescriptions of \cite{ms}, can be obtained in a straight forward
manner. The T-duality invariance is manifest when we are in the string
frame metric with shifted dilaton ${\hat \phi}-{1\over 2}{\rm ln~det}G_{\alpha
\beta}$. However, when one considers the S-duality properties of the theory,
it is convenient to go over to the Einstein frame metric,$g_{\mu\nu}$
 through the
conformal transformation,
$ g_{\mu \nu}= e^{-\phi}{\cal G}_{\mu\nu}$. 
In string theory, all the coupling constants are related to the VEV of the
dilaton and therefore, in order to identify the parameters of S-duality
group, we have to choose the field whose VEV will coincide with the
$\theta$ parameter. Notice that the field strength of antisymmetric tensor,
$H_{\mu\nu\rho}$ has only one degree of freedom in four dimensions when we
fix all gauge freedoms. In fact, if we take dual of this field, it is a 
pseudoscalar particle and that is what we need, an axion.
The starting point is the four dimensional effective action \cite{axjs} with
Einstein frame spacetime metric
\be
\label{duals}
S^{(4)} = \int_M dx ~ \sqrt{-g} \bigg \{ R - {1 \over 2} g^{\mu \nu}
\partial_{\mu} \phi \partial_{\nu} \phi + {\cal L}_2 + e^{- \phi} {\cal
L}_3 + e^{-2 \phi} {\cal L}_4 \bigg \}\, \ee
with ${\cal L}_2$, ${\cal L}_3$, and ${\cal L}_4$  defined as follows 
\be
{\cal L}_2 = {1 \over 8} {\rm tr} (\partial_\mu M^{-1} \partial^\mu
M)\,\,. \ee

\be  {\cal L}_3= - {1 \over 4}
{\cal F}^i_{\mu \nu} (M^{-1})_{ij} {\cal F}^{\mu \nu j}  \ee

\be  {\cal L}_4 = - {1 \over 12}
        H_{\mu \nu \rho} H^{\mu \nu \rho} \ee 
Here we closely follow the notation of \cite{ms} and \cite{axjs}.
The next step is to perform a duality transformation, which replaces
the field $B_{\mu \nu}$ by a scalar field $\chi$.  This is achieved by
first forming the $B_{\mu \nu}$ equation of motion
\be
\partial_{\mu} \big ( \sqrt{- g}~ e^{-2 \phi} H^{\mu \nu \rho} \big
) = 0 \ee 
and solving it by setting
\be
 \sqrt{ -g} ~ e^{-2 \phi} H^{\mu \nu \rho} = \gamma  \epsilon^{\mu
\nu \rho \lambda} \partial_{\lambda}  \chi \ee 
where $\chi$ is the ``axion" and $\gamma$ is a constant to be fixed
later.  In the language of differential forms,
\be
H = \gamma  e^{2 \phi} * d \chi \ee 
or, using $H= dB - {1 \over 2}  \eta_{ij}  {\cal A}^i_{\wedge} {\cal
F}^j$,
\be 
dB = {1 \over 2}  \eta_{ij}  {\cal A}^i_{\wedge} {\cal F}^j + \gamma
e^{2 \phi} * d \chi  \ee 
The Bianchi identity $(d^2 B = 0)$ now turns into the
$\chi$ field equation
\be 
{1 \over 2}  \eta_{ij} {\cal F}^i_{\wedge} {\cal F}^j + \gamma d
\big(e^{2 \phi} * d \chi \big) = 0 \ee 
or, in terms of components, (choosing a convenient value for $\gamma$)
\be
 \partial_{\mu} (e^{2 \phi} \sqrt{ -g } ~ g^{\mu \nu}
\partial_{\nu} \chi ) - {1 \over 8} \eta_{ij} \epsilon^{\mu \nu \rho
\lambda} {\cal F}^i_{\mu \nu} {\cal F}^j_{\rho \lambda} = 0 , \ee

This is an equation of motion if we replace the ${\cal L}_4$ term in $S^{(4)}$
by
\be
S_{\chi} = - \int dx \sqrt{ -g} \bigg ( {1 \over 2} e^{2 \phi}
g^{\mu \nu} \partial_{\mu} \chi \partial_{\nu} \chi + {1 \over 4} \chi
{\cal F}\cdot \tilde{\cal  F} \bigg )\,\, ,\ee
where
\be
{\cal F}\cdot \tilde{\cal  F} \equiv {1 \over 2 \sqrt{-g}}
\epsilon^{\mu \nu \rho \lambda} {\cal F}^i_{\mu \nu}  \eta_{ij}
{\cal F}^j_{\rho \lambda} \,\,.\ee

Let us briefly recapitulate the steps we have taken to modify the four
dimensional action in the Einstein frame. The field strength $H_{\mu\nu\lambda}
$ appearing in ${\cal L}_4$ is traded for the pseudoscalar axion, $\chi$.
The resulting action (3.51) contains not only the kinetic energy term of the
axion, but also the topological term which is like the $\theta$ dependant term
of the Yang-Mills action if the VEV of $\chi$ is identified with that
parameter.

Let  us now regroup the terms in the dual action in the following way:
\be
\tilde S^{(4)} = \int_M dx \sqrt{ -g} \big ( R + {\cal L}_2) + S_D +
S_F \,\, ,\ee
where
\be S_D = - {1 \over 2} \int_M dx \sqrt{ -g} g^{\mu \nu} \bigg (
\partial_{\mu} \phi \partial_{\nu} \phi + e^{2 \phi} \partial_{\mu} \chi
\partial_{\nu} \chi \bigg ) \ee
\be S_F = - {1 \over 4} \int_M dx \sqrt{ -g} ~\bigg ( e^{- \phi} {\cal F}^2
+ \chi {\cal F}\cdot \tilde{\cal  F} \bigg ) \ee
and ${\cal F}^2 \equiv  ~ g^{\mu \rho}
g^{\nu \lambda} {\cal F}^i_{\mu \nu}(M^{-1})_{ij}
{\cal F}^j_{\rho \lambda}$. 
Note that ${\tilde S}^{(4)}$ contains the usual Einstein-Hilbert action and
the part coming from kinetic energy term of the M-matrix. We have rearranged
the actions coming  from dilaton kinetic energy, gauge field part
and the axionic part (together with the 'topological' term) to define
$S_D$ and $S_F$ so that dilaton and axion are put together and the gauge field
kinetic energy along with the topological term are clubbed together. This
is very useful to study the S-duality properties of the action.
In order to  describe the $SL(2,R)$ symmetry of the dilaton and axion
kinetic terms, let us introduce  a complex modular parameter (recall
the case of Yang-Mills)
\be      \tau = \chi + ie^{-\phi} \,\, ,\ee
which has the nice property that under a linear fractional
transformation
\be \tau \rightarrow {a\tau + b\over c\tau + d} \ee 
the combination
\be       {g^{\mu\nu}\partial_\mu \tau \partial_\nu \bar\tau
\over ({\rm Im } ~ \tau)^2}
= g^{\mu\nu} (\partial_\mu \phi \partial_\nu \phi + e^{2\phi}
\partial_\mu \chi \partial_\nu \chi)    \ee
is invariant.
 It immediately follows that
\be
S_D = -{1 \over 2} \int_M dx \sqrt{ -g} ~
{g^{\mu\nu}\partial_\mu \tau \partial_\nu \bar\tau
\over ({\rm Im } ~ \tau)^2}\ . \ee

Now we consider the gauge field action , 
$S_F$.
Notice that the
$SL(2,R)$ transformations give rise to an electric-magnetic duality
rotation. 
Let us  define
\be  {\cal F}^{\pm}_{\mu\nu}= M\eta{\cal F}_{\mu\nu}\pm i \tilde
{\cal F}_{\mu\nu} \, . \ee
Then, using the identity ${\cal F}^{+\mu\nu}M^{-1}{\cal F}^-_{\mu\nu}
=0$, we can express  $S_F$ as
\be 
S_F=-{1\over 16i}\int_M dx \sqrt{-g} \bigg(\tau {\cal F}^{+\mu\nu}M^{-1}
{\cal F}^{+}_{\mu\nu} - \bar\tau{\cal F}^{-\mu\nu}M^{-1}
{\cal F}^{-}_{\mu\nu}\bigg) \, . \ee 
The ${\cal A}_{\mu}$ equation of motion is
\be
\label{eqa}
\nabla^{\mu} \big( \tau {\cal F}^{+}_{\mu\nu} -\bar\tau {\cal
F}^{-}_{\mu\nu}\big) =0 \ee 
and the Bianchi identity is
\be 
\label{bianc}
\nabla^{\mu} \big( {\cal F}^{+}_{\mu\nu} -
{\cal F}^{-}_{\mu\nu}\big) =0 \, . \ee 

To exhibit $SL(2,R)$ symmetry it is necessary to have ${\cal A}_{\mu}$
transform at the same time as $\tau$.
The appropriate choice is to require that
${\cal F}^{\pm}_{\mu\nu}$ transform as modular forms as follows
\be
\label{modtr}
   {\cal F}^+_{\mu\nu} \rightarrow (c\tau +d)
{\cal F}^+_{\mu\nu}\, , \quad {\cal F}^-_{\mu\nu}
\rightarrow (c\bar\tau +d) {\cal F}^-_{\mu\nu}\, . \ee 
This implies that
\be  \tau{\cal F}^+_{\mu\nu} \rightarrow (a\tau +b)
{\cal F}^+_{\mu\nu}\, , \quad \bar\tau{\cal F}^-_{\mu\nu}
\rightarrow (a\bar\tau +b) {\cal F}^-_{\mu\nu}\, . \ee 
Thus the equation of motion (\ref{eqa}) and the Bianchi identity (\ref{bianc})
transform into linear combinations of one another and are preserved.
In particular, the negative of the unit matrix sends
${\cal F}^{\pm}_{\mu\nu}\rightarrow - {\cal F}^{\pm}_{\mu\nu}$.
This result is acceptable if we identify the symmetry as $SL(2,R)$.
Note that $SL(2,R)$ is not a symmetry of the
action. The transformation in (\ref{modtr})
is a nonlocal transformation of ${\cal A}_{\mu}$, and
such transformations can do strange things to the action. For example,
the total derivative ${\cal F}\cdot \tilde{\cal  F}$ transforms into
an expression that is not a total derivative.\\
Thus far we have focused the attention to dilaton-axion system and the
gauge field part of the action. The explicit checks show that the rest of
the equations of motion are invariant under S-duality transformation. While
checking the invariance of the Einstein equation we must ensure that
that the contribution of $S_F$ to the energy--momentum tensor is
$SL(2,R)$ invariant. After a short calculation one finds that only terms of
the structure $e^{-\phi} {\cal F}^+{\cal F}^-$ survive,
and these are invariant since
$e^{-\phi} \rightarrow |c\tau + d|^{-2} e^{-\phi}$.
%%%%%% DISCUSS SL(2,Z)%%%%%%
 The symmetry of the equations motion is $SL(2,R)$. Notice that the axion
couples to the topological density term, product of $\cal F$ and its dual.
We can argue qualitatively that the part of the $SL(2,R)$ group which gives 
rise to the translation symmetry of the axion (VEV of $\chi$ is the $\theta$
angle) should break down to discrete group of translations due the
instanton effects. A more careful analysis is necessary \cite{senrev}
to show that $SL(2,R)$
breaks to $SL(2,Z)$.\\
The low energy string effective action, in four dimensions, contains
graviton, antisymmetric tensor, dilaton and nonabelian gauge bosons.
Furthermore, the Poincare dual of the three form field strength is a
pseudoscalar and this field can be identified as the axion. One can 
combine dilaton and axion to form a doublet of the S-duality group $SL(2,R)$.
It was argued \cite{ib,sjrey} that S-duality is an exact symmetry of
the string theory. Schwarz and Sen \cite{johnas} provided a general formulation
of S-duality in string theory.
%%%%%%%%% S-duality  in 4-dimensions %%%%%%%
Indeed the heterotic string compactified on $T^6$ has the effective action
of $N=4$ supersymmetric theory. How one can test S-duality in this case. One
of the important results in this direction was first derived by Sen \cite{senm}
when he showed that there are certain dyonic states in the theory whose
existence can be demonstrated using S-duality transformations on heterotic
string actions. These states precisely coincide with the ones we expect from
Montonen-Olive conjecture. The theory has electrically charged states and
magnetically charged states and each is 28-dimensional vector for the
heterotic string. Due to nonrenormalization theorem of $N=4$ supersymmetric
theory, the electric charges are not renormalized. Moreover, the spectrum of
the magnetic charges are fixed by the generalized Dirac quantization
condition; the magnetic charges are not renormalized either. Thus, spectrum
of theses charges will be same as in the tree level theory. Indeed, the
multimonopole moduli could be computed for the 
heterotic string \cite{senm}. 
In fact, the study of nonperturbative aspects of
supersymmetric Yang-Mills theories took new directions through the works
of Seiberg and Witten \cite{ne} in sequel to Sen's work.\\
It is interesting to look for extended objects which appear as  solution to
equations of motion of string effective action. Simplest extended object is
a string which is one dimensional. Let us denote the worldsheet coordinates
of this string as $\xi ^0 ~~{\rm and}~~\xi ^1$ and the spacetime coordinates
as $\{x^{\mu}\}$. This should appear as solution to string effective action.
Suppose, we consider a frame where $(\xi ^0,\xi ^1)$ lie along the spacetime
coordinates $(x^0,x^1)$ respectively. We look for a `spherically symmetric'
solution such that the solution is static and it depends only on the
magnitude of the transverse distance, $r={\sqrt {y_1^2+...+y_8^2}}$
where $x^2...x^9$ are denoted as $y_i$'s.The effective action 
 has graviton, dilaton and antisymmetric tensor fields. In the Einstein
frame the action
has the form
\be S_E={{1\over {\kappa ^2}}}\int d^{10}x{\sqrt {-g}}[R-{1\over 2}(\partial
\phi )^2-{1\over {12}}e^{-\phi}H^2] \ee
The macroscopic string solution which was identified with the heterotic
string \cite{dh} is obtained for following background configurations
\be ds^2=f^{-{3\over 4}}(-dt^2+(dx^1)^2)+f^{1\over 4}dy^idy^i \ee
\be B_{01} ={1\over f} \ee
The rest of the components of $B_{\mu \nu}$ are set to zero and 
\be f=1+{q\over {3r^6}} \ee
Here Q is the charge carried by the string and it is 
associated with antisymmetric tensor field. The field equations one needs to
satisfy are: Einstein equation, dilaton field equation and axionic charge
conservation which follow from field equation of  H. 
If we look at field equation carefully
there is a delta-function singularity at $r=0$ in the Laplace equation
$\nabla ^2 f $. Therefore, it was proposed \cite{dh}to resolve this problem by 
introducing a source for the string which will be the $\sigma$-model
action
 \be S_{\sigma}= {-T\over 2}\int d^2\xi[\partial ^aX^{\mu}\partial _aX^{\nu}
G_{\mu \nu} +\epsilon ^{ab}\partial _aX^{\mu}\partial _bX^{\nu}B_{\mu \nu}]
\ee
Here of course the metric $G_{\mu \nu}$ is the string frame metric. This
is the string solution carrying `electric' charge  and this charge
can be obtained from the conservation law. Indeed, $q=\kappa ^2 T/ {\omega _7}$
where $\omega _7$ refers to the volume of $S^7$. In the supersymmetric case,
there are BPS saturating solutions and here mass per unit length is equal
to the charge. \\
In four dimensions the dual of electromagnetic field tensor is also a two form,
thus if we have point particles, the dual objects are point-like 
('t Hooft-Polyakov monopoles look point like at large distances). However, if
we have a string in ten dimensions it couples to 3-form field strength the
dual of that field strength is 7-form. Therefore, the solitonic object for
the string is a 5-brane, extended in five spatial dimensions. In fact the
p-brane solutions were found in sequel to the string solutions \cite{ag}. 
As in case of monopole solution, we do not have magnetic source term while
looking for field equations (W-bosons carry electric charge), the solitonic
five-branes solutions are derived without adding a source term. Moreover,
if $e_2$ is `electric' charge of the string and $g_6$ is `magnetic' charge of
soliton, the Dirac quantization condition is
\be e_2g_6=2\pi n \ee
One has to be careful in deriving strong-weak duality relation here. The
coupling constant is determined in terms of dilaton expectation value.
The relations are $e_2 =e^{\phi _0 /2}$ and $g_6=e^{-\phi _0 /6}$. \\
There are special type of extended objects, the Dp-brane (D-branes), 
which carry R-R charges \cite{pod}. 
The type II theories admit gauge fields from the
RR sector. The corresponding effective contain these fields. If one look for
p-brane solutions with these gauge fields: strings, membranes and so on, they
have interesting properties. These are hypersurfaces or spacetime defects
on which the open strings can end. In D-dimensions, if there is a Dp-brane,
there are Neumann boundary conditions satisfied in (p+1)-directions, these are
the directions of the worldvolume coordinates of Dp-brane and we have
Dirichlet boundary conditions along the remaining transverse directions that
is $(D-p-1)$ coordinates. Written explicitly,
\be \partial _{\sigma}X^{\mu} =0, ~~for ~~ \mu =0,...p  \ee
\be X^{\mu}(\sigma=0,\pi)=a^{\mu}_0, ~~for ~~\mu=p+1,...9 \ee
A Dp-brane will couple to $(p+2)$-form RR field strength; therefore, D0-brane
is a particle,  D1-brane is a string and so on. The corresponding fermions
satisfy boundary conditions in accordance with the bosonic fields in order to 
maintain the worldsheet supersymmetry. The BPS saturated solutions, then
preserve half of the supersymmetry. From our earlier discussions, we note
that type IIA admits D0-brane and D2-brane (their dual objects too) and IIB
string has D-string, D3-brane and D-instantons, along with the duals. Thus,
we conclude that IIA has even D-branes and odd D-branes belong to IIB theory.
Of course, we are discussing the 10 dimensional case. The D-branes are
dynamical objects and there are excitation of such extended objects since
open string ends are attached to them.\\
Consider a situation when two D-branes are separated from each other. Since
open string ends can get attached to this surface, 
they will be connected
by open string/strings. The farther apart the two branes, it will cost more
energy to stretch the open string. More interesting is the configuration
when D-branes lie on top of each other. Then we can visualize an open
string starting from a brane and ending on it, open string starting from
one brane and ending on another coincident brane. In  this situation we
have massless states since there is no stretching of strings. 
Open strings contain massless vector state  in their spectrum. One can
incorporate nonabelian gauge symmetry for such a theory by introducing
the Chan-Paton factors. We can imagine a scenario where a quark belonging
to representation i of U(n) is attached to one end of the string and an
antiquark in representation $\bar j$ attached to the other end. Thus
the gauge field will carry index $i ~and ~ j$ like usual Yang-Mills fields and
these are called Chan-Paton factors.
 This characteristic of open string turned out to be
useful when we consider coincident D-branes. 
Therefore, if there are N
coincident branes, we get $U(N)$ Yang-Mills action, in fact we  get
supersymmetric gauge theory on the worldvolume of the brane.\\ 
Let us discuss some of the implications of dualities in the context 
of the branes we just introduced.  The experience from
monopole solution is that the charged particle couples to the field strength
tensor and the soliton couples to the dual tensor in four dimensions. In
ten dimensions, the solitonic counter part of string is five brane and
we saw that couplings are not really reciprocals of each other. If we
consider six spacetime dimensions, then we note that dual of 3-form field
strength is also another 3-form tensor and string couples to this tensor.
Therefore, the conjecture is  that in six dimensions there is string/string
duality. If there is a fundamental string the solitonic counter part is a 
string  too and their coupling constants satisfy the reciprocal relation.
For simplicity, consider a six dimensional reduced action, with only metric,
antisymmetric tensor field and the dilaton \cite{duffd}.
\be 
\label{sugra1}
I_6={1\over {2\kappa ^2}}\int d^6x{\sqrt {-G}}e^{-\phi} 
[R_G+(\partial \phi)^2
-{1\over {12}}H^2] \ee
Where $G_{MN}$ is six dimensional metric in string frame and $H_{NMP}$ is
the 3-form field strength associated with $B_{MN}$ and it is understood that
H is defined up to Chern-Simons terms. We can go over to Einstein metric
by the relation $G_{MN}=e^{\phi /2}g_{MN}$; $\phi$ being the  dilaton in 
six dimensions. Let us consider the dual six dimensional action
\be
\label{sugra2}
{\tilde I}_6={1\over {2\kappa ^2}}\int d^6x{\sqrt {-\tilde G}}e^{-{\tilde \phi}}
[R_{{\tilde G}} +(\partial {\tilde \phi})^2 -{1\over {12}}{\tilde H}^2] \ee
Here $\tilde \phi$ is the corresponding dilaton and $\tilde H$ is the
field strength of the  $\tilde B$, 2-form potential of the dual theory.
The two actions (\ref{sugra1}) and (\ref{sugra2}) are related if we identify
\be
\label{pduality}
\phi =-{\tilde \phi} ~~~{\rm and } ~~~{\tilde H}=e^{-\phi}*H \ee
The two metric being identified to be equal.
Here $*$ stands for Hodge dual. Note that just as in case of gauge field
kinetic energy term in four dimensions is conformally invariant, the $H^2$
term is also conformally invariant in six dimensions and it is immaterial
which metric we use while taking Hodge dual. As noted earlier, the
fundamental string solution with action(\ref{sugra1}) can be obtained by
adding a $\sigma$-model source term with coupling of  the G and B backgrounds.
The solution is given by
\be
\label{fstring}
ds^2=(1-{{q ^2}\over {r^2}})[-dt^2+(x^1)^2+(1-{{q ^2}\over {r^2}})^{-2}
dr^2+r^2d\Omega _3 ^2] \ee
\be  e^{\phi}=1-{{q ^2}\over {r^2}} \ee
\be e^{-\phi}*H_3=2q ^2\epsilon _3 \ee 
with
\be q^2={{\kappa ^2T}\over {\Omega _3}} \ee
Of course we have the BPS saturated mass relation
\be M=T<e^{\phi \over 2}> \ee 
Therefore, the mass density gets heavier as string coupling proceeds towards
strong coupling domain. The source free action (\ref{sugra1}) also admits 
solitonic string which is nonsingular and the solution is
\be
 ds^2=-dt^2+(dx^1)^2 +(1-{{{\tilde q}^2}\over {r^2}})^{-2}
dr^2+r^2d\Omega _3^2 \ee 
\be e^{-\phi}=1-{{{\tilde q}^2}\over {r^2}} \ee
\be H_3=2{\tilde q}^2\epsilon _3  \ee
Where ${\tilde q}^2 ={q ^2{\tilde T}\over {\Omega _3}}$ 
The mass density is
\be  {\tilde M}={\tilde T}<e^{- \phi\over 2}>  \ee
 In the weak coupling regime this string is heavier as one expects of 
a solitonic string. Notice that the solitonic string differs from the 
fundamental string by the replacement $\phi \rightarrow -\phi ,G_{MN}
\rightarrow {\tilde G}_{MN}, H\rightarrow {\tilde H}= e^{-\phi}*H,
{\alpha }'\rightarrow {\tilde \alpha }'$.
The Noether charge and the topological `magnetic' charge are respectively
given by
\be
e_2={1\over {{\sqrt 2} \kappa}}\int _{S^3}*H_3, ~~and ~~ g_2={1\over {{\sqrt 2}
\kappa}}\int _{S^3}H_3 \ee
The Dirac quantization rule for charges: $e_2g_2 =2\pi n$ gets translated to
relation between tensions. Moreover, the fundamental string and dual string
saturate Bogomolnyi bound for mass densities and break half of the supersymmetry
as expected. These solutions have the interpretation of being limiting cases
of more general solutions. They can be viewed as extreme mass equals charge
limit of two-parameter black string solutions.
\\
Again the question arises where can we test the string/string duality ? It
has been conjectured that \cite{hp,md,dk} heterotic string compactified
on $T^4$ is S-dual to type IIA theory compactified on $K_3$. When heterotic
string is compactified on $T^4$, the theory has charged states saturating
Bogomolnyi bound. On the IIA side,  elementary string states are neutral
since the gauge fields arise from RR sector. Moreover, for type IIA,
the analysis of the Bogomolnyi formula tells us the charged states (under
gauge fields) have their masses as ${1\over {g_{str}^{II}}}$, implying that
these are solitonic states. The duality between heterotic and type IIA
is understood in the  following sense\cite{senhe,hrs}: 
In type IIA theory, there are nonsingular
soliton solutions and these carry quantum numbers of fundamental heterotic
string. The properties of those strings are consistent with those of the
heterotic string. On the other hand the hetetotic string admits solitonic
solutions carrying the quantum numbers of type IIA string. Moreover, we know
that the moduli of heterotic string compactified on $T^4$ parameterize the
coset ${O(4,20)}\over {O(4) \times O(20)}$.  When type IIA is compactified on
$K_3$, the moduli also turns out to  be exactly the same. Therefore, there
is a very good evidence for this heterotic - type IIA duality conjecture.
Another duality relation, that has been verified,  is toroidal compactification
of IIA and IIB theory via  T-duality. Again the simplest one being 
compactification on $S^1$. If one theory is compactified on circle 
of radius $R$, it 
is equivalent to the other theory compactified on circle of 
reciprocal radius \cite{abc}, 
although in ten dimensions these are two different theories.  
Some of the important consequences  of S-duality  can be examined in type IIB
theory. It is conjectured that type IIB theory is self-dual and the effective
action can be cast in a manifestly $SL(2,Z)$ invariant form. We shall study
this aspect in the next section. The two heterotic strings i.e. $SO(32) {\rm
~and }~ E_8 \times E_8$ when compactified  on $S^1$ are T-dual to
each other in the reciprocal radius sense 
that one theory compactified on a circle of 
radius $R$ is equivalent to the other which is compactified on a circle
of radius $1\over R$. Finally, we comment that
heterotic string with $SO(32)$ gauge group is S-dual to type I theory
with $SO(32)$ group. The heterotic string effective action, with $SO(32)$
gauge group has the following form
\be
\label{heto}
S_{het}=\int d^{10}x{\sqrt {-g}}[R-{1\over 8}(\partial \phi)^2 -{1\over 4}e^{-
{\phi \over 4}}{\rm Tr}(F_{\mu \nu})^2 -{1\over {12}}e^{-{\phi \over 2}} H^2] 
\ee
Here $F_{\mu \nu}$ is the nonabelian field strength and $H =dB$. 
We work in the Einstein frame as it is the most
convenient frame to study S-duality properties, since this metric remains
invariant under S-duality.
This action is  obtained after rescaling the backgrounds and the slope 
parameters. The type I string has graviton and dilaton coming from the closed
string NS sectors and closed string RR sector gives the antisymmetric tensor.
The gauge fields
come from NS sector of the open string and they have to be in the adjoint
representation of $SO(32)$.
Again with appropriate scalings the effective  action can be brought 
to the following form
\be
\label{oneo}
S_I=\int d^{10}{\sqrt {-\bar g}}[{\bar R}-{1\over 8}(\partial {\bar \phi})^2
-{1\over 4}e^{{\bar \phi}\over 4}{\rm Tr}({\bar F}_{\mu \nu})^2-{1\over {12}} 
e^{{\bar \phi}\over 2} {\bar H}^2] \ee
Here all the fields of type I theory are defined with `bar' to distinguish from
those of heterotic string theory and the metric is in Einstein frame.
 Now, the comparison between the two actions shows
that they will be identical if
\be
\phi =-{\bar \phi},~~~g_{\mu \nu}={\bar g}_{\mu \nu},~~~ H_{\mu \nu \rho}={\bar
H}_{\mu \nu \rho}, ~~{\rm}~~A_{\mu}={\bar A}_{\mu}  \ee
Thus, if we compare the two actions,  (\ref{heto}) and (\ref{oneo}),
 we see that the two theories are related to each other by strong-weak
duality in 10-dimensions, since $g_{str}^2 =e^{\phi}$.
 There are host of duality relations among various string theories in diverse
dimensions; we refer the interested reader to large number of review articles
in this area.
%%% This is M-theory Compactifications... Tests  
\section{M-theory and Unified String Dynamics}

\setcounter{equation}{0}

\def\theequation{\thesection.\arabic{equation}}
We have briefly introduced some of the essential features of string theory
and their symmetry properties. There are five perturbatively consistent string
theories and one of their most attractive attributes is that they describe
 quantum gravity which is perturbatively finite and unitary. The dualities
are powerful symmetry properties which provide important information about
intimate connections between string theories. We have seen that one string
theory, in  a spacetime dimension, is related to another string theory either
through T-duality or by the S-duality. When two theories are S-dual to each
other, we can study strong coupling regime of one theory by going over to
the weak, perturbative domain of its dual theory. Therefore, the 
nonperturbative
aspects of some of the string theories could be investigated  
by these powerful tools. However, we still have five string theories.
Therefore, the natural goal is to search for a theory which will provide a
unified description of all the five string theories. The zero slope limits
of the string theories yield all the known  10-dimensional supergravity
theories. However, there is the $D=11$ supergravity theory consisting of
graviton and 3-form potential, endowed with total 128 bosonic degrees of 
freedom,  and the 128 fermionic degrees of freedom. It was shown several years
ago \cite{elcom}
that compactification of 11-dimensional theory on a circle gives rise to
$N=2$ supergravity theory in 10-dimensions. It was not possible to
establish any  relation 
between the  
11-dimensional theory  and any string string theory  for a long time. 
The connection of $N=2$, 10-dimensional  supergravity with string theory is
rather transparent since the supergravity actions can be obtained 
in the zero slope limit
of corresponding type II string theories. There was no string theory that
could be related in some such limit  to 11-dimensional supergravity.
Therefore, if 11-dimensional supergravity were to have any connection with
one of the string theories, then only the nonperturbative regime of a theory
will show the inter-relation. Moreover, when one views from the 11-dimensional
perspective, the supergravity theory does not have any small parameter, like
$e^{\phi}$, in string theory, which can be chosen to take 
 small value as an expansion 
parameter.\\
The connection between type IIA string theory and 11-dimensional supergravity
were recognised by Witten \cite{dk} and Townsend \cite{pkm}
following the developments in string dualities. The massless bosonic 
sector of the 
type IIA theory, we might recall from our discussions of Section II, consists 
of 
dilaton, $\phi$,
graviton, $G_{\mu\nu}$ and gauge field, $A_{\mu}$, antisymmetric
tensor, $C_{\mu\nu\lambda}$  coming from the NS and Ramond sectors respectively.
The effective action of type IIA theory
\bea
\label{twoa}
 S_{IIA}& =& {1\over {2\kappa _{10} ^2}}\int d^{10}x{ \sqrt {-G}}
\big [e^{-\phi}
(R+ (\partial \phi)^2
-{1\over {12}}H^2 )-({1\over 4}F^2 -{1\over {48}}{F_4'} ^2 )\big  ]\nonumber\\
& & -{1\over {4\kappa_{10}^2}}\int F_4 \wedge F_4 \wedge A   \eea
We have suppressed the Lorentz indices of the field strengths and we shall
define them now: R is the scalar curvature, $H_{\mu \nu \rho}$ is the field
strength of $B_{\mu\nu}$ 
from the NS sector, $F_{\mu \nu}$ is the field strength of RR gauge 
potential $A_{\mu}$ and in form notations, 4-form field strength, 
 $F_4'=dC_3+A\wedge dB$; $C_3$  being
the 3-form potential coming from the RR sector and B is the 2-form potential
whose field strength is H. Last term in (\ref{twoa}) is the Chern-Simons term.
and $F_4=dC_3$ is the antisymmetric 4-form field strength of potential $C_3$
A few remarks are in order at this point: the metric used in action (\ref{twoa})
 is the string frame metric. Note that the factor $e^{\phi}$ multiplies only
R and $H^2$ piece; fields coming from the NS sector. The reason is that in the
worldsheet supersymmetric formulation of NSR type II theories the R-R sector
fields through local worldsheet interactions (in NS sector the worldsheet
fields couple to potentials),  couple via bilinears of spin fields (in
fact to field strengths).  As a consequence, there are cuts and the usual
arguments that tree level term starts with ${1\over {g_{str}}}$ does not go
through. Thus we see this mismatch of $e^{\phi}$ between NS and RR fields
in the effective action. 
Now, it is easy to see that this theory will admit D0-brane and D2-brane and
their duals will be D6-brane and D4-brane from RR sector and a string
and its dual five brane from the NS sector.\\
Let us consider the bosonic part of the eleven dimensional supergravity action
\be
\label{elesug}
S_{11}={1\over {2\kappa _{11}^2}}\int d^{11}x{ \sqrt {-\tilde G}} \big[{\tilde
R}-{1\over {48}}{\tilde F}_4 ^2 \big ] -{1\over {12\kappa _{11}^2}} \int
{\tilde C}_3\wedge {\tilde F}_4\wedge {\tilde F}_4 \ee
Here the field with tilde belong to bosonic components of 11-dimensional
supergravity. Let us compactify one of the spatial dimensions on $S^1$,
following the procedure outlined in the last section. There will be a gauge
field and a scalar field,  when the metric is expressed is decomposed in terms
of the metric of the 10-dimensional theory. The 3-form potential will
decompose into  a 3-form potential but with additional piece according
to the procedure of \cite{ss,ms} and a two form potential will appear
as  well. It is most convenient to express the 11-dimensional metric in the
following form
\be
\label{compe}
{\tilde G}_{MN}=e^{-{1\over 3}\phi}\pmatrix {G_{\mu \nu}+
e^{\phi}A_{\mu}A_{\nu} & 
e^{\phi}A_{\mu} \cr
e^{\phi}A_{\nu} & e^{\phi} \cr} \ee
The dimensional reduction of (\ref{elesug}) goes over exactly
to the type IIA action (\ref{twoa}). Note that if we had not adopted this
form of the decomposition of the 11-dimensional metric with the over all
factor of $e^{-{1\over 3}\phi}$ and the factors of $e^{\phi}$ in various
places inside the matrix; but had compactified on a circle of radius, say,
 $\cal  R$;  we would have obtained a reduced action with 10-dimensional metric,
the moduli $\cal R$ and the antisymmetric tensor potentials (2-form and 3-form)
with appropriately modified C-S terms. The resulting action in ten dimensions 
would need some field redefinitions to match with the type IIA action.
Let us see how the radius of compactification $R_{11}$ is related to type IIA
string coupling constant $g_s^{(A)}$. Note from (4.3) that
$R_{11}^2= (e^{{({2\over 3})}\phi})^2$ and by definition 
$e^{\phi} = (g_{str}^{(A)})
^2$. Therefore, we conclude that
\be R_{11}=(g_{str}^{(A)})^{2\over 3} \ee
Therefore, in the perturbative regime of the type IIA theory, the radius of 
compactification of the 11-dimensional theory is very small. When we want to
go over to the decompactification regime i.e. large radius limit of 
11-dimensional theory, we can't realise that domain since it is the strong
coupling phase of the type IIA theory and perturbation theory does not provide
any clue for the existence of the 11th dimension in the ten dimensional theory.
The correspondence established between type IIA theory and 11-dimensional
theory is at the level of the effective action. The 11-dimensional supergravity
has a 3-form potential in the bosonic sector and the natural extended object
is a membrane. The 10-dimensional theory admits a string as a fundamental
object and supergravity action is zero slope limit of the string theory.
How can one establish the relation between membrane and the string ? The
idea of double dimensional reduction provides an important clue. One can
envisage a situation where we start from a membrane in eleven dimensions
and compactify 11th dimension on a circle. Then, according to the 
prescription of double
dimensional reduction \cite{dubld}, 
the membrane wraps around the compact direction so
that the end result is the ten dimensional string.\\ 
We have described in the  previous  section how one can establish 
connections among the five string theories various dimensions through duality
transformations in different spacetime dimensions; although there are five
distinct ten dimensional theories when viewed in the perturbative frame work.
The 11-dimensional theory is also recognised to play an important role in
string dynamics. It is believed that there is an underlying fundamental
theory, yet to be discovered, so that the  manifestations of the theory in its
various phases are realized through the string theories. It is postulated
that in the low energy limit, we should derive  the 11-dimensional supergravity
action as an effective theory. The unknown fundamental theory is named U-theory.
Since the 11-dimensional theory naturally admits membrane as a fundamental
extended solution, it has been argued that the underlying fundamental theory
is a theory of membranes. The M-theory is taken to be the underlying theory.
We shall illustrate, with a few examples, that starting with an eleven
dimensional theory with membrane, how one can obtain a host of relations
about the structure of branes in various string theories.\\
Since the BPS states do not get any quantum corrections, it is interesting to
look for BPS states and then propose tests for the theory. When we compactify
M-theory on a circle, the momenta in that direction will be quantized and
we shall get towers of KK massive states. These states will fall into 
representations of the 11-dimensional supergravity. In fact they are BPS states.
In the KK reduction, the charge of a state (in the lower dimensions) is related
to the momentum along the compact direction (thus automatically quantized) and
in some suitable units the charge is proportional to $m\over R_{11}$, $m$
being an integer. This is the charge associated with the gauge field $A_{\mu}$
as a result of $S^1$ compactification (\ref{compe}). From the type IIA point of
view this charge is that of gauge field coming from RR sector and the whole
tower should exist as BPS state. We know already that elementary string
states are RR-charge neutral and those massive towers belong to RR sector.
We can identify the state with unit charge, $m=1$ as a D0-brane
 of type IIA theory. The open string ends can get
attached to D0-brane and act as the collective coordinates to give excitations.
One can show that IIA theory has those BPS states belonging to the ultra-short
multiplets and these also correspond to the states counting done from
M-theory side. Therefore, we notice that duality between type IIA theory
and M-theory is established for such states. In case of $m>1$, the test is
not so simple. One of the properties of BPS states is that the binding 
energy for composite BPS states is zero. That means, 
if we have a single D0-brane, 
a BPS state with $m$ units of charge, we can't distinguish it from collection
of $m$ BPS particles each carrying  unit charge. 
Thus a test for the general case is 
rather difficult.\\
The relation between M-theory and type II theories can be established by
exploiting the duality relations. Note that type IIA and type IIB theories
are T-dual to each other when one of the directions is compactified. Since
M-theory with one compact direction, $S^1$,  is related to type IIA, 
therefore, M-theory with two compact dimensions, compactified on $T^2$,   is
expected to be intimately connected \cite{jhsm} to type IIB with one
direction compactified to $S^1$. 
We shall see that one  needs to  
exploit the $SL(2,Z)$ S-duality symmetry of type IIB theory in this context
\cite{jhstb}. The type IIB theory has graviton, 2-form antisymmetric potential,
$B^{(1)}_{\mu\nu}$ and dilaton, $\phi$ in the NS sector and 2-form potential,
$B^{(2)}_{\mu\nu}$, axion, $\chi$ and 4-form potential, $D_{\mu\nu\rho\lambda}$
in the RR sector; the field strength of D-field is self dual. For our purpose,
it suffices to drop the D-field from considerations presently. The action
is
\bea
\label{twob}
S_{str}& =& {1\over {\kappa ^2}}\int d^{10}x { \sqrt {-G}}
\big [e^{-\phi} \big (
R+(\partial \phi)^2 - {1\over {12}} {H^{(1)}} ^2 \big )
- {1\over 2}(\partial \chi)^2\nonumber\\
& &  -{1\over {12}}{\chi }^2{H^{(1)}} ^2 -{1\over 6}
\chi H^{(1)}\cdot H^{(2)} -{1\over {12}}{H^{(2)}} ^2 \big ] \eea
This action is written in the string frame metric. It is useful to go over to
the Einstein frame by the conformal transformation. Furthermore, to write
the Einstein frame action in a manifestly $SL(2,Z)$ invariant form, let us
define
\be
{\cal M}=\pmatrix {\chi ^2+e^{-\phi} & \chi e^{\phi}\cr
\chi e^{\phi} & e^{\phi}\cr}, ~~~~{\bf H}=\pmatrix {H^{(1)}\cr
H^{(2)}\cr } \ee
Then the action,
\be
S_E={1\over {2\kappa ^2}}\int d^{10}x  [R_g+{1\over 4}{\rm Tr}(\partial _{
\mu}{\cal M}\partial ^{\mu}{\cal M}) -{1\over {12}} {\bf H} ^T{\cal M} {\bf H}
\big ] \ee

This action is invariant under the transformations
\be
{\cal M} \rightarrow \Lambda {\cal M}\Lambda ^T, ~~{\bf H} \rightarrow 
(\Lambda ^T)^{-1}{\bf H} ~~~{\rm and }~~ g_{\mu\nu} \rightarrow g_{\mu\nu} \ee
If one looks for a string solution in this theory then the solutions
will be of three kinds: strings carrying NS charge, strings with RR charge
and ones with both NS and RR charge. The procedure adopted in \cite{jhstb} is
as follows: first look for a string solution with NS charge such that
asymptotic values of axion, $\chi _0=0$ and that of dilaton $\phi _0
=0$. In the language of complex moduli introduced earlier, asymptotic value
of $\tau _0 =i$. Moreover, one starts  with $H^{(2)}=0$, since one is
looking for a string carrying NS charge only. Next introduce a specific
$SL(2,Z)$ transformation such that the resulting string carries both types
of charges; the relevant matrix is
\be \Lambda ={1\over { { \sqrt {q_1^2+q_2^2}}}} \pmatrix {q_1 & -q_2\cr 
q_2 & q_1\cr } \ee

Although the string carries both types, NS and RR charges, still the modulus
preserves the  asymptotic value, $\tau _0=i$. Finally, introduce a general
$SL(2,Z)$ transformation so that $\tau _0$ will take arbitrary value as a
result of the duality transformation. The matrix is $\Lambda =\pmatrix {e^{-
\phi _0 /2} & \chi _0e^{\phi _0 /2}\cr 0 & e^{\phi _0 /2}\cr } $. As a 
consequence of the $SL(2,Z)$ transformation, not only we have strings
which carry charges $(q_1,q_2)$, but also the tensions of the strings
depend on these charges; after all these are BPS strings. The formula for
the tension of string with $(q_1,q_2)$ charges is
\be 
\label{btenson}
T_q=[e^{\phi _0}(q_2\chi _0-q_1)^2+e^{-\phi _0}q_2^2]^{1\over 2} T \ee
where T is the tension of the NS string one started with for which $\phi _0
=\chi _0 =0$ i.e. $\tau _0=i$. Since we consider $SL(2,Z)$ transformations,
$q_1 ~{\rm and }~ q_2$ should be integers. For stable strings, $(q_1,q_2)$
should be relatively prime; otherwise these string will decay into multiple
strings. If we compactify this theory on $S^1$, more interesting results
follow.The spectrum of the nine dimensional theory is governed by the mass
formula
\be M_B^2= ({m\over {R}})^2+(2\pi RT_qn)^2 +4\pi (N_L+N_R) \ee
Here we have explicitly kept the tension term showing that when the string
winds `n' times it stretches by its perimeter and energy is  obtained by
multiplying the tension $T_q$. The last term is sum of contributions from left
and right oscillators. The level matching condition tells us $N_R-N_L=mn$.
The BPS saturating multiplets have either $N_L=0$ or $N_R=0$; ultrashort
corresponds to both being zero. If we choose $N_L=0$, then mass formula is
(also level matching relation is used)
\be 
\label{mfm}
M_B ^2=(2\pi nRT_q+{m\over R})^2 \ee
We have a rich spectrum and these masses should remain protected from any
quantum corrections.\\
We can describe the same phenomena by compactifying the M-theory on $T^2$.
There is membrane in M-theory with tension $T_{11}$ and if it wraps m times
on a torus of area $A_{11}$, then the contribution to mass will be of the
form $mA_{11}T_{11}$. But the area is $A_{11}=(2\pi R_{11})^2\rho _2$, 
$\rho _2$ being the modular parameter of the torus and $\rho=\rho _1+i\rho _2$
the area is computed using 11-dimensional metric.
Since we are considering compactification of the M-theory on $T^2$,
the wave function of the two dimensional Laplacian (corresponding to
two coordinates on the torus) must satisfy periodicity property 
appropriate to the torus
and the mass formula should be suitably generalised with respect to  mass
formula for a string compactified on a circle.
\be M_{11}^2= \big [m(2\pi R_{11})^2\rho _2 T_{11} \big ]^2+{1\over {R_{11}}}
\big [l_2^2 +{1\over {\rho _2}^2}(l_1-l_2\rho _1)^2 \big] \ee
$l_1,l_2$ are integers which enter the mass formula as the contribution to the
KK part since the two dimensional Laplacian $-\partial _x^2-\partial _y^2$
acting on the wave function
\be
\label{mthm}
\psi_{l_1,l_2}(x,y) =exp\{{i\over {R_{11}}}[xl_2+{1\over {\rho _2}}y(l_1-l_2
\rho _1)] \} \ee
It is easy to see the periodicity property of the wave function by defining
$z=(x+iy)/2\pi R_{11}$, since the invariance now translates to
$z\rightarrow z+1$ and $z\rightarrow z+\rho$
In order to compare the above mass formula, obtained from M-theory with $T^2$
compactification, with the corresponding one (\ref{mfm})
from type IIB in 9-dimensions, we should recognize that (\ref{mthm}) is derived
using 11-dimensional metric. Therefore they could differ from each other by
a multiplicative constant: $M_{11}=C M_B$. Now the exact  matching of the mass
formula implies that the the modular parameters of $T^2$, denoted as 
$\rho$, should
be identical to the parameters of $SL(2,Z)$, $\tau$. Thus the modular
group appearing in $T^2$ compactification of M-theory , $SL(2,Z)$ is 
identical to the duality group, $SL(2,Z)$,  of type IIB theory.The 
following relations should be satisfied for matching of (\ref{mfm}) and
(\ref{mthm}) 
\be R^{-2}=TT_{11}A_{11}^{3\over 2},~~{\rm and} ~~C^2={{2\pi R_{11}
e^{-\phi _0/2}
T_{11} }\over T}  \ee
Since type IIA theory, in 9-dimensions, is related to IIB theory by T-duality,
we can also get some insight into D0-branes in IIA theory. The mass spectrum
of these (point) particles can viewed from two perspectives. 
One way is to  identify
the winding modes of the family of type IIB strings on the circle with the
KK modes of the torus;  and the other way of looking is to identify KK modes
of the circle with wrapping of the membrane on the torus. Again the mass 
formula matching relations can be used to relate parameters on
both sides.\\
There is also a five brane, the soliton counter part of fundamental membrane,
in the M-theory and these are the only two extended objects in the 
11-dimensional theory. Therefore, one 
expects that M-theory should be able to give description of NS branes
and Dp-branes in the lower dimensional string theories. Moreover, the membrane
tension of 11-dimensional theory, $T_2^{(M)}$, is the only parameter since
the 5-brane tension, $T^{(M)}_5$ is determined from the Dirac quantization
relation in terms of $T^{(M)}_2$. In order to study the branes in 9-dimensional
theory, we should see how different branes arise from M-theory and from type
IIB theory. The simplest is the 2-brane. In case of M-theory, the membrane
remains a membrane; but type IIB in 10-dimensions has no membrane (due to
the absence of  even RR 
field  strength), therefore, the D3-brane of ten dimensional IIB theory wraps
around $S^1$ to produce a membrane. But the type IIB string tension also
is related to $T^{(M)}_2$ since membrane wraps around torus to produce the 
string. So the simplest result is D3-brane tension and string tension of type
IIB are related: $T^{(B)}_3 ={1\over {2\pi}}(T_2^{(B)})^2$; this result involves
only IIB theory tensions derived through M-theory route. The 9-dimensional
IIB theory will have D3-branes too. They will arise, from M-theory view point,
as  wrapping of the 5-brane around $T^2$. There are 4-branes in 9-dimensional
IIB theory. Since 10-dimensional IIB theory has $SL(2,Z)$ pair of strings,
their solitonic partners 5-branes will come in same multiplets too. These
5-branes,  compactified on $S^1$,  will give the 4-branes in 9-dimensions. 
Similarly, one can discuss type IIA theory from both the perspectives in 
ten dimensions.  If we define $L=2\pi R_{11}$ as the perimeter of the
compact circle, then tension of IIA string gets related to $T^{(M)}_2$ since
IIA string, in 10-dimensions, will arise due to wrapping of membrane around
the circle. The relation is $T^{(A)}_1 =g_A^{-{2\over 3}}LT^{(M)}_2$. For
type IIA membrane we have $T^{(A)}_2 =g^{-1}_{A}T^{(A)}_2$. The 4-brane in
IIA theory will come from wrapping of M-theory 5-brane around the circle
and the relation becomes $T^{(A)}_4=g_A^{-{5\over 3}}LT^{(M)}_5$ and we 
must also have $T^{(A)}_5=g^{-2}_A T^{(M)}_5$. Then using relation
between $T^{(M)}_2 ~{\rm and}~ T^{(M)}_5$ together with the relation between
$T^{(A)}_4 ~{\rm and}~ T^{(M)}_5$, one can get an expression: $T^{(A)}_5=
{1\over {2\pi}}(T^{(A)}_2)^2$. \\
The purpose of above examples was to illustrate how one  can derive a large
number relations using the M-theory. In lower spacetime dimensions, the
theory provides a rich basis to understand branes coming from various theories.
We would like to record one important fact for our future considerations. Note
that the formula (\ref{btenson}) of general $(q_1,q_2)$ string is for a string
carrying NS and RR charges. For a string with only NS charge  the $(1,0)$ 
the tension
scales as $T\sim g_B ^{1\over 2}$ and for the one  
 carrying only unit RR charge it is $T\sim g_B ^{-{1\over 2}}$. In the
string
frame, after rescaling the metric,  we find that a string with one unit of
NS charge has tension of order 1 and the string with one unit of RR charge
has tension $g_B ^{-1}$. Therefore, the mass density also has same
dependence on coupling constant.\\
There are duality relations which relate compactified M-theory to other string
theories. One of the interesting cases is $E_8 \times E_8$ heterotic string
in ten dimensions \cite{hw}. 
There is no other string theory which can be related
to this one. So it is expected that $E_8\times E_8$ is connected to the
11-dimensional theory. But it cannot be $S^1$ compactified M-theory, because
that compactified theory is type IIA as we have seen. Moreover, 11-dimensional
theory as such is free from anomalies. However, if one considers 
compactification on ${{S^1 }\over {Z_2}}$ of 
M-theory it gets related to $E_8\times E_8$
ten dimensional theory. The orientation of $S^1$ is reversed under $Z_2$ and
it flips the sign of 3-form potential C. As a consequence of this projection,
we are left with the metric in ten dimension, the dilaton and 2-form potential.
The gauge boson and 3-form potential C are projected out. The surviving
fermions are Majorana-Weyl gravitino and Majorana-Weyl fermion. This is the
supergravity in the bulk. Actually, ${{S^1}\over {Z_2}}$ is a line segment with
fixed points at the boundary. These are two copies of 10-dimensional flat 
space. The states from twisted sector should be localized on these planes. It
was shown that half of the anomalous variation is localized in one plane
and the other half on the other plane. The possible gauge groups that can
cancel the anomaly are $E_8 \times E_8$ or $U(1) ^{496}$. It is obvious
the string theory to be identified is the $E_8 \times E_8$ heterotic string. 
There are
other duality conjectures \cite{d1,d2,d3} between M-theory and other string
theories in lower dimensions: M-theory on $K_3$ $\leftrightarrow {heterotic/
Type~I}$ on $T^3$. Compactification of M-theory on ${{T^5}\over {Z_2}}$ is
dual to type IIB on $K_3$. The lower dimensional compactifications
${{T^8}\over {Z_2}}~  {\rm and }~~{{T^9}\over {Z_2}}$ are related to
Type I/ heterotic on $T^7$ and type IIB on ${{T^8}\over {Z_2}}$ respectively.\\
There are attempts to construct gauge supersymmetric gauge theories by choosing
suitable combinations of intersecting branes and establish Seiberg-Witten
dualities from this M-theory point of view of SUSY Yang-Mills gauge theories. 
Undoubtedly, the dualities together with the proposal for M-theory has brought
us nearer to the goal of unified description of string theories. However,
the underlying fundamental theory is yet to be discovered although we have
seen many facets of that theory.

%%% BLACK HOLES %%%%
\section{ Black holes and String Theory}

\setcounter{equation}{0}

\def\theequation{\thesection.\arabic{equation}}

The physics of the black holes has many fascinating aspects. The classical
black hole is the final stage of a collapsing heavy star. As the name suggests,
matter falls into it and nothing comes out; there is an event horizon. However,
deeper investigations have revealed, almost a quarter of a century ago, 
that there
are strong similarities between thermodynamics and black hole mechanics 
\cite{b1,b2}. If M is mass of the black hole,
\be 
dM={1\over {8\pi G}}k dA ,~~~~\delta A\ge 0  \ee
Here G is the Newton's constant, A is the area of the event horizon and $k$ is
the surface gravity. This is to be compared with thermodynamical relation,
\be dE=TdS , ~~~~\delta S\ge 0  \ee
Hawking's startling discovery \cite{swh} that black holes radiate with a black 
body spectrum of temperature $T={{\hbar k}\over {2\pi}}$, when quantum effects
are accounted for, raised several important issues in black hole physics.
One can also associate entropy with a black hole
\be S_{BH}= {A\over {4G\hbar}} \ee
The thermodynamical relations used to describe macroscopic phenomena can be
derived from statistical mechanics starting with microscopic fundamental
laws of physics. Since $\hbar$ appears in the black hole entropy formula, it
is expected that the microscopic derivation of black hole entropy requires
quantum gravity calculations. Moreover, entropy of a system, when interpreted
from statistical mechanical point of view, counts the total number of degrees
of freed in the system. How do we count the number of degrees of freedom in
a black hole and  obtain the expression for entropy? There are more fundamental
issues related to quantum mechanics when we carefully examine the implications
of Hawking radiation. We can think of allowing some matter to go into the
black hole, prepare the initial state as  a pure quantum state to be the 
incident wave.
However, the emitted Hawking radiation has a black body distribution and
thus these are mixed states. Therefore, the
S-matrix that will describe the above process will loose its unitarity property.\\
In the perturbative regime, string theory can provide reliable results
for computations of processes involving graviton. The resulting S-matrix
elements respect the required unitarity and analyticity properties.
Thus, it is
pertinent to ask what string theory has to offer in resolving the issues 
alluded to earlier. Recently, one of the important 
achievements of the string theory
has been the microscopic derivation of the black hole entropy, for a 
special class of black holes that arise in string theory. We shall, initially,
not set $G=1$, to bring out a few salient points in discussions of stringy
black holes and some times we shall display presence of $\hbar$ in formulas.
Recall, that the Newton's constant is related to string coupling and tension
as $G\sim g_{str} ^2 /T$, in four spacetime dimensions. If we have a massive
string state, the gravitational field is $GM_s$, where $M_s$ is mass of a 
string state measured in units of T; also some times we shall denote it as M.
Thus, the field increases as string coupling increases. String states are
given my the mass formula $M^2=NT$ and it is well known that at a given mass
there are a lot of states and the degeneracy \cite{fubvn} 
grows exponentially with mass, i.e. 
$e^M$. Thus one might think that the excited states, if treated as black holes,
will reproduce the entropy formula; however, this simple argument in not
adequate since black hole entropy grows like $M^2$, whereas the naive argument
will give $S_{BH}\sim M$. There have been attempts to explain this discrepancy
saying that the mass that would appear in microscopic derivation of $S_{BH}$
is not the same as the one appearing in Bekenstein-Hawking formula and
there might be renormalization effects to be accounted for \cite{susk}. The
perturbative string states appear in infinite levels and thus, for high enough
mass,  the massive elementary string state will lie inside the Schwarzschild
radius associated with it. Consequently,  they will 
require black hole descriptions.
One  of the  ways to  derive black hole entropy 
microscopically is to consider
such BPS states, so that when string coupling gets strong, the state is
unchanged. In this approach \cite{ampl}, first step is to pick up appropriate
BPS state and compute the microscopic entropy. Next, compute the 
Bekenstein-Hawking entropy of the BPS state, it is also an extremal black hole,
and verify whether the two ways of calculating entropy are in agreement. This
is the first clue that string theory might explain black hole entropy in
microscopic way. However, the black holes constructed from the 
elementary string states had some short comings while computing the entropy. 
The area of the
event horizon, for such black holes, tends to zero as one approaches the
extremal limit; moreover, the dilaton also diverges at the horizon in this 
limit. This problem was encountered for string states in the NS sector.\\
The D-brane in RR sector can come as elementary states and there are 
corresponding
solitonic states contained in the full spectrum. 
We had argued in the context of type IIB $SL(2,Z)$ strings
that in string frame metric, NS states have tensions of order 1, whereas,
D-strings had mass density of the order of ${1\over {g_{str}}}$. For the
solitons of NS sector the mass goes  as ${1\over {g_{str} ^2}}$; but the
solitons for RR sector still have mass order ${1\over {g_{str}}}$. In the
weak coupling regime NS solitons and RR ones are heavy. We should account for
the gravitational fields they produce, which is 
$GM$. In view of above discussions,
(i) NS elementary states produce very low field and (ii) RR states also 
produce low field in weak coupling limit; field tends to 0 as 
$g_{str}\rightarrow 0$. We may argue that in this regime, flat spacetime is a
good description of the geometry. Since we are dealing with BPS states, as
string coupling increases the mass remains unchanged, but the gravitational
field keeps increasing and after some critical 
coupling, the spacetime is not flat any
more; we must employ general theory of relativity. If these states describe
black holes, then we should be able to compute 
the degrees of freedoms associated with them.
It is possible to construct black hole configuration such that 
the area of the horizon in not  zero nor the dilaton diverges at
the horizon, when we take the extremal limit. For five dimensional black
holes, we need at least three charges to have  nonzero area for the 
horizon together with constant value for the dilaton at the horizon.
In case of the four dimensional
black hole needs four charges in order to satisfy the requirement of nonzero
horizon area and finite value of dilaton (at the horizon).\\
The black holes which we shall consider now have some special characteristics.
They can be thought of as composites of many D-branes carrying Ramond charges.
We have mentioned before that the BPS states have the property that mass of
composite BPS state is the sum of the masses of the constituents. One starts
in the weak string coupling phase with such D-branes and proceeds towards
strong coupling domain when gravity becomes strong. In weak coupling regime,
the degeneracy of the level can be estimated reliably and microscopic entropy
can be computed. In the strong coupling domain, the D-brane is inside the
horizon and one can treat this like a black hole and compute the ratio $A\over
{4G}$, which is independent of string coupling $g_{str}$ since both area
and Newton's constant grow like $g_{str} ^2$.\\
Let us discuss how the five dimensional black hole configuration is constructed
with D-branes \cite{ascv}. We start with 
type IIB theory in 10-dimensions. We know that  
it will admit D1-string and  D5-brane.
We want to make the composite object heavy; therefore, we put $Q_5$ 
number of D5-branes and $Q_1$ number of D1-strings together. 
Let us compactify this theory on $T^5$ such that the $Q_5$ number of D5-branes
are wrapped around $T^5$, the $Q_1$ D1-strings wrap along one of the directions
of the torus. Then put some momentum along the direction in which the
D-string wrapped; this momentum will be quantized in units of inverse radius
of $S^1$. The aim is to evaluate the microscopic entropy by counting
number of degrees of freedom for this system and it involves some
detail technical steps \cite{wittg,senm,vafa1,vafa2}; but we shall outline
only essential points.
We expect to have a $U(Q_5)$ supersymmetric Yang-Mills theory
on  the D5-brane worldvolume. This will be a gauge theory in $5+1$ dimensions
which is derived by dimensional reduction of $N=1$ supersymmetric
Yang-Mills theory from ten dimensions \cite{wittg}. The D-string is inside
this pack of D5-branes ($Q_5$ of them). The D-string can be viewed as an
instanton in this six dimensional spacetime, since an instanton in 6-dimensional
theory with no time dependence and extension in one direction is a string.
There are $Q_1$ such strings in the D5-brane configuration. Their low
energy dynamics is described by two dimensional supersymmetric sigma
model in $4Q_1Q_5$ dimensional hyper Kahler manifold. Every boson contributes
factor 1 and every fermion contributes $1\over 2$ to the  central charge as
we noted in Sec. II. Thus, total central charge is 
\be c=6Q_1Q_5 \ee
Since we are dealing with BPS states, for these states $L_0=0$ and
the momentum given along $S^1$ is related to the difference $L_0-{\bar L}_0$.
If we take momentum to be large i.e. $P_s=-{n\over R}$, $n$ large; then
using Cardy's result (relating degeneracy to central charge), one
gets
\be d(Q_1,Q_5,n) ={\rm exp}(2\pi {\sqrt {Q_1Q_5n}}) \ee
The black hole entropy computed from the microscopic view point is given by
\be S_{microscopic} =2\pi {\sqrt {Q_1Q_5n}} \ee
In order to derive the black hole entropy, $S_{BH}$, from Bekenstein-Hawking
formula, we have to specify the metric, the charges and then compute the area
of the event horizon in the extremal limit.\\
There is way to visualize the physical processes that lead to microscopic 
\cite{sdsm}
derivation of the entropy formula. The D-string is inside D5-brane and the 
low level excitations are the lowest lying modes of the open strings attached
to this one. If we think of the physical degrees of freedom, these are
8 transverse vectors and their super partners. Since these have to satisfy the
Dirichlet boundary condition, they are constrained to move along the D-string.
We are dealing with BPS state, therefore, these move only in one direction (say
left). Since the D-string is wrapped around one circle of $T^5$, we choose
$x_1$, then length is winding number times the radius of the circle. But
the momenta of individual open strings moving on this unidirectional path on
the circle is quantized. Moreover,  sum of their momentum is constrained too by
the total momentum we have put on that direction. Therefore,
 this is analogous to
solving statistical mechanics of a one dimensional system on a circle where
total energy (momenta are same as energy) is fixed.\\
The next step is to define the metric for the above configuration of the branes
and the obtain the harmonic functions that are necessary to satisfy the
equations of motion for the brane configurations \cite{cmalda,maldath}.
\bea
\label{metric} 
ds^2& =& H_1^{1/2} H^{1/2}_5  \bigg\{[ H_1^{-1} H_5^{-1}
(-K^{-1}dt^2 + K (dx_1 - (K^{-1}-1)dt)^2)\nonumber\\ 
& & +H_5^{-1}(dx_2^2 + \cdots + dx_5^2)
+ dx_6^2 + \cdots + dx_9^2 \bigg\}
\eea
We specify the compact directions as follows: the $Q_5$  number of D5-branes
are wrapped in $x_1,.....x_5$ directions, D-string is wrapped in $x_1$ and
the momentum is along $x_1$ too. Since we toroidally compactify to five
dimensions $x_i,~ i=1,...5$ are periodic and the radius of compactification
is $R_i$ along $i$th direction. 
and
\be
e^{-2 \phi} = H_1^{-1} H_5,~~~~ 
B_{01}=H_1^{-1}-1 \ee
\be H_{ijk}={1\over 2} \epsilon _{ijkl} \partial _l H_5, 
~~~~~ i,j,k,l=6, ......,9 \ee
\be r^2=x_6^2 + \cdots +x_9^2
\ee
The harmonic functions are equal to
\be 
\label{har1}
H_1=1+C_1 {Q_1 \over r^2}, ~~~~~ C_1={ g_{str} {\alpha '}^3 \over V}
\ee
\be 
\label{har2}
H_5=1+C_5{Q_5 \over r^2}, ~~~~~ C_5= g_{str} {\alpha }' \ee 
\be  K=1+C_K{Q_K \over r^2}, ~~~~~ C_K={n g_{str}^2 {\alpha '}^4 
\over {R_1^2 V}}
\ee
where $V=R_2 R_3 R_4 R_5$, we displayed the $\alpha '$ dependence to show
how the dimensionality of the charges appear, but now on we set the slope
to unity as usual. Let us briefly note how the charges arise in this black
hole. There is electric charge $Q_1$ coming from $B_{01}$ which is a gauge
field now, after compactification of $x_1$ coordinate. The $Q_5$ charge is
magnetic type originally attributed to D5-brane in 10-dimensions. After
compactification the  Poincare dual of that 3-form RR field strength is two
form field strength and it becomes an electric charge counting D-brane 
charges. Of course, the third charge comes from momentum given along $x_1$
direction and is quantized. When any one of these charges vanishes, the area
of the event horizon vanishes too. 
The  dimensional reduction \cite{ms} over the periodic coordinates
$x_1, ...., x_5$, yields the 5-dimensional effective action. The metric
in the five dimensional space takes the following form   
\be
ds^2 = \lambda ^{-2/3} dt^2 + \lambda ^{1/3}(dr^2 + r^2  \Omega _3^2)
\ee
where
\be
\lambda = H_1 H_5 K =(1+C_1{Q_1 \over r^2}) (1+C_5{Q_5 \over r^2}) 
(1+C_K{Q_K \over r^2})
\ee
This corresponds to an  extremal charged black hole and the horizon is 
located at $r=0$. However, the area of the horizon is nonzero and it is
proportional to the product of the charges. The expression for the
area is 
\be
A_5=(r^2 \lambda ^{1/3})^{3/2} \big| _{r=0}  = \sqrt{C_1Q_1C_5 Q_5 
C_KQ_K} (2 \pi ^2) = {{g_{str}^2}\over {R_1V}} \ee
The Newton's constant in five dimensions gets related to the ten dimensional
Newton's constant after we compactify on $T^5$ and the relation is
\be
G_N^{(5)}={G^{(10)}_N \over (2 \pi)^5 R_1 V}={1\over 4}{{\pi g_{str}^2}\over {R_
1V}}  \ee

Therefore, the entropy is equal to
\be \label{extS}
S_{BH}={A_5 \over 4 G_5} = 2 \pi \sqrt{Q_1 Q_5 n}
\ee
This expression exactly agrees with the expression for $S_{microscopic}$.
A few comments are in  order to discuss the constraints on the parameters
for the above relation to be valid. The string effective action  adopted
to obtain the brane solutions is valid when string loop corrections and
$\alpha '$ corrections are nonleading. The string loop corrections are small
when $g_{str} \rightarrow 0$ with the values of the charges held fixed. The
charges correspond to characteristic scales of the system. If we want ignore
$\alpha '$ correction terms then the charges should be larger than string scale
i.e. $Q_1, ~ Q_5 ~{\rm and} ~  n$ are much larger than $\alpha '$.
If the compactification radii of the torii be taken as order of string length
scale, then we should have $g_{str}Q_1,~ g_{str}Q_5,~ g_{str}^2n~~ >>~1$.
This tells us that $n~>>~~Q_1~ \sim ~Q_5 >>~ 1$.
\\
The entropy of nonextremal black holes can be considered in a similar manner;
however, we must keep several points in mind. First of all, the extremal black
holes are BPS stated and they get no quantum corrections. Therefore, whereas the
microscopic entropy is computed in the weak coupling phase, the 
Bekenstein-Hawking entropy is obtained after we go over to the strong coupling
domain so that the composite D-brane configuration lies inside the horizon. In
case of nonextremal black holes, we have no theorem against quantum corrections
and therefore, passage to strong coupling limit is not so simple. It is argued,
that a black hole which is slightly away from extremality might allow smooth
increase of the coupling constant as one starts from weak coupling limit.
This type of black holes configuration can be achieved by allowing some
low level right moving oscillators compared to the high left moving levels
(note that for extremal case $N_R=0$).
We shall not discuss the properties of these black hole in detail here.\\
The BPS extremal black holes are stable and they have zero temperature; 
therefore, 
they will not emit Hawking radiation. If we intend to understand the Hawking
radiations from black holes in string theory, we have to look for those ones
which are excited states and can decay into lower energy state. The starting
point is to consider a nonextremal black hole. Since there will be left and
right movers, the open string states will be going in opposite directions
on the D-string. Again, it is a one dimensional problem where one can
imagine that two oppositely moving open string 
states collide to give a closed string
state. If we were to calculate the S-matrix element for such a process, we
shall consider initial state, final state and a suitable interaction 
Hamiltonian for our computational purpose. 
In order to get the emission rate, one will take modulus  square
of this amplitude, average over initial states, sum over final states and
divide by usual phase space factor. The state of the initial nonextremal
black hole is given by occupation numbers $N_L ~{\rm and}~N_R$ and the amount
of momentum we give on the compact circle which are going in opposite
directions. The momenta are quantized as $n\over R $ in either direction and
thus the closed string state will carry momentum $2n\over R$. As we have seen
 there are  
$4Q_1Q_5$ bosonic and fermionic oscillators. The string theory calculation
gives the amplitude for emission of a closed string state from these initial
state \cite{cmalda}. The sum over final state and averaging over initial
states leads to a factor $\rho _L\rho _R$,  where for example
\be \rho _R={1\over {N_i}}\sum _i \langle i|N_R|i\rangle \ee
where  $N_i$ is the total number of initial states and $N_R$ is the number
operator of right movers. We might carry out the averaging over all possible
initial states with a given value of $N_R$ by adopting the statistical 
mechanical prescription. The problem actually maps to the case of one 
dimensional gas and the microcanonical ensemble can be used since we are
holding $N_R$ fixed; energy is held constant.  The configuration of the
black hole is such that $N_L~>>~N_R ~>~1$.  
If $k_0$ is the momentum of out going massless closed string the final 
calculation give the decay rate as
\be d\Gamma \sim ({\rm Area}) {{e^{-{{k_0}\over {T_R}}}}\over {1-e^{-{{k_o}\over
{T_R}}}}}d^4k \ee
A more careful calculation \cite{sdsm1} reveals a surprising result that
not only the form of thermal distribution is recovered, but also
the numerical coefficients match with semi-classical results of Hawking. The
result has been derived for four dimensional black holes as well \cite{gk}. It
is an interesting question to ask whether one can calculate the absorption
cross section of an extremal black hole for a closed string massless
scalar and then relate that cross section to the decay rate of a nonextremal
black hole  by using the principle of detailed balance in quantum mechanics
taking into account all the subtelities. Indeed explicit verification
shows that such a check yields the correct result \cite{dmw}.

%%% M(ATRIX) MODEL %%%%

\section{M-theory and the M(atrix) model}

\setcounter{equation}{0}

\def\theequation{\thesection.\arabic{equation}}

Our present understanding of string dynamics together with duality symmetries
strengthen the belief that there is a fundamental theory and the five 
perturbatively consistent theories are different phases of that underlying
theory. However, we do not know what this theory is except the conjecture
that the low energy limit of this theory is the 11-dimensional supergravity
action. There are deep questions about the structure of this theory. We
shall call it the M-theory. We recall that strong coupling limit of the
type IIA theory is identified with 11-dimensional supergravity. When viewed
from type IIA perspective, the existence of D0-branes as nonperturbative
RR point like objects is quite important for our discussion. They are BPS
states and their mass is of the order $1\over {g _{str}}$ and scaled by 
10-dimensional length scale $l_s$. These being BPS states, one could assume
that there are threshold bound states of many, say N, D0-branes which satisfy
the properties of bound BPS states.  Now
if we take the strong coupling limit, then it is found that the low energy
spectrum is same as the spectrum of the 11-dimensional supergravity. This
is an important evidence. Furthermore, the 11-dimensional theory is known to
admit membrane and five brane and we have argued how one can study
properties of various brane configurations in string theories after
compactifying the M-theory. The M(atrix) model 
\cite{matrix,r12,r13} can describe perturbation  
expansions of various string theories. There is  a limit in which the theory
provides connection with 11-dimensional supergravity theory. However, one
would like to seek answers to several questions from this theory. For
example, the general prescription for the compactification of the theory
is not known. Similarly, the complete set of degrees of freedom of this theory
is to be obtained. The M(atrix) theory, nevertheless, provides insight into
nonperturbative definition of string theory and it also exhibits string
dualities \cite{ab}. One can also go over to various string theories by adopting
different limiting prescriptions.\\
The model resorts to infinite momentum frame (IMF) technique boosted along a
compact direction. The momenta along compact direction is quantized; and
one starts with N units of these momenta and then  
$N\rightarrow \infty$ limit is taken. 
Since one is working in the light-cone frame
while constructing M(atrix) theory, the theory is not manifestly Lorentz
invariance. Thus Lorentz invariance might be recovered in the large N limit.
In the M(atrix) model formulations one encounters parameters which have
the interpretation of being expectation values of scalars when viewed
from  the string theory  side. But in the M(atrix) model when we have IMF
formulation, these constant modes have infinite frequency and they are frozen
into fixed configuration. The theory in its  present formulation 
is not background
independent. Moreover, one encounters problems while compactifying the theory on
an arbitrary d-dimensional torus. We may remind the reader that the M(atrix)
theory provides  a rich structure to study various aspects of string
theory from M-theory stand point.\\
The infinite momentum frame (IMF) technique played a very useful role in
current algebra \cite{curr}. In field theoretic calculations it simplifies
perturbation theory calculations \cite{stw,lsjk}. When we have to deal with
a collection of particles, we can define IMF to be a frame where the total 
momentum is taken to be very large. If we designate particles by index I, J...
then
\be P_I=\eta _IP+P_{TI} \ee
where T stands for 'transverse' and $P\cdot P_{TI} =0, \sum P_{TI} =0
~{\rm and} ~ \eta _I \le 1 $.  For a highly boosted coordinate system we could
have all $\eta _I$ positive. Particularly, for the case at hand, 
we deal with massive
particles and we can choose an appropriate frame to satisfy our requirement.
Energy of any particle satisfies relativistic relation
\be 
\label{imf}
E_I={\sqrt {P_I^2+m_I^2}}=\eta _IP +{{P_{TI}^2+m_I^2 }\over {2\eta _IP}}+..
 \ee
it is understood that there are terms higher order in $1\over P$ denoted
by dots. The expression for energy is similar to that of a nonrelativistic
particle in a lower dimension with mass term taking a modified form. When we
use a light-cone (LC) frame, a spatial direction is identified and designated
as longitudinal. The longitudinal 
momentum is $P_{LI}=\eta _IP$  and one defines $P_{\pm I}=
E_I\pm P_{LI}=E_I\pm \eta _IP$. The mass shell condition translates to
$P_{+I}P_{-I} -P_{TI} ^2 =m_I ^2$ and we can rewrite this relation as
\be E_I-\eta _IP={{P_{TI} ^2+m_I^2} \over {P_{I+}}} \ee
In the limit of large P, we have $\eta _I P$ large and therefore,
$E_I \rightarrow \eta _IP$ with $P_{I+} \sim 2\eta _I P$. When M-theory is 
envisaged in IMF, let us designate the momenta as $p_0, p_i, i=1,..9 ~{\rm and}
~ p_{11}$. One compactifies 11th direction with  and this is also boosted,
therefore $\{p_i \}$ are collectively denoted as $p_T$. Thus for collection
of the D0 particles
\be 
\label{lce}
E-p_{11}^{total}=\sum _I{{p_{TI}} ^2 \over {2p_{I11}}} \ee   
We note that there are 32 real supercharges in the theory. When one adopts
IMF description, it is convenient to split them into two groups each having
16 of them. The charges in every group transform as spinors of $SO(9)$.
Let us denote charges as $Q_{\alpha}, \alpha =1,...16, ~{\rm and } ~
q_A, A=1,.. 16$. The algebra of these charges  are
\be  \{ Q_{\alpha}, Q_{\beta} \} =\delta _{\alpha \beta} H,~~
\{ q_A,q_B \}=\delta 
_{AB}P_{11} \ee
\be \{ Q_{\alpha},q_A \}=\gamma_{A\alpha}^iP_i \ee
Here H is the Hamiltonian operator, P's are the corresponding momentum 
operators and $\gamma _i$ are 16 dimensional gamma matrices.\\
We have discussed earlier, how D0-brane has a natural interpretation from
the 11-dimensional theory with a compact coordinate and the RR charge is
related to quantized momenta along this direction. The relation between mass
and charge is satisfied since these are BPS states. There exists a sector
with N units of D0-brane charge,  
carrying Kaluza-Klein momentum $N\over {R_{11}}$. If we hold N fixed
and take a limit $R_{11} \rightarrow 0$, we go over to the weak coupling
phase of string theory; however, in the passage to this limit, the string
scale is not held fixed. The aim is to study the phenomena in the
11-dimensional theory and thus $l_{11}$ is to be kept fixed. We recall
that 
\be R_{11} ^3 = g_{str} ^2 l_{11} ^3 \ee
 and the string length scale, $l_s ^2=
{{l_{11} ^3}\over {R_{11}}}$. Thus as the compactication radius tends to zero
string scale diverges. We have also seen earlier, as the radius shrinks, the
mass of D0-brane tends to infinity, when measured in 11-dimensional Planck units
in ten dimensions. In other words the mass of D0-brane is
\be {1\over {g_sl_s}}={1\over {R_{11}}} \ee
and therefore,  it is appropriate to identify them as the KK modes.
Thus, when we consider mass of the these particles in
10-dimensions, in scales of eleven dimensional theory, the particles become
very heavy and a nonrelativistic description is quite adequate. If we were
to describe M-theory in terms of type IIA zero branes, then we have a
scenario where M-theory is equivalent to $N\rightarrow \infty$ limit of
the nonrelativistic quantum mechanics of N D0-branes which are in weak coupling
phase of type IIA theory. Furthermore, as Witten has argued \cite{wittg} the
physics of N coincident D0-branes is described by dimensionally reducing
ten dimensional $U(N)$ supersymmetric Yang-Mills theory to $0+1$ dimensions
\cite{dsusy}. 
Let us consider supersymmetric quantum mechanics of a single D0 particle.
The starting point is the action 
\be 
\label{dzero}
\int dt {\rm Tr}\bigg ( {1\over {2g_{str}}}(D_0X^i)^2-i\theta ^TD_0\theta
+{1\over {4g_{str}}}([X^i,X^j])^2+\theta ^T\gamma _i[X^i,\theta] \bigg ) \ee
This is the action obtained from 10-dimensional super Yang-Mills theory reduced
to one dimension. Here $i=1,...9$ stands for transverse directions and $\theta
$ are real spinors with 16 components. Since $X^i ~{\rm and}~ \theta$ come
from the gauge groups, they are in the adjoint representations of $U(N)$.
Since they carry only time dependence, these are $N\times N$ matrices. $D_0
= \partial _t+[A_0,]$ is the covariant derivative and this can be converted to
ordinary derivative with the gauge choice $A_0=0$. The mass of D0-brane is
order $1\over {g_{str}}$, thus the first term in (\ref{dzero}) can be 
written as $\int dt {M\over 2}({{d X^i}\over {dt}})^2$.   
Note that the action (\ref{dzero}) contains parameters of type IIA theory. It
is convenient to scale $X^i={g_{str}}^{1\over 3} Y^i$ which amounts to 
rescaling of the metric to that of 11-dimensional theory.
Moreover, one scales the time variable as $t={g_{str}}^{2\over 3} \tau$
and denotes the $\tau$ derivative by a dot. 
The action is rewritten as
\be
\label{sqm}
S=\int d\tau {\rm Tr}\bigg ({1\over {2R_{11}}}({\dot Y}^i)^2 -
i \theta ^T{\dot \theta}
+{{R_{11}}\over 4}([Y^i,Y^j])^2 +R\theta ^T\gamma _i[Y^i,\theta] \bigg )
\ee
If $\Pi _i={{{\dot Y}^i}\over R_{11}} ~{\rm and} ~ \pi= -i \theta ^T$ 
are conjugate
momenta of $Y^i ~{\rm and } ~ \theta $ respectively, the corresponding
Hamiltonian is given by
\be
\label{mham}
H=R_{11} {\rm Tr}\bigg ({1\over 2} \Pi _i ^2-{1\over 4} ([Y^i,Y^j])^2 -\theta ^T
\gamma _i[Y^i,\theta] \bigg ) \ee
One can define $H\equiv R_{11}\bar H$ for convenience factoring out over all
$R_{11}$.  
Notice also that the potential energy term ${1\over 4} R_{11}{\rm Tr}([Y^i,Y^j]
)^2$ is non-negative. When $R_{11} \rightarrow \infty$, we are in 
decompactification phase of M-theory. Thus, the finite energy states of H
are those for which the Hamiltonian $\bar H$ has vanishing eigenvalues.
One seeks those states for which $\bar H |\psi \rangle ={{\epsilon \over N}}|
\psi \rangle$, which is  equivalent to seeking a solution$H|\psi \rangle
={R_{11}\over N}|\psi \rangle $
where $\epsilon$ is finite. We know that, for collection of N number of 
D0-branes, the total momentum $p_{11}={N\over {R_{11}}}$ and therefore, the
energy is given by $E={\epsilon \over {p_{11}}}$. We have to identify $\epsilon$
with ${1\over 2}P_T^2$ if we recall (\ref{lce}). The $N\times N$ matrices
$X^i$ can be interpreted as the location of N D0-branes. When we consider
the potential term in $Y^i$ variables (\ref{sqm}), we notice that there are
flat directions when $[Y^i,Y^j]=0$. Here we deal with a quantum mechanical
system and $Y^i$ have are the collective coordinates. In such situation as
mutually commuting $Y^i$, we can diagonalize $Y^i= {\rm diag}~(y^i_1,y^i_2...
y^i_N)$. Thus $y^i_n$ is the $i$th coordinate of the $n$th D0-brane. It is
easy to see that there is invariance under Galilean translation, $Y^i
\rightarrow Y^i+d^i {\bf 1}$ and the Galilean boost $Y^i\rightarrow Y^i+v^it
{\bf 1}$ as is expected of a nonrelativistic system, here ${\bf 1}$ is the 
unit matrix. The boost will affect the center of mass momentum; but neither 
the relative momenta nor 
interaction term are affected by these transformations.\\
We can consider two clusters separated from one another. This is familiar
in composite model of hadrons where quarks are the basic constituents. In
the parton picture, the proton is made of large number of partons with very
small binding energy and one could describe photon-hadron deep inelastic
scattering in IMF \cite{dick}. In this case we can think of configurations
where the $N\times N$ matrices $Y^i$ can be decomposed to block diagonal form
of say n blocks of $N_1\times N_1, N_2\times N_2,....N_n\times N_n$ such that
$\sum _mN_m =N$. This decomposition can be interpreted as if we have
$n$ separated clusters of D0-branes where each of the clusters has $N_1, N_2,..
N_n$ number of particles. The distance between two clusters can be defined as
\be r_{ab}=\bigg |{1\over {N_a}}{\rm Tr}~ Y_a-{1\over {N_b}}{\rm Tr}~Y_b \bigg |
\ee
where $a$ and $b$ are the two clusters. Now we can visualize how the
potential will arise. It comes from ${\rm Tr}([Y^i,Y^j])^2$ and this goes like
modulus squared of the off diagonal block elements multiplied by the minimum of
the $r_{ab}^2$ and an appropriate numerical constant. Thus, if we consider
well separated cluster of D particles, the off diagonal elements are required
to be small; otherwise, the potential will grow like $r_{ab}^2$. We should
keep in mind that the system is supersymmetric and having a harmonic oscillator
type potential does not imply ground state energy is that of the oscillator.\\
The supersymmetric quantum mechanical system has a very rich structure. This
becomes transparent if we consider a single D0-brane, i.e. $N=1$.
\be H={{R_{11}\over 2}}\Pi _i^2\equiv {{R_{11}\over 2}}P_T^2
={{P_T}\over {p_{11}}} \ee
When we look at this  equation from 11-dimensional point of view, this 
corresponds to the relation between energy and momentum of a massless particle
in IMF. When we take into account the 16 component fermions, $\theta$ we
eventually get the supermultiplet with 256 total degrees of freedom
and this agrees with the massless degrees of freedom of $N=1$ supergravity in
eleven dimensions. In fact the bosonic components are 128 equal to fermionic
degrees of freedom.
As is well known, there 44 components from graviton and 84 from the 
antisymmetric tensor field in 11-dimensions. When we have $N>1$, it is 
necessary to separate the center of mass motion and define the relative
coordinates and the decomposition is as follows:
\be Y^i=Y^i_{r}+Y^i_{cm}{\bf 1},~~~ Y^i_{cm}={1\over N}{\rm Tr}~ Y^i \ee
\be \Pi _i=\Pi _{r~i}+{1\over N}P_{cm~i}{\bf 1},~~~P_{cm~i}={\rm Tr}~\Pi _i  \ee
and ${\rm Tr} ~Y^i_{r}={\rm Tr }~\Pi _{r~i} =0$. Now the total Hamiltonian will 
be written as a  sum of two terms
\be H=H_{cm}+H_r \ee
with
\be H_{cm}={{R_{11}\over {2N}}}(P_{cm~i})^2={1\over {p_{11}}}(P_{cm~i})^2 \ee
Note the appearance of the factor ${R\over N}={{1\over {p_{11}}}}$ as expected.
We have defined the center of mass coordinate, canonical momentum and
the Hamiltonian by taking trace over $U(N)$ matrices. Therefore, the relative
Hamiltonian is a function of $\{Y^i_r,~\Pi _{i~r} \}$. Thus $H_r$ is quite
similar to the original Hamiltonian; however, all the variables are $SU(N)$
matrices, they are traceless since the trace part is separated out.
It has been shown that the relative Hamiltonian has zero energy bound states
due to the presence of supersymmetry \cite{wittg,seth,prz}. The total energy
is due to the center of mass energy: $E=E_{cm}={{1\over {2p_{11}}}}(P_{T~cm})^2
$. In this case one also gets the supergravity multiple which has 256 states.
Therefore, for any N, we see that the spectrum  
contains supergravitons. Suppose we decompose $Y^i$ to various blocks 
which describe clusters of D-particles.
In the simplest case, if the submatrices are exactly block diagonal so that
off diagonal elements are zero, then the total Hamiltonian will be given by
sum of n separate Hamiltonians without any interactions amongst them. If we
let the off diagonal elements appear (give them small values), that will
amount to switching on interactions between the clusters. The physical picture
is that we have several  clusters, each cluster will have its supergraviton in
the spectrum. There could be arbitrary number of them and therefore, we let
N go to infinity. Thus the matrix model contains the full Fock space of
supergravitons. The interaction among the supergravitons is described due to
the presence of off diagonal elements and one should be able to describe
various processes involving supergravitons in this picture.\\
In order to compute S-matrix element for scattering of two supergravitons when
their transverse velocities are small, we have to determine potential 
between them. One starts by considering the classical configurations and
the fluctuations over them to compute the effective action \cite{bsist}. 
Suppose we give transverse velocity $v$ and define the impact parameter as b 
and expand the coordinates around their backgrounds as follows: 
\be X^9={1\over 2}b\sigma _3+{\sqrt {g_{str}}}\delta X^9~,~~X^8={1\over 2}vt
\sigma _3+{\sqrt {g_{str}}}\delta X^8  \ee
\be  X^i={\sqrt {g_{str}}} \delta X^i,~~~~i\ne 8,9  \ee
Here $\delta X^i$ denotes the fluctuations and $\sigma _3$ is the Pauli matrix.
When we  have vanishing fluctuation, the classical configuration is such that
total transverse center of mass momentum and position vanish. The $2\times 2$
matrices are block diagonal which describes two clusters of D0-branes and in
this case we have $N_1=N_2=1$. Now the separation between the two particles
is given by $r_{ab}={\sqrt {v^2t^2+b^2}}$. The effective action can be
computed using the standard techniques and the order $\hbar$ term will contain
determinant of (basically) propagators when we restrict to one loop level.Thus
\be S_{eff}= S_0+\int d\tau V_{eff}(r(\tau)) \equiv \int d\tau V_{eff}
({\sqrt {v^2\tau ^2+b^2}})  \ee
For large impact parameter, the long range  part of the potential in the
leading order is given by \cite{bsist}
\be V_{eff}(r)= - {{15}\over {16}}{{v^4}\over {r^7}} +~~{\rm higher~ orders} \ee
The result is striking in the sense that this form of the potential can be
derived from the supergravity action at the tree level i.e. considering
graviton exchange. Thus starting from s simple M(atrix) model
description, one could extract a result of 11-dimensional supergravity.\\
The 11-dimensional supergravity admits supermembrane. It is worthwhile to
ask how much the M(atrix) model can tell us about the underlying membrane theory. The membrane is extended object in two spatial directions as the name 
suggests. Moreover, the dimension of spacetime in which the supermembrane
can exist is quite restricted \cite{restr,r6}. The reason for such constraints
lies in the fact that the action contains Wess-Zumino-Witten term and the
supersymmetry invariance of the full action restricts the spacetime 
dimensions to 4,5,7 or 11. The membrane is described by $Z^{\mu}(\sigma ,\xi ,
\tau)$, where $\sigma ,\xi ~{\rm and }~ \tau$ are the worldvolume coordinates. 
When one adopts a Hamiltonian formalism, a fixed $\tau$-slice is chosen and
thus the explicit $\tau$ dependence in $Z^{\mu}$ does not appear and the
derivatives with respect to worldvolume time are traded for canonical momenta
${\cal P}_{\mu}$. The light-cone gauge is a convenient description to see
the physical degrees of freedom and in this gauge the membrane Hamiltonian 
takes the following form \cite{mmham}
\be 
\label{hmembrane}
H_M={{1\over {2p_{11}}}}\int {{d\sigma d\xi}\over {(2\pi )^2}}{\cal P}
_i^2 +{{(2\pi T_2)^2}\over {4p_{11}}}\int d\sigma d\xi(\{Z^i(\sigma ,\xi),
Z^j(\sigma ,\xi) \})^2 +~{\rm fermionic~ terms} \ee
where the brackets appearing is the second term are defined as
\be \{A,~B \}=\partial _{\sigma}A\partial _{\xi}B -\partial _{\xi}A
\partial _{\sigma}B  \ee
and $T_2$ is the membrane tension. Let us assume  that the worldvolume of the
membrane can be written as $\Sigma \times R$, where $\Sigma$ has the topology
of a torus. For this topology, $Z^i (\sigma , \xi)$ is a double periodic 
function and we can expand $Z^i$ in double Fourier series with $Z^i_{mn}$ as
the Fourier coefficients. Thus we have nine $\infty \times \infty$
matrices and same would be the case if we had considered nine $Y^i$'s in
the $N\rightarrow \infty$ limit. In order to establish relation with the
membrane Hamiltonian (\ref{hmembrane}), we have show how the commutator
$[Y^i,Y^j]$ goes over to the bracket $\{Z^i, Z^j \}$. For arbitrary finite N,
introduce two $N\times N$ matrices $U$ and $V$, satisfying the properties
\be U^N=V^N={\bf 1},~~{\rm and } ~~ UV=e^{{{{2i\pi}\over N}}}VU \ee 
This can be realized if U and V have the following special form
$U_{j,j+1}=U_{N,1}=1$ and $V_{j,j}=e^{{{2i\pi (j-1)}\over N}} $ and all
other matrix elements set to  zero. A more abstract, 't Hooft, representation
is
\be U=e^{ip},~~ V=e^{iq},~~~[p,q]={{2\pi}\over N} i   \ee
This is the canonical commutation relation between position and momentum when
the space is taken to be compact and discrete. It is worthwhile to
point out that the above commutation relation will not hold good for
finite dimensional matrices. However, acting on states with low wave number,
the error on the $r.h.s$ of the commutator $[p,q]$ is further
down by power of N and therefore, ${{2\pi}\over N}$ is the leading
term.  Thus as  N assumes higher and higher values, the error gets smaller
and smaller. From $U^N=V^N={\bf 1}$ we can
conclude that $p ~{\rm and}~ q$ take eigenvalues $m{{2\pi}\over N}$, where
$m$ takes values 0,1,2...$(N-1)$. Moreover ${\rm Tr} U^nV^m =N\delta _{n,0}
\delta _{m,0}$, where $0$ in both the Kronecker delta are to be understood
as $mod$ N.
Now we can expand any $N\times N$ matrix in terms of Fourier modes.
\be 
\label{fdbl}
A=\sum _{n,m=N/2-1}^{N/2} A_{nm}U^nV^m =\sum_{n,m=N/2-1}^{N/2} A_{nm}e^{inp}
e^{imq} \ee
Since commutator of p and q is order $1\over N$, in the $N\rightarrow \infty$
limit, they will commute. The eigenvalues of these two operators will fill the
interval $[0,2\pi]$ and 0 is to be identified with $2\pi$ since we have toric
geometry. The double Fourier expansion (\ref{fdbl}) takes the form
\be A(p,q)=\sum _{n,m=-\infty} ^{\infty} A_{nm}e^{inp}e^{imq} \ee
and the Fourier coefficients with the double index are defined as
\be A_{nm}=\int _0^{2\pi} \int _0^{2\pi}{{dp}\over {2\pi}}{{dq}\over {2\pi}}
A(p,q)e^{-inp}e^{-imq} \ee
Also ${\rm Tr}~A=NA_{00}$, when we take $N\rightarrow \infty$ limit, ${\rm Tr}~
A \rightarrow N\int _0^{2\pi}\int _0^{2\pi}{{dp}\over {2\pi}}{{dq}\over {2\pi}}
A(p,q) $. One can show with some algebra that the commutator of two matrices
in the infinite N limit goes over to the $\{ , \}$. Finally bosonic part of
the M(atrix) model Lagrangian goes over to a form (identify ${{dp}\over {2\pi}}
=d\sigma ~{\rm and} ~ {{dq}\over {2\pi}}=d\xi$)
\be L_{m}\rightarrow {N\over {2R}}\int d\sigma d\xi ({\dot {Z}}^i
(\sigma ,\xi))^2 -
{{R\over {4N}}}\int d\sigma d\xi({Z^i(\sigma ,\xi), Z^j(\sigma ,\xi)})^2 \ee
Note that ${N\over R}=p_{11}$, therefore conjugate momentum of $Z^i$ is $p_{11}
{\dot Z}^i$. Thus passage to the Hamiltonian (in light-cone gauge) gives the
membrane Hamiltonian (\ref{hmembrane}).
 This is indeed a remarkable result
that a simple supersymmetric quantum mechanical system encodes the dynamics of
the supermembrane.\\
It is natural to ask whether one obtain a string starting from the M(atrix)
model. First, one compactifies the theory to ten dimension. When the 
compactification radius is small, the theory contains the Fock space of the 
type IIA string. As the radius tends to zero the string becomes free \cite{dvv}
and correct leading order string interactions could be reproduced. In order
to carry out compactification, we replace the matrices by infinite dimensional
operators. The compact coordinate is represented as $X^a \rightarrow 
-i{{\partial}\over {\partial \sigma ^a}}{\bf I}_{N\times N} - A_a(\sigma)$.
Here A is a $U(N)$ gauge potential. The rest of the variables are taken to be
matrix valued function of $\sigma$. If we use this ansatz, the resulting
Hamiltonian is that of maximally supersymmetric $1+1$ dimensional 
Yang-Mills theory. In the limit when radius goes to zero and N is 
taken to be infinity, the moduli space  of this model coincides with the Fock
space of type IIA theory.\\ Indeed, the M(atrix) model has opened up new
avenues to study dualities between compactified  model on torus and Yang-Mills
theory on dual space. Moreover, there are applications of the M(atrix) model
to study black holes we refer the interested reader to the review on the
subject \cite{r13}. Another interesting development has been to understand
type IIB theory and its dualities from a matrix model formulation. In this 
approach
one adopts procedure of Eguchi and Kawai to consider reduced 10-dimensional
super Yang-Mills theory and it is a theory of $N\times N$ matrices which
even carry no time dependence \cite{ikkt}. We refer the reader to the review
article of Makeenko \cite{iibmak}. 
%%% THIS IS x.AdS/CFT SECTION %%%%% 

\section{Anti-de Sitter Space and Boundary field Theory Correspondence}

\setcounter{equation}{0}

\def\theequation{\thesection.\arabic{equation}}

Recently, attentions have been focused in constructing 
supersymmetric gauge theories by considering various configurations of branes
in string theories as well as in  M-theory. 
When we have
N coincident Dp-branes, a supersymmetric $U(N)$ gauge theory lives in 
worldvolume of the branes. The $1\over N$ expansion proposed by 't Hooft 
\cite{onen} revealed several aspects of $SU(N)$ Yang-Mills theory. According
to 't Hooft, one should consider large N limit of the theory keeping $g_{YM}^2
N$ fixed, $g_{YM}$ being the gauge coupling constant. Then a Feynman diagram
is designated by the topological factor $N^{\chi}$, $\chi$ being the Euler
characteristic of the Feynman diagram. When we consider, expansion in 
$1\over N$, rather than in coupling constant, each order in $1\over N$, contains
diagrams to all orders in coupling constant and the leading order corresponds
to the planar diagrams. Maldacena \cite{juanad} has made remarkable conjecture
regarding large N conformal gauge theories. The proposal states that large
N limit of a conformally invariant theory in $d$ dimensions is determined by
supergravity theory on $d+1$ dimensional Anti-de Sitter space times a compact
space (for a sphere it is maximally supersymmetric). The AdS/CFT connection
has led to the generalization of the holography principle in this
context \cite{edhol,ewsus}
which was first introduced in black hole physics  
\cite{gth,lessus} in order to understand the Bekenstein entropy bound
and the area law for black hole entropy. Thus the conjectures of Maldacena
led to reveal deeper connections between string theory and superconformal
gauge theories.\\
We have emphasized earlier that gravity is an integral part of string theory 
since graviton is a part of the spectrum.
Moreover, gauge fields also invariably appear  in string theories.  
 Let us 
recapitulate a few points in order to get a perspective of AdS/CFT connections.
We have seen that the heterotic strings,  through their constructions,  contain
nonabelian gauge groups and graviton in their massless spectrum. The type II
theories have graviton, coming from NS sector,  in their perturbative spectrum.
However, with the discovery of Dp-branes, we know that supersymmetric gauge
theories can arise if we consider coincident Dp-branes in type II theories.
Type I string theory admit nonabelian gauge field since Chan-Paton
factors can be attached to the end points as was discussed earlier. 
Furthermore, consistency of the
theory requires that we have to incorporate closed string sector in order
to account for nonplanar loop corrections; therefore there is gravity coming
from the  closed string spectrum. For this theory,  when we take $\alpha '
\rightarrow 0$ limit Yang-Mills theory appears automatically and since
consistency requires inclusion of closed string states, gravity also will
appear in the zero slope limit. In view of preceding remarks, one might
conclude that, in string theory, gravity and gauge theory invariably appear
simultaneously. Thus the important question to answer is that how the string
theory can describe the strong interaction among quarks and gluons. The
recent developments \cite{juanad,igpol,edhol,spoly} have provided connections
between string theory and gauge theories.\\ 
The configuration under consideration is N coincident Dp-branes and open 
strings can end on these hypersurfaces. When we look into the dynamics in
the worldvolume we have collection of these open strings and their 
excitations. Moreover, the worldvolume fields have their interactions and also
there exists interaction with the bulk. An interesting limit to consider
is when dilaton remains at a fixed value and the slope parameter tends to zero
value. Then, at low energies, the gravity decouples; but to keep the
interactions in the worldvolume in tact, we should have gauge coupling finite, 
for the $U(N)$ gauge theory. In fact, if we ignore the center
of mass part, then we need to consider the $SU(N)$ gauge theory. It is
necessary to go near the horizon, $r \rightarrow 0$, to see the connection 
between AdS and  CFT. In the near horizon  limit, recall eq.(5.11) and
eq.(5.12), that the 
factor 1 appearing in the definition of the harmonic function of the Dp-brane
can be neglected. To be specific let us first consider the metric in the
case of N coincident branes.
\be
\label{nrhor}
ds^2=H_p^{-{1\over 2}}(r)\eta _{\mu\nu}dx^{\mu}dx^{\nu}+H_p^{1\over 2}dy_idy_i
\ee
where, $\{ y_i \}$ are  the transverse coordinates and $r={\sqrt {y_iy^i}}$. 
The indices $\mu ,\nu ....$ are for tensors on the worldvolume.
The dilaton and the  $(p+1)$-form potential, coming from the RR sector,  are given by
\be
e^{-(\phi -\phi _0)}=H_p(r)^{{(p-3)}\over 4},~~{\rm and}~~ A=[H_p(r)]^{-1} \ee
and
\be H_p(r)=1+{{C_pN}\over {r^{7-p}}}, ~~~ C_p={{(2\pi {\sqrt{\alpha '}})^{7-p}}
\over {(7-p)\Omega _{8-p}}} g_{str} \ee
Here we have suppressed the indices of the $(p+1)$-form gauge potential and
$\Omega _r ={{{2\pi ^{{(q+1)} \over 2}}}\over  {{ \Gamma [{{(q+1)}
\over 2}]}}}$ and 
$\phi _0$ is the asymptotic constant value of the dilaton.
When we have N coincident D-branes, the worldvolume action is the generalised
Born-Infeld action proposed by Tseytline \cite{tark}
\be
\label{brni}
S_{BI}= -\tau _p^{(0)}\int d^{p+1}\xi e^{-\phi}{\rm STr}{\sqrt {-{\rm det}[
G_{\mu\nu}+2\pi \alpha 'F_{\mu\nu}]}} \ee
Here $G_{\mu\nu}$ is  the pullback of the metric $G_{MN}$ to the world volume
and $F_{\mu\nu}$ is the gauge field strength on the brane. The tension of the 
brane is
\be T_p={{(2\pi {\sqrt {\alpha '}})^{(1-p)}}\over {2\pi \alpha ' g_{str}}} =
{{\tau _p^0}\over {g_s}} \ee
and $g_{str}$ is the string coupling constant. The action (\ref{brni}) under
the square root can be expended and keeping the second order term in gauge
field strength one can write the action in more familiar form
\be
S_{gauge}=-{{1\over {4{g_{YM} ^2}}}}\int d^{p+1}\xi {\rm Tr}~F_{\mu\nu}
F^{\mu\nu} \ee
where ${\rm Tr}$ is taken over the gauge group matrices and the gauge coupling
constant is identified as $g_{YM}^2=2g_{str}(2\pi)^{(p-2)}(\alpha ')^{{(p-3)}
\over 2}$. We know from the  solutions discussed in previous  
section (recall eq.(5.12) and eq.(5.13)) that, in
the limit,  when $r\rightarrow \infty$, the metric is flat. Here one is looking
for the behaviour of the solution in the $r\rightarrow 0$ limit and
one chooses a brane for which the dilaton is constant at the horizon. If we 
consider D3-branes, then we find that not only the dilaton is independent of 
$r$, but also the Yang-Mills coupling constant is  dimensionless.
As mentioned above, one examines the configuration of N coincident
branes in the following limit
\be
r\rightarrow 0,~~~ \alpha ' \rightarrow 0, ~~{\rm and}~~ U\equiv {{r\over 
{\alpha '}}}={\rm fixed} \ee 
Therefore, we can neglect 1 appearing in the harmonic function. and the
D3-brane metric goes over to
\be
{{ds^2}\over {\alpha '}} \rightarrow {{U^2}\over {{\sqrt {4\pi Ng_{str}}}}}
(dx_{3+1})^2+{{{\sqrt {4\pi Ng_{str}}}}\over {U^2}}dU^2+{\sqrt {4\pi Ng_{str}}}
d{\Omega _5}^2 \ee
The last term is the line element of five sphere and the metric describes
the manifold $AdS \times S_5$ The radius  of AdS is the same as that of $S_5$
and the radius is given by
$R_{AdS}=(\alpha '{\sqrt {4\pi Ng_{str}}})^{1\over 2} $
Since the Yang-Mills coupling constant satisfies the relation $g_{YM}^2=
4\pi g_{str}$, the radius of the AdS gets related to the Yang-Mills coupling
constant as
\be {{R_{Ads}^2}\over {\alpha '}}= {\sqrt {Ng_{YM} ^2}} \ee
We know that the worldvolume theory of N coincident Dp-branes is supersymmetric
Yang-Mills theory in $p+1$ dimensions and therefore, in this  case the
$N=4$ SUSY gauge theory will appear. This is known to be a conformally 
invariant theory. From the supergravity side, we could describe the theory
even for large radius; but that will amount to taking $Ng_{YM}^2$ to large
values. Maldacena's conjecture states that strongly coupled $N=4$ super
Yang-Mills theory is equivalent to 10-dimensional supergravity compactified on
$AdS _5 \times S_5$. However, the consistency of the supergravity theory
requires string theory at a deeper level. Thus supersymmetric four dimensional
Yang-Mills theory is equivalent to type IIB theory compactified on $AdS_5 
\times S_5$. The relations among the parameters are
\be 
\label{2bpar}
g_{YM}^2\equiv {\lambda \over N}=4\pi g_{str}, ~~ {\rm and }~~
R_{AdS}^2= \alpha '{\sqrt {\lambda}} \ee

Let us very briefly recall some essential features of the Anti-de Sitter space. The
Einstein-Hilbert action in the presence of cosmological constant term is
\be S_{EH}={{1\over {16\pi G_D}}}\int d^Dx{\sqrt {|g|}}[R+\Lambda] \ee
We consider D-dimensional spacetime with Minkowski metric. The field equations
are
\be R_{\mu\nu}-{1\over 2}g_{\mu\nu}R={1\over 2}g_{\mu\nu} \Lambda  \ee
Taking the trace of this equation, we can determine curvature scalar $R$
in terms of $\Lambda$, and then derive the relation
\be R_{\mu\nu}={{\Lambda}\over {2-D}}g_{\mu\nu} \ee
In this case the Ricci tensor is proportional to the metric and these are
Einstein spaces. This is also maximally symmetric space \cite{weins}
with the property that
\be R_{\mu\nu\rho\lambda}={{R\over {D(D-1)}}}(g_{\nu\lambda} g_{\mu\rho} -
g_{\nu\rho} g_{\mu\lambda}) \ee
The example of such space, with nonzero curvature, are de Sitter, 
Anti-de Sitter and D-spheres. In this sign convention,  AdS space has 
{\it positive} cosmological constant. The AdS space is best 
described by an embedding.
We start with $D+1$ dimensional pseudo-Euclidean embedding space with
coordinates $\{y^a=y^0,y^1,...y^{D-1}, y^D \}$ and metric $\eta =~{\rm diag}~
(+,-,-.....+)$ and the distance squared is
\be y^2\equiv (y^0)^2+(y^D)^2-\sum _{n=1}^{D-1} (y^n)^2 \ee 

Note the appearance of two time coordinates from the form of the metric. The
length remains invariant under $SO(D-1,2)$ global transformations
\be y^n \rightarrow {y'}^n= L^n_my^m \ee
where $L^n_m$ is an $SO(D-1,2)$ matrix. If we consider the locus of 
\be y^2=b^2={\rm constant} \ee
and  that defines $AdS_D$. It is worth noting that the invariance group for
theories defined on $AdS_D$ is same as that of the D-dimensional flat space
that is D generators corresponding to translations and ${1\over 2}D(D-1)$, generators from
Lorentz rotations.\\
Next, let us consider what is the conformal group in D-dimensional Euclidean
space $E^n$. In this case the Poincare group has altogether ${1\over 2}D(D+1)$
generators (D translations and rest from Lorentz group). Then we have following
extra generators:
\be {\vec x} \rightarrow \lambda  {\vec x} \ee
this is {\it dilation} and $\lambda$ is a real number. Furthermore, there is
special conformal transformation
\be 
\label{sct}
{{{x'}^{\mu}}\over {{x'}^2}} ={{x^{\mu}}\over {x^2}}+\alpha ^{\mu} \ee
This transformation involves n parameters $\alpha ^{\mu}$ The transformation
(\ref{sct}) can be rewritten as
\be {x'}^{\mu}={ {x^{\mu}+\alpha ^{\mu} x^2}\over {1+2x^{\mu}\alpha _{\mu}
+\alpha ^2x^2}} \ee
Thus we see that the total number of generators are: ${1\over 2}D(D+1) +
1+D = {1\over 2}(D+1)(D+2)$. This is the same number of generators that
$AdS_{D+1}$ space has. Indeed, in view of the recent developments, one can
establish the connection that the isometry group of $AdS_{D+1}$, $SO(2,D)$
acts on the boundary as the conformal group acting on Minkowski/Euclidean
space.   We list below the generators of conformal group and their algebra
\be
[M_{\mu\nu}, P_{\lambda}]=i(g_{\nu\lambda} P_{\mu}-g_{\mu\lambda} P_{\nu}) \ee
\be
[M_{\mu\nu},M_{\lambda\rho}]=i(g_{\mu\rho}M_{\nu\lambda}-g_{\mu\lambda}M_{\nu
\rho}+g_{\nu\lambda}M_{\mu\rho}-g_{\nu\rho}M_{\mu\lambda}], \ee
\be
[M_{\mu\nu}, K_{\lambda}]=i(g_{\nu\lambda}K_{\mu}-g_{\mu\lambda}K_{\nu} \ee
\be
[D, P_{\mu}]=iP_{\mu} \ee
\be
[D, K_{\mu}] -iK_{\mu} \ee
\be
[P_{\mu}, K_{\nu}]= 2i(g_{\mu\nu}+M_{\mu\nu}) \ee
The generators of conformal transformation have the following representations,
when we choose Cartesian coordinate system and consider transformation
properties of a real scalar field: 
$P_{\mu}=-i\partial _{\mu}$, $M_{\mu\nu}=x_{\nu}P_{\mu}-x_{\mu}P_{\nu}$, $D=
x^{\mu}P_{\mu}$ and $K_{\mu}=x^2P_{\mu}-2x_{\mu}D$, corresponding to
translation, Lorentz transformation, dilation and special conformal
transformations respectively.\\
Let us discuss the evidences in support of Maldacena's conjecture. When we 
consider collections of D3-branes of the type IIB theory we note that
D3-branes couple to the 5-form field strength and N units of this flux will
pass through the five sphere of the $AdS_5 \times S_5$ manifold. The isometry
group of $S_5$ is  $SO(6)$ and the $AdS_5$ is endowed with isometry group
$SO(4,2)$ as we have just mentioned. The IIB theory has fermions and therefore,
it is more relevant to consider the covering groups $SU(4)$ and $SU(2,2)$ of
$SO(6)$ and $S(4,2)$ respectively. We also know that type IIB theory has 32
Majorana supercharges. These supersymmetries are preserved by the background
under consideration. The invariance group is the super Lie group $SU(2,2|4)$
for this theory. On the super Yang-Mills part, one has to examine how the
above symmetry appears on the boundary theory. We have mentioned how the
conformal group, for the case at hand, is to be identified as $SO(4,2)$ or
$SU(2,2)$. It is well known that $N=4$ super Yang-Mills theory is conformally
invariant in four dimensions, since the theory has vanishing  $\beta$-function 
 \cite{smdl},  and thus the origin of the conformal group is well understood. 
Let us now focus our attention on the other symmetries present in type IIB 
theory. The ten dimensional super Yang-Mills has gauge bosons, $A^a_{\mu}, \mu
=0,1,.. 9$, a being $U(N)$ group index and thus there are 8 physical states corresponding to
each gauge field.
The superpartners are Majorana Weyl gauginos having matching numbers. The
theory has 16 Majorana supercharges in $D=10$. When we consider the
4-dimensional action, dimensionally reduced from ten dimensions \cite{ss,ms}
physical degrees of freedom of each of the ten dimensional 
gauge field decomposes into 
2  (corresponding to physical degrees of freedom of gauge field in $D=4$)
 and six scalars, $\phi ^a _i, i=1,2...6$, a is group index suppressed
from  now on. The number of, gauginos
are given by the Weyl spinors, $\lambda ^A_{\alpha}, ~A=1,2,3,4,~\alpha =1,2$.
 One of these fermions, together with the gauge field can be grouped define
a vector superfield. The rest  of the three spinors can be grouped with the
scalars (which appeared after dimensional reduction) to define 3 chiral 
superfields. The 16 supercharges can be grouped into 4 sets of complex Majorana 
charges $Q^A_{\alpha}, {\bar Q}^A_{\alpha},~A=1,2,3,4 ~{\rm and}~ \alpha =1,2$
These two supercharges transform as $\{ \bf 4 \}$ and $\{ {\bf {\bar 4}} \}$
of the R-symmetry group $SU(4)$. The scalars $\phi _i$ transform as 
$\{ \bf 6 \}$ of the $SO(6)$, since we deal with the covering group $SU(4)$, the
scalars transform in the antisymmetric, rank 2 representation of the $SU(4)$.
We see that type IIB theory has 32 supercharges, but the super Yang-Mills has
only 16 of them. We know from discussions in Sec.IV that in the presence of the
coincident D3-branes, half of the supersymmetries are preserved. When we
consider the superconformal algebra the rest also appear as the extension of
the superconformal group \cite{shls}.\\
Another important nonperturbative symmetry of type IIB theory is the $SL(2,Z)$
symmetry where dilaton and axion define the moduli. In the Yang-Mills
sector the S-duality symmetry is robust and is known to be, again, $SL(2,Z)$.
In this  case, the modular parameter $\tau ={{\theta}\over {2\pi}}+{{4i\pi}
\over {g^2_{YM}}}$ whereas in the former case it is $\tau =\chi +ie^{-\phi}$.\\
The preceding discussions were focused to show that the symmetry properties
of the type IIB theory and those of $N=4$ super Yang-Mills are the same.
It is important to investigate which physical properties are common to
both the theories. Indeed, if the two theories are equivalent, it should be
possible to identify a physical field $\Psi$ in the bulk theory and find
the corresponding object on the boundary theory. Then, one of the tests
will be to compute the correlators involving relevant objects in each of the
theories and check the consistencies. Thus it is important to identify the
physical quantities (operators) in both the theories. In the case of the
boundary theory, one obvious criterion will be to choose gauge invariant
operators while computing the correlators. One could formally express the
equivalence between the theories through the relation  among the generating
functionals. 
\be 
\label{fgen}
e^{-{S_{II}[\Phi (J)]}} =\int {\cal D}Ae^{-(S_{YM}[A]+{\cal O}_{\Delta}[A]J}
\ee
The $l.h.s.$ of the above equation is to be identified as the generating 
function for the supergravity theory (rather low energy limit of IIB theory).
The action $S_{II}$ is determined in terms of the massless states of the
supergravity and the Kaluza-Klein towers and these are collectively denoted
as $\Phi (z,\omega)$. Here the coordinates $z^N \equiv (x^{\mu} ,r)$ and $\mu$
taking values 0,1,2,3 are to be identified as the AdS coordinates and $\omega$
is the coordinate on five sphere. Moreover, it is implied due to the presence
of $J(x)$ that it also depends on the boundary data of the bulk fields. The
$r.h.s.$ defines the generating function for $N=4$ super Yang-Mills theory;
however, one only computes the correlation functions of gauge invariant 
composite operators denoted by ${\cal O}(A)$ with couplings to $J(x)$. In this
general setting \cite{igpol,edhol,ffz}, one will be able to compute the
correlation functions from both the theories and establish the correspondence
between the two theories. Let us consider a simple example as illustration
for the case of minimally coupled scalar in the bulk theory which could be 
identified with the dilaton. The action on the bulk for the dilaton on
$AdS_5 \times S_5$ is
\be
\label{adsd}
S= {{\pi ^3b^3}\over {4G_{10}}}\int d^5x {\sqrt {|g|}}g^{\mu\nu}\partial _{\mu}
\phi \partial _{\nu}\phi \ee
The factor $\pi ^3b^3$ comes from the volume of $S_5$, through implicit
assumption that $\phi$ has no dependence on coordinates of five sphere. The
metric is $g_{\mu\nu}= {{b^2}\over {x_0 ^2}}\delta _{\mu\nu}$, is metric on
$AdS_5$, now in the Poincare coordinates. For large $\lambda >>1$, the classical
supergravity can be taken to be a good approximation (\ref{2bpar}). 
The dilaton equation of motion is given by
\be \partial _{\mu}({\sqrt {g}}g^{\mu\nu}\partial _{\nu}\phi)=0 \ee
Of course, this equation can be solved by the standard Green's function method.
The purpose is to determine the generating function with value of dilaton
computed on  the boundary, call it $\phi _0$ which is value of $\phi$ as  $x_0 \rightarrow 0$.
Thus we can write
\be 
\phi (x_0, {\vec x})=\int d^4{\vec z}K(x_0,{\vec x},{\vec z})\phi _0({\vec z})
\ee
the vectors refer to four dimensional vectors on the boundary space and
the Green's function is defined as,
\be 
K(x_0,{\vec x},{\vec z}) \sim {{x_0^4}\over {[x_0^2+({\vec x}-{\vec z})^2 ]^4}}
\ee
Now, one can insert the solution for $\phi$ into the action to determine it
at the classical value of dilaton
\be S={{\pi ^3 b^8}\over {4G_{10}}}\int {{d^4{\vec x}}\over {x_0^3}}\phi
\partial _0\phi | ^{\infty}_{\epsilon} \ee
$\epsilon $ is the cut off for the lower limit of integration. 
Once expression for $\phi$ is inserted into the action, then  it is possible to 
take cut off to zero and everything is finite. The action is given by 
\be
S \sim -{{\pi ^3 b^8}\over {4G_{10}}}\int d^4{\vec x}\int d^4{\vec z}
{{\phi _0 ({\vec x})\phi _0({\vec z})}\over {({\vec x}-{\vec z})^8}} \ee
Then the generating function can be obtained by exponentiating this action.
On the super Yang-Mills side, since it is a conformal field theory in four dimensions, the quadratic of Yang-Mills field strength has dimension 4 and product
of two of the $F^2$ terms behave as
\be
\label{bcorr}
 \langle F^2({\vec x})F^2({\vec z}) \rangle \sim {{N^2}\over {({\vec x} -{
\vec z})^8}} \ee
If we want to determine the dilaton correlation function on boundary, we compute

\be 
\label{bulkcr}
{{\delta ^2 Z_{II}(\phi _0)}\over {\delta \phi _0({\vec x})\delta \phi _0({
\vec z})}} \sim {{N^2}\over {({\vec x} -{\vec z})^8}} \ee
Now comparing (\ref{bcorr}) and (\ref{bulkcr}) we find that they are in 
agreement. If one considers, metric perturbation of the form $g_{\mu\nu}=
\eta _{\mu\nu} +h_{\mu\nu}$ and then computes the two point correlation of this
perturbation on the brane taking the boundary limit; this correlation is
identical to the correlation of stress energy momentum tensors 
(product of a pair of them;
just as  we took correlation of two $F^2$ terms while identifying  the dilaton
two point functions).\\
Let us recall that the 't Hooft coupling $\lambda \equiv Ng_{YM}^2$ and
the length parameter $b^4 =l_s^4\lambda = 4\pi l_s^4N g_{str}$ are related. If
we hold $\lambda$ fixed and let $N\rightarrow \infty$, 
then the string coupling tends to zero. Therefore, string 
perturbation theory can give reliable result
in this limit. Thus, one can get a full quantum theoretic description of the
Yang-Mills theory in the $N\rightarrow \infty$ limit. Instead of holding 
$\lambda $ fixed, if we allow it to take large values, then in the domain,
where AdS radius is kept constant, the relevant limit is $\alpha ' \rightarrow
0$.  We know that in the zero slope limit the string theory goes over to
supergravity theory. We saw the matching of AdS/CFT in this limit. But the
consequences of Maldacena conjecture is very interesting in this regime, it
tells us how the superconformal gauge theory in the $N\rightarrow \infty$
limit behaves in strong coupling domain. Of course, the example we have been
considering is the one where the $\beta$-function of the theory vanishes
identically and therefore, it is not a realistic theory if we want to
establish connection with supersymmetric gauge theories which have running
coupling constants leading to asymptotic freedom. There are attempts to
construct field theories which will have broken SUSY and conformal invariance
(for example classical SQCD is scale invariant, but in the  quantized theory
scale invariance is broken). Witten \cite{ed7}has proposed that one
should consider $AdS_7 \times S_4$. 
The resulting boundary theory corresponds to 
6-dimensional theory whose action is yet to be explicitly constructed. Then one
compactifies the theory on $T^2$ and require that fermions satisfy 
anti-periodic boundary condition around  a cycle of the two-torus. Then the
boundary theory is a four dimensional one. Conformal invariance and
supersymmetry are broken in this 4-dimensional theory and we have a pure gauge
theory with large N. \\
There has been rapid developments in studying the interconnection between
supergravity (rather type IIB) theory on AdS space and boundary gauge theory.
Several important issues pertaining to string theory and gauge theories
have been addressed in this context. We refer to some of the recent review
articles in this subject \cite{adss1,adss2,adsss3}.

\section{Cosmology and String Theory}

\setcounter{equation}{0}

\def\theequation{\thesection.\arabic{equation}}
The remarkable attribute of string theory is that it provides a unified
description of laws of Nature. Although, a connection with the phenomenological
aspects of elementary particle physics is not firmly established so far,
there are several indications that we are persuing the right path. We have
discussed, in Action V, how string theory has provided an adequate description
of the physics of the black holes from a microscopic point of view as
is expected from a theory describing gravity.\\
It is natural to address questions intimately related to evolution of
the Universe and its creation in the frame work of string theory.
Einstein's theory provides a very good description of classical gravity and
has been tested with precision measurements. The  principles of
equivalence, cosmological principle and the big bang hypothesis are fundamental
ingredients of the standard cosmological model \cite{scmr}. The experimental
data have verified the predictions of the standard cosmological model to a
great degree of accuracy. However, in order to understand some of the salient
features of our Universe such as its flatness, isotropy and
homogeinity, the horizon problem and the large scale structure; the
paradigm of inflation has been accepted as an integral part of the
theory of the cosmos. In simple words, our Universe underwent rapid
superluminary expansion after the big bang so that we can understand some
of the cosmological observations alluded to above. Indeed, considerable
efforts have been made, in recent times, to understand mechanisms of inflation
and to explore consequences of various inflationary models. It is necessary
to introduce a scalar field, in generic inflationary model, to explain
the mechanism; however, the inflaton field is introduced in an ad hoc manner.\\
We expect that string theory should  provide answers to the questions related to
the evolution of our Universe. In  the  cosmological scenario, in this approach,
one considers Einstein and matter field equations obtained from the string
effective action, (2.37), when the metric and background fields (corresponding
to dimensionally reduced 4-dimensional action) depend only on the time
coordinate; usually identified with the cosmic time.
In string theory, the scalar dilaton, appears naturally  in the massless
spectrum of theory and it is tempting to identify this field as the one
responsible for inflation in early epochs \cite{dilinfl}.
It is well known that a dilatonic
potential cannot be generated in a superstring theory
perturbatively. Furthermore, the VEV of dilaton determines the
Newton's gravitational constant, gauge coupling constant, Yukawa couplings
of the fermions amongst other parameters of string theory. In a cosmological
context, the dilaton acquires time dependence and it will roll with evolution
of the Universe. However, the dilaton must decouple at some appropriate time
in the history of the Universe \cite{tdpol},
otherwise, the nice (tested) predictions of
late time cosmology will be seriously affected due to the fact that
a time dependent dilaton controls masses and coupling constants. Notice that,
eventually, the dilaton potential becomes important and the dilaton settle down
at the bottom of the well. There are important consequences for a
massless dilaton:  it violates equivalence principle \cite{eqvp}. Moreover, the
dilaton mass is bounded as $M_{\phi} >10^{-4}$ eV, in order to fulfill the
observational constraints \cite{dilmass}.\\
There is a novel approach to describe inflation phenomena in string theory
\cite{pbb,pbb1}, known as the pre-big bang (PBB) proposal. Since, considerable
attention has been focused to investigate the consequences of the PBB scenario,
we shall discuss important features  of this proposal
and refer the interested
reader to some of the recent reviews in the subject 
\cite{pbb2,pbb3,pbb4,pbb5,lwe}.
The target space duality, often called T-duality, is a key ingredient leading
to a new mechanism for inflation in string cosmology. In this approach, one
does not need potential for the dilaton for accelerated expansion of the
Universe unlike the case of standard mechanism for inflation. It is the
coupled dilaton and metric evolution equations which are responsible for
inflation. It follows from the properties of these equations, as discussed
below, that there are two branches of solution, denoted as $\pm$ and in each
branch there are two sets of solutions. One of the solutions in the
$(+)$-branch is such that the corresponding Hubble parameter and the
second derivative of the scale factor are positive. Therefore, the
Universe has accelerated expansion for this case. This branch gets related,
by combined operation of duality and time reversal to a solution in the
$(-)$-branch which has the features of the FRW metric in the sense that
it is expanding but decelerating. If there is a mechanism for a smooth
transition from the afore mentioned $(+)$-branch to the $(-)$-branch, then
following interesting scenario emerges. In the pre-big bang scenario the
time begins somewhere in the infinite past, in contrast with the big bang model
in which we identify beginning of time with the big bang singularity. Thus
the Universe evolves from a low curvature regime proceeding towards strong
coupling, high curvature domain with accelerated expansion. Then the Universe
emerges into the FRW like post big bang phase in which standard cosmological
model applies. Of course, it is essential to understand the mechanism for
transition from one phase to another known as the problem of graceful exit. Let
us recall, very briefly, how T-duality and time reversal transformations
relate different solutions of the cosmological evolution equations of string
effective action.\\
The string effective action, in the cosmological scenario,  is given by
\bea S=-{1\over {2{\lambda _s}^{d-1}}}\int dt {\sqrt |g|} e^{-\phi}(R+
{\dot {\phi}}^2) \eea
where $\lambda _s$ is the inverse of the string scale. The $d+1$ dimensional
metric has the form $ds^2=dt^2-g_{ij}dx^idx^j$, $g_{ij}$ is the spatial part
of the string frame metric since we can always choose $g_{00}=1$ and $g_{0i}=0$
in the cosmological context. We have omitted the field strength of the
antisymmetric tensor field. In the presence of graviton, antisymmetric tensor
and dilaton the string effective action can be brought to manifestly $O(d,d)$
invariant form, d being the number of spatial dimensions \cite{oddmv}. As
we have mentioned above the duality symmetry of string theory plays a very
important role in relating the two epochs i.e. $t>0$ and $t<0$, in the
evolution of the Universe. Let us consider a homogeneous, isotropic Universe
so that we have only one scale factor $a(t)$  and the spatial
metric is diagonal:
$ g_{ij} =diag~ (a(t)^2,a(t)^2,a(t)^2,...)$. The resulting field equations
take the following form:
\bea 2{\ddot \phi}+2d{\dot \phi}H-{\dot \phi}^2-2d{\dot H}-dH^2-(dH)^2=0 \eea
\bea
\label{econ}
{\dot \phi}^2-2d{\dot \phi}H-dH^2+(dH)^2 =0 \eea
\bea {\dot H}^2+dH^2-H{\dot \phi}-d{\dot H}+{1\over 2}dH^2 -{1\over2}(dH)^2-
{1\over 2}{\dot \phi}^2 + {\ddot \phi}+d{\dot \phi}H=0 \eea
The first of the above three equation comes from variation of the dilaton
$\phi$, the second follows from the $0-0$ component of the Einstein's
equation which is the Hamiltonian constraint on this occasion. Here,
$H={{\dot a}\over a}$ is the Hubble parameter. The last
equation, in an expanded form, is the $(i,i)$ component of the Einstein's
equation and the off-diagonal space-space components are found to be
trivially satisfied. If we stare at the dilaton equation and the last equation,
the $(i,i)$ part, at the first sight, we note that both have second derivative
of time for dilaton and it might be a formidable task to solve the graviton
dilaton equation. However, the last five terms of the third equation coincide
with dilaton equation of motion and we are left with a simple equation
\bea
\label{dilsc}
{\dot H}-H{\dot \phi} -dH^2=0 \eea
Notice that (\ref{econ}) is quadratic in $\dot \phi$ and therefore,
the solution
has two roots. Moreover, (\ref{dilsc}) is an evolution equation for the Hubble
parameter with a term involving $\dot \phi$. Thus $\dot H$ will have two
equations corresponding to each root of $\dot \phi$. Therefore, there are
altogether four branches when we inspect the solutions of coupled
graviton-dilaton equations. Let us introduce the shifted dilaton
\bea {\bar \phi}(t)=\phi (t) - d~ln~a(t) \eea
It is obvious from the evolution equations that if $\{a(t),\phi (t) \}$ satisfy
the equations of motion, then the new set $\{ a(t)^{-1}, \phi -2d~ln~a(t) \}$
are also solutions to the equations of motion; indeed, this is a part
of the $O(d,d)$ group and the shifted dilaton, $\bar \phi$, is invariant
under this T-duality. Another property of the time evolution
equations for dilaton
and that for the scale factor is that they are invariant under time
inversion $t \rightarrow -t$ with $H \rightarrow -H ~{\rm and}~ {\dot {\bar
\phi}} \rightarrow  -{\dot {\bar \phi}}$.
Let us consider a simple illustrative example to demonstrate how one can
generate solutions in the four branches. We begin with a
specific isotropic solution
\be a(t)=t^{1/{\sqrt {d}}} ~~~~~~ {\bar \phi}(t) = -ln~t \ee
for $t>0$. We can generate new set of solutions by implementing duality
and time reversal transformations. In fact we can get the four solutions,
mentioned above, starting from the (seed) solution of the previous equation.
The solutions are
\bea a_{\pm}(t)=t^{\pm 1/{\sqrt {d}}}, ~~~~~{\bar \phi}(t)=- ln~t \eea
\bea a_{\pm}(-t)=(-t)^{\pm 1/{\sqrt {d}}}, ~~~~~{\bar \phi}(-t)=-ln(-t) \eea
Let us examine the characteristics of the four solutions in some
detail. $a_+(t)$ corresponds to decelerated expansion, whereas $a_-(t)$
is identified for decelerated  contraction. These two solutions lie in
the positive $t$ branch. In the negative time branch, $a_+(-t)$ and
$a_-(-t)$ are
identified to be accelerating, contraction, and accelerating
expansion respectively. If ${\dot a} >0$ (${\dot a }<0 $) the solution
describes expansion (contraction). Similarly, a solution is called an
accelerated one (decelerated) if $\dot a$ and $\ddot a$ have same sign
(opposite sign). Since the dilaton is the coupling constant in the theory,
it is important to extract time dependence of the dilaton in the four branches.
\be \phi _{\pm} (\pm t)=(\pm {\sqrt d}-1)~ln(\pm t) \ee
which can be obtained from definition of $\bar \phi$. Note that $a_+(t)$ has
the characteristics of FRW solution in the sense that it corresponds to
expanding Universe with deceleration and the singularity lies in its past. On
the other hand, $a_-(-t)$ is the one which is expanding with an acceleration
and the singularity is in its the future. Furthermore, these two solutions are
related to each other under simultaneous duality and time
reversal transformations as we mentioned earlier. If one introduces a dilatonic
potential, the full set of solutions continues to exhibit same
characteristics: there are two  branches and each branch
having two solutions. In view of remarks at the beginning of the section,
a dilatonic potential is very much desirable. According to the PBB proposal,
the Universe initially is flat, cold and is in the weak coupling regime
and therefore, the tree level string effective action is a good starting point.
Moreover, we can trust the perturbative vacuum. Subsequently, the Universe
evolves towards curved, hot and strong coupling domain under going
accelerated expansion. If the Universe could smoothly pass over to
the FRW like regime, then we would have not only resolved the mechanism
of inflation, but also the initial singularity problem could be circumvented.
It is well known that a smooth
transition from the $(+)$-branch to the $(-)$-branch is not possible
if we consider the tree level string effective action due to the no-go
theorems \cite{gex1,gex2}. Therefore, graceful exit is an important
issue when we
envisage pre-big bang scenario. One of the possibilities is to appeal to
quantum string cosmology in order to resolve the graceful exit problem
\cite{wgmv} and the other approach is to consider the string effective
action with higher derivative terms due to string look and/or $\alpha '$
corrections \cite{hgdr}. \\
The examples, illustrating mechanisms for  inflation in the
pre-big bang scenario,
correspond to spatially flat, homogeneous solutions of the equations of
motion. It is not desirable  to start with a homogeneous solution from the
onset, if we want to solve the homogeneity and flatness problem in
cosmology. Therefore, a more appropriate approach will be to consider generic
initial condition in the vicinity of the perturbative vacuum. One could
consider a scenario where, long before the big bang, the Universe
was inhomogeneous so that the fields (dilaton in this case)
were spacetime dependent and moreover,
the time derivative and spatial gradients were comparable. Furthermore, if
we want to follow the PBB approach, these derivatives were small, to begin
with, so that initially we are in the perturbative regime.
If one looks at the evolution, it is noted that certain patches
develop where time derivatives dominate over spatial gradients. If the
kinetic energy of the dilaton contributes a good part to  the critical density,
 the dilaton driven inflation sets in so that the patch gets blown up
and it becomes homogeneous, isotropic and spatially flat. Another important
aspect is to investigate situations when the spatial curvature is nonzero
and examine how various cosmological criteria are affected \cite{twein}.
In view of above discussions, there have been considerable activities to study
the evolution of the Universe taking into account effects of spatial curvature,
inhomogeneity and examine various aspects of pre-big bang string cosmology.\\
Recently, the principle of holography has attracted considerable attentions
 in cosmological
context  and especially in string cosmology. The
Bekenstein-Hawking entropy formula states that
the entropy associated with a black hole
is proportional to the area of the horizon. According to holography principle,
when we are dealing with a system with gravity, the degrees of freedom of that
system is bounded by the surface area of the volume, V, in which the
system resides. Recently, Fischler and Susskind \cite{holfs} examined the
issue of holography in the cosmological frame work. If we consider the
Universe as a whole and explore the consequences of holography as applied to
the black holes, we encounter difficulty in the following way. Let us
consider the FRW metric as an example. The entropy per unit comoving volume
is constant as can be inferred from the covariant conservation of stress energy
momentum tensor. Thus, if we take a large enough value for the scale factor,
the holographic bound ${S\over A} \le 1$ will be violated. It was proposed
in \cite{holfs} that one should consider the following situation while
applying holography in the cosmological scenario.
Let us consider a four dimensional spacetime manifold,M. Suppose B is the
two dimensional boundary of a spatial region ${\cal {\bf R}}$. We define
the light surface ${\cal {\bf L}}$  to be the one bounded by B and generated by
the past light rays from B towards the center of ${\cal {\bf R}}$. The
cosmological holography principle,  expounded by Fischler and Susskind,  states
that the {\it entropy} passing through ${\cal{\bf L}}$ never exceeds the
area of the bounding surface B. In the special case of adiabatic evolution
of the Universe, the total entropy of the matter in the horizon should be
smaller than area of the horizon. The issue of holography, in cosmology,
has been addressed in a general frame work recently \cite{holrev}. Moreover,
the consequences of the the principle has been studied in the cosmological
context \cite{holomod}.
Let us consider a PBB scenario in the Einstein
frame with graviton and dilaton \cite{bmphol}.
The entropy per comoving volume remains
constant when the Universe undergoes adiabatic expansion (or contraction) and
the location of the horizon is determined from the condition $ds ^2 =0$.
It follows from the covariant conservation of the stress energy tensor
that
\bea {\sqrt g}{{dp}\over {dt}} ={{d\over {dt}}}({\sqrt g}(\rho +p)) \eea
where $\rho$ and $p$ are defined from the diagonal of $T^{\mu}_{\nu}$
in the cosmological context, $g$ being the determinant of the spatial part
of the metric. It is easy to see that the comoving entropy density
remains unchanged with time through out PBB and is given by
\be S^c={{(\rho+p){\sqrt g}}\over T} \ee
where T is identified as the temperature of the fluid. For the dilatonic
matter $\rho =p$ and then
\be S^c=Const. ~(\rho g)^{1/2} \ee
where the constant is related to the Stefan's constant of the dilaton.
Now, consider a simple homogeneous PBB cosmological case where
\bea ds^2=-dt^2+\sum ({t\over {t_0}}-1)^{2\lambda _a}(dx ^a)^2 \eea
and
\bea \phi (x,t)=\phi _0-{\sqrt 2}{\sqrt{ (1-{\sum {\lambda _a ^2}})}}{\rm ln}
({t\over {t_0}}-1) \eea
Here $\{\lambda _a \}$ are independent of $x$ and they satisfy Kasner
condition
\bea \sum {\lambda _a}=1, ~~~~ \sum {\lambda _a ^2}=\rho ^2, ~~~~~ {1\over 3}\le
\rho ^2 \le 1 \eea
Now the volume is given by
\be V^c_H={\Pi}_a(X^a_H) \ee
and $X^a_H$ is given by
\bea X^a_H={{t_0({t\over {t_0}}-1)^{1-\lambda _a}}\over {(1-\lambda _a)}} \eea
The area of the surface bounding the volume enclosed by the horizon
is given by
\bea A_H=\{ \Pi_a(X^a_H)({t\over {t_0}}-1)^{\lambda _a} \}^{2\over 3} \eea
Having obtained the expressions for the volume factor and the area, we
compute the holographic ratio to be
\bea {S\over A}={{\sigma}^{1\over 2} \over {l_p ^2}} {{1\over {2\sqrt \pi}}}
{{(1-\rho ^2) ^{1\over 2}}\over {\Pi (1-\lambda _a)^{1\over 3}}} \eea
We have now introduced the Stefan's constant, $\sigma$, for dilaton
explicitly and  the Planck length makes its appearance in the above
equation so that $\sigma$ is dimensionless. Since the exponents
$\{\lambda _a \}$ appearing in the definition of the line element satisfy
the two constraints, the holographic ration $S\over A$ is function of
only one exponent. Therefore, by eliminating two of the $\lambda$'s
we can write the ratio as an expression involving $\sigma, \rho$ and one
of the $\lambda _a$'s, call it Y.
\bea
{S\over A} ={{ \sigma ^{1\over 2}}\over {l_p ^2}}} {{1\over {2\sqrt \pi}}}
{{(1-\rho ^2)^{1\over 2}}\over {[(1-Y)Y^2+{{(1- \rho ^2)}\over 2}]^{1\over 3}}
\eea
The minimization and maximization of the denominator as a function of Y will
give us an upper bound and lower bound on the holographic ratio,
respectively  as given below,
\bea
{S\over A} {\le} {\sqrt {{\sigma}\over {\pi}}}{{1\over {2l_p ^2}}}
{{{\sqrt {1-\rho ^2}} }\over {
{[(11/3 -3\rho ^2 -{{(3\rho ^2 - 1)^{3/2}}\over {3{\sqrt 2}}})/9]^{1/3}}}}
\eea
\noindent and
\bea
{S\over A} {\ge} {\sqrt {{\sigma}\over {\pi}}}{{1\over {2l_p ^2}}}
{{{\sqrt {1-\rho ^2}} }\over {
{[(11/3 -3\rho ^2 +{{(3\rho ^2 - 1)^{3/2}}\over {3{\sqrt 2}}})/9]^{1/3}}}}
\eea
Note the appearance of $\sigma$, the Stefan constant for the dilaton, the
holographic ratio which could be determined in principle from string
theory. It is quite interesting that the ratio could be bounded from
above as well as from below in this case.\\
The field of string cosmology and especially PBB string cosmology is still
developing and there are important issues to be resolved. We would like
to make a few comments before closing this section. One would like to
study phenomenological aspects of PBB cosmology and compare and contrast
the results of PBB scenario with standard inflationary models. A lot
of work has been done in this direction and we refer the reader to
the review article for more detailed references. It might be worth while
to point out a few features of PBB cosmology. In the standard inflationary
scenario, the Universe goes through de Sitter type phase where the curvature
remains constant; however, in the PBB inflation the curvature changes
with time. Thus the quantum fluctuation of background fields are amplified
in different modes with different spectra.  Therefore, some of the
distinct features of PBB cosmology could be experimentally observed
in gravitational wave detectors in future. Similarly, the axion spectrum
has been computed in PBB \cite{axion} and if detected, it will be another
test of this scenario. There have been attempts to provide an understanding
of galactic and intergalactic magnetic fields from the point of string
cosmology. Since dilaton couples to the gauge field the mechanism for
amplification of the magnetic field is related to the growth of the
dilaton. It is quite possible that some of the predictions of PBB cosmology
will be tested by  on going experiments and/or by the experiments planned
in the near future.

\section{Summary and Conclusion}

\setcounter{equation}{0}

\def\theequation{\thesection.\arabic{equation}}

We have made some efforts to convey to the reader some of the interesting
and important developments in string theory through this article.  It is
not possible to include all  developments in the field in diverse directions 
in an article of this nature. A global perspective of string theory
is contained in the article of John Schwarz \cite{johnrev} in this volume.
We may recall that the research in string
theory has stimulated progress in other fields such as mathematics,
quantum field theory and statistical mechanics of lower dimensional systems
to mention a few areas. We have seen that string theory has made very
important contributions to our understanding of the physics of the black holes.
As we have mentioned, for a special class of black holes, the 
Bekenstein-Hawking entropy formula could be derived from an underlying
microscopic theory. Similarly, the nature of the Hawking radiation from a
stringy black hole, slightly away from extremality, could derived from 
the theory.\\
We have noted that, there are intimate connections between the five string
theories. Some of them are inter related through
dualities in ten dimensions and some are related in lower dimensions. Thus
it is recognized that dualities have a special role in our understanding
of string dynamics. Moreover, there are increasing evidence that
there is a unique, fundamental theory and the five perturbatively
consistent string theories are various phases of the fundamental theory. It
is argued that M-theory might be that theory and the low energy effective
action of M-theory is to be identified with the eleven dimensional
supergravity theory. In this context, we discussed the M(atrix) model
proposal to show that the model captures many important features
of M-theory.   \\
Recently, the conjecture due to Maldacena has attracted considerable attention
since it provides an important connection between supergravity on the bulk
and the supersymmetric gauge theories living on the boundary. The connection
between type IIB theory on $AdS_5 \times S_5$ and $N=4$ supersymmetric gauge
theory on the boundary has been at the center of attention. Furthermore,
there are interesting developments in the study of theories on $AdS_3$ and
corresponding two dimensional conformal field theories.\\
One of the most important achievements of string theories has been to
address important issues in quantum gravity and provide answers to
some of the puzzles. However, the theory is yet to provide a satisfactory
answer to the cosmological constant problem. The cosmological constant
is a parameter in physics which is measured to be closest to zero. It plays
a dual role. When we look at it from the point of view of macroscopic physics,
the smallness of the constant conveys to us that the Universe is very large
and it is flat. On the other hand, it is expected that, the cosmological 
constant, like other parameters in Nature, should be explained from a 
microscopic theory and the short distance physics, i.e. quantum gravity,
will explain the smallness of the cosmological constant. Therefore, one
expects that string theory will be able to resolve this outstanding problem
\cite{cosed,bband}.
The author along with his collaborators had made an attempt in this
direction \cite{kms}. It is expected that string theory will provide us
clues to understand the creation of the Universe and the evolution of the
Universe in early epochs. Indeed, string cosmology has attracted considerable
attention is recent years; however, we have not included discussions on
this topic in this article due to limitations of space. Indeed, string
cosmology makes several predictions which might be subjected to 
experimental tests in next few years \cite{cosmoven}.\\

\centerline{{\bf Acknowledgments}}
\vspace{.3in}

I would like to thank Professors P. Majumdar, S. Panda, B. Sathiapalan and 
J. H. Schwarz and A. Sen
for their suggestions  and advice. I would like to thank the Yukawa Institute
for Theoretical Physics, Professors T. Maskawa, M. Ninomiya and R. Sasaki for 
their very warm hospitality where most of this article was written.

%%%% REFERENCES BEGIN HERE %%%%%%%
%%%%%%%%%%%%%%%%%%%%%%%%%%%%%%%%%%%%%%%%%%%%%%%%%%%%%%%%%%%%%%%%

\end{document}